\documentclass[
reprint,
aps,
prc,
nofootinbib,
floatfix
]{revtex4-2}
\usepackage{amsmath,amssymb}
\usepackage[pdftex]{graphicx}
\usepackage{subcaption}
\usepackage{comment}
\usepackage{placeins}
\usepackage{caption}
\graphicspath{{./Images/}}
\usepackage{dcolumn}
\usepackage{bm}
\usepackage{slashed}
\usepackage{physics}
\usepackage{xfrac}
\usepackage{color,xcolor}
\usepackage[scr=boondox,  
            cal=esstix]   
           {mathalpha}
\usepackage[colorlinks = true,
            linkcolor = blue,
            urlcolor  = magenta,
            citecolor = blue,
            anchorcolor = blue]{hyperref}%

\allowdisplaybreaks[1]
\newcommand{\new}[1]{}

\newcommand{\reorg}[1]{}

\begin{document}

\preprint{APS/123-QED}

\title{The Effect of Hadronic Matter on Parton Energy Loss}

\author{Ritoban Datta}
\affiliation{Department of Physics and Astronomy, Wayne State University, Detroit, MI 48201.}

\author{Abhijit~Majumder} 
\affiliation{Department of Physics and Astronomy, Wayne State University, Detroit, MI 48201.}

\date{\today}

\begin{abstract}
Modified thermal distributions (dispersion relations) are introduced within both the \texttt{MATTER} and \texttt{LBT} event generators used to describe jet modification in a heavy-ion collision, within the \textsc{Jetscape} framework. 
Hard partons, propagating through dense matter, scatter off the partonic substructure of the medium, leading to stimulated emission, accompanied by recoiling medium partons. 
We introduce a simple modification, a multiplicative $(1 + a/T)$ correction to the dispersion relation of quarks and gluons (equivalent to an effective fugacity). 
This leads to calculated transport coefficients (e.g. $\hat{q}/T^3$) showing the expected behavior of depreciating at lower temperatures, including within the hot hadronic gas. 
This simple modification recovers the light-like dispersion relations at high temperatures, and introduces an excess depreciation factor for parton populations at lower temperatures, allowing partonic energy loss and recoil calculations to be extended into the hadronic phase.
This modified distribution, in combination with initial state cold nuclear matter effects (shadowing), is used to simultaneously describe the nuclear modification factor and elliptic anisotropy of jets and leading hadrons, over multiple centralities and collision energies. 
\end{abstract}


\maketitle



\section{Introduction}
\label{sec:Introduction}
The quenching of hard jets as they propagate through the dense medium created in a relativistic heavy-ion collision is expected to yield information on the partonic substructure of the produced quark-gluon plasma (QGP)~\cite{Majumder:2010qh,Cao:2024pxc}. There is now somewhat wide agreement that, at temperatures at least an order of magnitude above the transition temperature $T_c$, the thermodynamics of the QGP can be described using Hard Thermal Loop (HTL) perturbation theory~\cite{Braaten:1989kk,Braaten:1989mz,Frenkel:1989br}. This notion of agreement is based entirely on the comparison of the results of higher order HTL perturbation theory~\cite{Andersen:2011sf} with simulations from Lattice QCD~\cite{Bazavov:2009zn,Borsanyi:2010cj}. As a result, simulations of jet modification which incorporate some part of HTL dynamics~\cite{He:2015pra,Zigic:2021rku,JETSCAPE:2022jer}, should (and do) show considerable agreement with 
experimental data~\cite{JETSCAPE:2023hqn}.

To simultaneously describe jet and high transverse momentum (high-$p_T$) leading hadron suppression, requires a multi-stage energy loss framework~\cite{JETSCAPE:2017eso,Caucal:2018dla,Modarresi-Yazdi:2024vfh}. 
This involves an earlier (high virtuality) stage where scattering and stimulated emission are rare~\cite{Wang:2001ifa,Kumar:2019uvu,Majumder:2013re,Cao:2017qpx}, compared to vacuum like emission, and a later (lower virtuality) stage dominated by scattering and stimulated emission~\cite{Baier:1994bd,Baier:1996kr,Arnold:2002ja,Jeon:2003gi,Majumder:2009ge,He:2015pra,Cao:2016gvr,Schenke:2009gb} (some hard partons may experience vacuum like emission on exit from the medium).
The scattering of hard partons in the medium, leading to stimulated emission, is typically quantified using the transverse momentum diffusion transport coefficient:
\begin{align}
    \hat{q} = \frac{1}{N_{events}} \sum_{events} \frac{1}{L} \sum_{i = 1}^{N \in L } (k_\perp^i)^2,
    \label{eq:qhat-general}
\end{align}
which represents the event averaged transverse momentum squared gained by a parton per unit length traversed in the plasma (where the parton may scatter an arbitrary $N$ times, per event, within the length $L$). 

        In addition, allowing for scale evolution of the transverse momentum exchange transport coefficient $\hat{q}$~\cite{Baier:2002tc,Majumder:2012sh,Kumar:2020wvb}, that modulates stimulated emission~\cite{Kumar:2019uvu,Kumar:2025rsa}, yields simulations that can simultaneously address jet and leading hadron nuclear modification factors, over a range of centralities ($0 - 50$\%), from top RHIC to all LHC energies~\cite{JETSCAPE:2022jer}, for light and heavy flavors~\cite{Cao:2017crw,JETSCAPE:2022hcb}. 
These same simulations, with parameters unchanged, also compare well with jet substructure observables~\cite{JETSCAPE:2023hqn}, photon triggered jet distributions~\cite{JETSCAPE:2024nkj}, and even photon triggered jet substructure observables~\cite{JETSCAPE:2025rip}.

All of the above simulations invoke the requirement of consistent recoil and scattering in jet modification, first introduced in Refs.~\cite{Zapp:2012ak,He:2015pra,He:2018xjv}: Thermal partons, which stimulate radiation from jet partons via scattering, may get converted into new jet partons, if these thermal partons are struck with sufficient momentum transfer from the evolving jet. 
Thus, the thermal partons in the medium modify jet distributions in 3 ways: By 4-momentum exchange between jet and thermal partons, by inducing more radiation from the developing shower, and by adding partons from the medium to the shower. 
These ``recoiling" partons may further undergo induced radiation, and excite more recoil partons out of the medium. 

In most calculations, the scattering in the medium that induces radiation is parameterized using the transport coefficient $\hat{q}$. In calculations with recoil, $\hat{q}$ is derived from the scattering kernel. 
If an HTL distribution is used to compute $\hat{q}$~\cite{Arnold:2008vd}, this should also yield the distribution of recoil partons after the scattering with the jet. 
These recoil partons, may undergo further scattering and energy loss, leading to the generation of more recoils. 
These are hadronized along with the original shower partons, and then clustered within jets. 
The distribution of incoming thermal partons, prior to scattering off the jet, are then subtracted as energy-momentum holes within each jet cone (thereby conserving energy-momentum). 
Thus, the distribution of recoils and holes, in addition to stimulated radiation, affects \emph{jet} modification, 
while high-$p_T$ leading hadrons are only affected by scattering (which, on average, depends on the  $\hat{q}$ calculated using the distribution of holes and recoils). 
In this way, leading hadron and jet modification are described simultaneously, from the same simulation.

In spite of the success of the multi-stage recoil based framework of jet modification, a close examination of these efforts highlights slight tensions in the comparison with data, particularly when jets and leading hadrons (over a range of collision energy and centrality) are compared simultaneously to results from the same simulations~\cite{JETSCAPE:2024cqe}. The majority of the tension originates from leading hadrons with a $p_T  < 30$~GeV. 
The results of these simulations also tend to gradually fail in comparison to data at centralities above $50\%$, when the hadronic phase in a heavy-ion collision is as large as, or larger than, the deconfined phase. 
In addition, these simulations have typically produced a lower elliptic anisotropy $(v_2)$ for both leading hadrons at high $p_T$ and jets~\cite{Park:2019sdn,JETSCAPE:2020hue}. 

In this paper, we will demonstrate that all three issues above may be resolved if one simply introduces initial state nuclear modification of parton distribution functions (shadowing)~\cite{Eskola:2009uj,Eskola:2016oht} and a slight (realistic) modification to the distribution of quark and gluon quasi-particles, allowing for the extension of the recoil based jet modification formalism past the deconfined phase, into the hadronic phase.  
Remarkably, none of the prior efforts by the \textsc{Jetscape} collaboration~\cite{JETSCAPE:2021ehl,JETSCAPE:2022jer,JETSCAPE:2022hcb,JETSCAPE:2023hqn,JETSCAPE:2024cqe,JETSCAPE:2024nkj,JETSCAPE:2025rip} included any nuclear shadowing effects. 
In this effort, we will further demonstrate that the inclusion of partonic energy loss in the dense hadronic stage provides an overall improvement of the nuclear modification factor ($R_{AA}$), for both hadrons and jets, particularly for hadrons with $p_T<30$~GeV. It improves the comparison with jet and hadron $R_{AA}$ in more peripheral events, as well as for the elliptic anisotropy for high $p_T$ hadrons and jets.

In order to clear any confusion early, we distinguish the concept of \emph{parton energy loss in the hadronic phase}, from \emph{hadronic energy loss in the hadronic phase}. 
The latter is the modification of the portion of the jet that has already hadronized within the hot hadronic gas, due to hadronic rescattering. 
While this can also be calculated, and is expected to affect the modification of low $p_T$ hadrons and jets, it will be ignored in the current effort, in order to focus on the modification of leading hadrons and jets due to initial state shadowing and partonic energy loss in the hadronic phase. 

The highest $p_T$ where hadronic energy loss in a hadronic medium can affect a jet can be estimated by a simple formation time argument: 
Say a hard parton, with energy $E$, starts from the center of a central heavy-ion collision. %
It will hadronize when its virtuality reaches $Q_{had} \gtrapprox \Lambda_{QCD}$. 
This happens after a time (= distance for a light-like parton) of  
\begin{align}
    \tau_{had} = \frac{2E}{Q_{had}^2}.
\end{align}
Based on fluid dynamical simulations used in the current effort, we assume a maximal radial size of the interacting medium (QGP + interacting hadron gas) prior to freeze-out of $L_{fo} \simeq 10$~fm. We require $c\tau_{had} > L_{fo}$ for there to be minimal effect of hadronic energy loss.
This gives, 
\begin{align}
    E_{min} \gtrsim \frac{L_{fo} Q_{had}^2}{2c} \simeq \frac{L_{fo} \Lambda_{QCD}^2}{2c} \sim 6 {\rm GeV}. 
\end{align}
To obtain the above estimate, $\Lambda_{QCD}$ is ascribed a somewhat larger value of $0.5$~GeV. 
Lower and more typical values of $\Lambda_{QCD}$ will lead to an even lower $E_{min}$.

In addition to this, we point out that recombination effects~\cite{Fries:2003vb,Molnar:2003ff,Greco:2003xt,Hwa:2002tu,Hwa:2004ng} can extend up to $10$~GeV at LHC energies. 
To steer clear of these effects that are not included in our calculation, we will only consider hadrons with $p_T > 10$~GeV and jets with $p_T > 50$~GeV. 
While jets receive contributions from hadrons at all energies, asymmetric partonic branching ensures that high-$p_T$ jets are less sensitive to the modification of the population of hadrons with $p_T \lesssim 10$~GeV. 
One should also note that recombination and hadron energy loss are competing contributions.
Recombination draws energy into a jet, from the bulk, as hadrons are formed from the fusion of constituent quarks from the hard jet and the bulk. 
In contrast, hadronic energy loss in a hadronic medium obviously causes energy flow out of the jet. 

The inclusion of all the above effects represents a major enhancement in the simulation of jet modification in heavy-ion collisions, and is beyond the scope of the current effort. 
In this paper, we will focus solely on the effect of partonic energy loss in the hadronic phase and shadowing in cold nuclear matter prior to the hard scattering. 
Interestingly, both of these are hadronic effects. 
The remainder of the paper is organized as follows: In Sec.~\ref{sec:Kinetic_Description}, we outline the basic \texttt{MATTER+LBT} model of multi-stage energy loss within the \textsc{Jetscape} framework, in Sec.~\ref{sec:Reduced_Temperature} we outline the new reduced parton distributions which will allow the default \textsc{Jetscape} simulations to be extended within the hadronic phase, in Sec.~\ref{sec:simulation} we outline the basic components of our simulations, highlighting the addition of nuclear shadowing, in Sec.~\ref{sec:results} we demonstrate the improved agreement between our simulations and experimental data, highlighting the first simultaneous description of the $R_{AA}$ and $v_2$ of jets and hadrons, followed by a summary and outlook towards future work in Sec.~\ref{sec:summary}.


\section{Basic Model}
\label{sec:Kinetic_Description}


In this section, the basic ingredients of the multi-stage energy loss model are outlined. 
We include this in the interest of completeness. 
All descriptions will be kept very brief, as most of this material has appeared elsewhere. 
The informed reader, familiar with Refs.~\cite{JETSCAPE:2019udz,JETSCAPE:2022jer}, can skip this section. 
The first subsection describes the inclusion of nuclear effects on parton distribution functions, and the ensuing modification to the hard scattering cross section. 
The next subsection will describe the high virtuality stage of the outgoing partons, followed by a section on the low virtuality stage of parton evolution.

\subsection{Nuclear Effects and hard scattering}
The production of a high-momentum parton in a nucleus–nucleus $(A+B)$ collision is described within perturbative QCD using the factorized expression:
\begin{align}
    \frac{d\sigma_{AA\rightarrow k}}{d^2p_T dy} &= \sum_{ij}\int dx_1 dx_2 f_{i/A}(x_1,Q^2) 
    f_{j/B}(x_2, Q^2)\nonumber \\
    & \times \frac{d\hat\sigma_{ij \rightarrow k}(Q^2)}{d \hat{t}} .
    \label{eq:FactorizedCrossSection}
\end{align}
where $f_{i/A}(x,Q^2)$ denotes the nuclear parton distribution function (nPDF) for parton flavor $i$, in nucleus $A$, evaluated at momentum fraction $x$ and factorization scale $Q$. 
In a nuclear environment, the parton distributions differ from those in free protons due to shadowing effects~\cite{Arneodo:1992wf}. 
These modifications are generally encoded as a nuclear mass number dependent multiplicative factor,
\begin{equation}
    f_{i/A} ( x , Q^2 ) = R_i^A ( x , Q^2 ) f_{i/p} ( x , Q^2 )
 \label{eq:PartonModification}
\end{equation}
where the nuclear shadowing factor $R^A_i(x,Q^2)$ is extracted from global analyses of an ever growing data set, \emph{e.g.} EPS09~\cite{Eskola:2009uj}, EPPS16~\cite{Eskola:2016oht}, nCTEQ15~\cite{Kovarik:2015cma} \emph{etc.} 
These analyses constrain nuclear corrections over a wide range of Bjorken-$x$ and $Q^2$ using deep-inelastic scattering, Drell–Yan production, neutrino scattering, and $p$-$A$ data.

Nuclear modifications significantly alter the distribution of partonic momentum fractions involved in the hard scattering, particularly in the small-$x$ region relevant at LHC energies. 
Shadowing reduces the parton densities at $x \lesssim 10^{-2}$, while anti-shadowing enhances them at intermediate $x \sim 10^{-1}$. 
At larger $x$, not relevant for the current work, EMC and Fermi motion effects become important. 
These effects modify both the longitudinal momentum fractions of the incoming partons and the transverse-momentum spectrum of the outgoing hard partons produced in the initial collision, thereby reshaping the distribution of partons that seed the subsequent jet evolution. 
Such modifications to the initial flux of hard partons are essential for a consistent interpretation of nuclear modification observables, including $R_{AA}$, jet yields, and high-$p_T$ hadron suppression.

To gauge the type of nuclear modification, we make very simple estimates of the $x$ range probed by either experiment considered: 
Given a minimum hadron $p_T \gtrsim 10$~GeV, the minimum $x$ probed at LHC ($5.02$~TeV) is given as $x \gtrsim 10/2510 \sim 4 \times 10^{-3}$. 
The largest value of $x$ is probed in the highest $p_T$ jets with $p_{T,jet}^{max} \sim 1000$~GeV, yielding an $x$ value of $x \lesssim 0.4$ (given the larger range in rapidity at the LHC, one can access even higher values of $x$). 
Alternatively, at top RHIC energies of $0.2$~TeV, the $x$ range that contributes is obtained from the lowest $p_T$ hadron that is calculated ($\sim 10$~GeV) to the highest $p_T$ of jets possible $\sim 40$~GeV. 
As a result, RHIC collisions probe a much smaller range of $0.1 \lesssim x \lesssim 0.4$.

In the present study, the nuclear modified parton spectrum generated through 
Eqs.~(\ref{eq:FactorizedCrossSection},\ref{eq:PartonModification}) forms the initial condition for the multistage in-medium jet evolution. 
As in Refs.~\cite{JETSCAPE:2019udz,JETSCAPE:2022jer}, we run a \texttt{Pythia} hard event with initial state radiation (ISR) and multi-particle interaction (MPI) turned on, but final state radiation (FSR) turned off. The evolution of the jet in the final state is described in the subsequent subsection.
Unlike in some other \textsc{Jetscape} simulations~\cite{JETSCAPE:2024dgu}, we do not allow for final state energy loss in $p$-$A$ collisions.

\subsection{Inelastic Scattering and Medium-Induced Radiation}

Exiting from the hard interaction, multiple hard partons from the \texttt{Pythia} MPI routines enter the multi-stage evolution routines in \textsc{Jetscape}. 
In this effort, we allow for a high virtuality routine represented by \texttt{MATTER}, which simulates the radiative (and scattering based) evolution of partons with a virtuality much larger than the saturation scale of the medium ($\hat{q}\tau$, where $\hat{q}$ is the transverse broadening transport coefficient~\cite{Baier:2002tc,Majumder:2012sh,Kumar:2020wvb} and $\tau$ is the formation time of each parton), and a lower virtuality module represented by \texttt{LBT} that simulates the evolution of partons at the saturation scale.

The first step in final state jet modification requires each parton to be processed by the \texttt{MATTER} generator. 
Each parton is then ascribed a virtuality $\mu$ which lies between $Q_{min} = 1$~GeV and $Q_{max} = p_T/2$, by sampling the ratio of Sudakov form factors:
\begin{align}
    R (\mu^2, Q_{max}^2, Q_{min}^2 ) &= 
    \frac{ \Delta( Q_{max}^2, Q_{min}^2 )   } { \Delta ( \mu^2 , Q_{min}^2 ) },
\end{align}
where, the in-medium Sudakov form factor is given as, 
\begin{align}
    \Delta (\mu_1^2 , \mu_0^2 ) &= \int\limits_{\mu_0^2}^{\mu_1^2} \frac{ d \mu^2 }{ \mu^2} \int\limits_{ \sfrac{\mu_0^2}{\mu_1^2}  }^{1 - \sfrac{\mu_0^2}{\mu_1^2} } dz \frac{\alpha_S(\mu^2)} {2 \pi } \tilde P (z,\mu^2,p^+).
\end{align}
In the equation above, the light-cone momentum of the parton is given as $p^+ = \sfrac{ [ p^0 + |\vec{p}| ]}{\sqrt{2}}$ and $\tilde{P} (z,\mu^2,p^+)$ is the full medium modified splitting function, given as 
\begin{align}
   \tilde{P} &= P(z) + \!\!\!\int\limits_{0}^{3\tau_f^+} \!\!\!d \zeta^+  
    \frac{\hat{q} ( r^+ + \zeta^+ ) }{\mu^2 z(1-z)} 
    \left[ 4\sin^2{ \left(\frac{\zeta^+ }{2\tau_f^+} \right)} \right].
    \label{eq:medium_modified_split}
\end{align}
In the equation above, $P(z)$ is the leading order (un-regulated) Altarelli-Parisi splitting function (including the color factor). 
The positive light-cone direction is defined by the direction of the hard parton, hence $\zeta^+ = \sfrac{ ( \zeta^0 + \vec{\zeta}\cdot\vec{p} \,)}{\sqrt{2}} $. Similarly, the light-cone formation time is given as,
\begin{align}
    \tau_f^+ = \frac{2 p^+}{ \mu^2}.
\end{align}
It was pointed out in Ref.~\cite{Sirimanna:2021sqx}, that there are several corrections to Eq.~\eqref{eq:medium_modified_split}, however, when $\zeta^+$ is integrated over $3 \tau_f^+$, the distance integral is approximately equal to that from Eq.~\eqref{eq:medium_modified_split}.

A close examination of Eq.~\eqref{eq:medium_modified_split}, shows that the medium portion of the equation diminishes at large values of the virtuality $\mu$. 
There are two contributions here; one is the explicit factor of $\mu^2$ in the denominator, and the other is the shrinkage of $\tau_f^+$ at large $\mu$. 
At each time step of the calculation, partons with a virtuality $\mu \geq Q_{\rm SW} = 2$~GeV are retained in \texttt{MATTER} for further evolution, while those partons whose virtualities are below this cut-off are transferred to the \texttt{LBT} module. 
If partons within the \texttt{LBT} module exit the dense medium (at $T_0$), with a virtuality $\mu \geq Q_{min}$, they are reintroduced within the \texttt{MATTER} module for further evolution. 
All partons continue to evolve, i.e., split (with or without scattering), until they reach $\mu \leq Q_{min}$. 

Given a start time and location for each parton, and a formation time $\tau_f^+$, a Gaussian distribution of split times is sampled and assigned to each parton (see ~\cite{Majumder:2013re,Cao:2017qpx,JETSCAPE:2022jer} for details). 
The entire simulation evolves in time steps of $0.1$~fm/c. 
Once the simulation time reaches the split time of a parton in \texttt{MATTER}, the parton is split by sampling the medium modified splitting function in Eq.~\eqref{eq:medium_modified_split} to obtain a value of $z$. 
The daughter partons are assigned virtualities between $Q_{min}$ and $z\mu$ or $(1-z)\mu$. 
The remainder of the parent parton's virtuality generates the transverse momentum between the daughter partons, 
\begin{align}
    \ell_\perp^2 = z(1-z) \mu^2 - z \mu_1^2 - (1-z) \mu_2^2. 
\end{align}

Along with the $1\rightarrow 2$ process, the \texttt{MATTER} generator also incorporates a scattering with an in-medium parton. 
A thermal distribution at the ambient temperature $T$ is sampled to obtain the momentum of an on-shell parton. One of the outgoing virtual daughter partons then undergoes a scattering off this on-shell parton. 
The framework tracks both the outgoing \emph{recoil} parton, which becomes part of the shower, and the incoming thermal parton, now designated as a \emph{hole}. 
Holes are hadronized separately from the shower and recoil parton. 
If a hole is found within a jet, its energy momentum are subtracted from the jet. 
This ensures one aspect of medium response and energy momentum conservation~\cite{He:2015pra}.

As the virtuality of the partons drops below the transition scale of $Q_{\rm SW} \simeq 2$~GeV, the parton enters the transport-dominated regime, where the Linear Boltzmann Transport (\texttt{LBT}) model governs both elastic momentum exchange and inelastic, medium-induced radiation. 
The full evolution is described by the Boltzmann equation,
\begin{equation}
    p^\mu \partial_\mu f_a(x,p) 
    = C_{\mathrm{el}}[f] + C_{\mathrm{inel}}[f],
    \label{boltzman}
\end{equation}
where $C_{\mathrm{inel}}[f]$ encodes $1\!\rightarrow\!2$ parton splittings triggered by scatterings with thermal quasi-particles. 
\texttt{LBT} implements an inelastic rate inspired by the higher-twist~\cite{Wang:2001ifa} formalism,
\begin{equation}
    \frac{d\Gamma^{a\to bc}}{dz \, dl_\perp^2 d\zeta}
    =
    \frac{\alpha_s\,C_a}{2\pi}
    \frac{P_{a\to bc}(z)}{l_\perp^4}
    \,\hat{q}\, 4
    \sin^2\!\left(\frac{\zeta}{ 2\,\tau_{f} } \right),
    \label{eq:diff_Gamma}
\end{equation}
where $z$ is the momentum fraction of one daughter, $l_\perp$ is the transverse momentum generated in the splitting, and $P_{a\to bc}(z)$ is the vacuum DGLAP splitting function. 
The factor $\hat{q}$ is determined by elastic scatterings, and the $\sin^2$ term encodes Landau–Pomeranchuk–Migdal (LPM) interference. The time/distance traveled by the hard parton from its origin is $\zeta$.

The probability for generating a medium-induced splitting during a time interval $\Delta t$ is obtained by integrating the differential rate,
\begin{equation}
    P_{\mathrm{inel}}(E,T)
    = 
    \Delta t 
    \sum_{b,c}
    \int dz 
    \int dl_\perp^2\,
    \frac{d\Gamma^{a\to bc}}{dz\, d l_\perp^2 d\zeta}.
\end{equation}
Whenever a splitting is generated, the momenta of the daughter partons are sampled from sampling Eq.~\eqref{eq:diff_Gamma} above. 
Along with the $1 \rightarrow 2$ process, the \texttt{LBT} simulator also generates a scattering with an in-medium parton. 
The treatment of recoils and holes is the same as in the \texttt{MATTER} case above. 
%

As in Ref.~\cite{JETSCAPE:2022jer}, in both \texttt{MATTER} and \texttt{LBT}, we use running couplings $\alpha_S(\bar{\mu}^2)$, such that,
\begin{align}
  \alpha_S(\bar{\mu}^2) = \frac{12 \pi}{ 11N_C - 2n_f} \frac{1}{\log{ \frac{ \bar{\mu}^2 }{\Lambda^2}   } } ,  
\end{align}
where, $\Lambda$ is chosen to ensure $\alpha_S ( 1 GeV^2 ) = \alpha_S^{\rm fixed}$, which is a parameter of the model, representing the coupling at the medium scale. In addition, in the \texttt{MATTER} stage, $\hat{q}$ has an additional scale dependence: $\hat{q} (\mu^2) = f(\mu^2) \hat{q} (\mu^2 = Q^2_{\rm SW})$, where $f(\mu^2 \leq Q_{SW}^2) = 1$, and continues to drop at larger values of $\mu^2 > Q_{\rm SW}^2$, with the same form as in Ref.~\cite{JETSCAPE:2022jer}. Similarly, 
$\hat{q} (\mu^2 = Q^2_{\rm SW})$ is the $\hat{q}$ in the low virtuality stage with a formula discussed in the subsequent section [see Eq.~\eqref{eq:qhat-htl}]. 

In the context of this work, the inelastic rate depends sensitively on the density and momentum distribution of thermal scatterers in the medium. 
The modified thermal distributions that will be introduced in Sec.~\ref{sec:Reduced_Temperature} therefore directly influence both $\hat{q}$ and the phase-space integral governing medium-induced radiation.


\subsection{Elastic Scattering and Medium Response}


Elastic $2\!\to\!2$ scatterings between a hard parton and thermal constituents of the medium play a central role in determining transverse momentum broadening, the transport coefficient $\hat{q}$, and the redistribution of energy and momentum into the surrounding medium. Although emissions in the \texttt{MATTER} stage are dominated by virtuality-ordered, vacuum-like branchings, rare elastic scatterings with medium quasi-particles are included through the transverse-momentum broadening kernel and the medium-modified Sudakov factor. These interactions contribute small but finite transverse kicks that broaden the splitting kinematics; however, their probability is strongly suppressed at high virtuality. Once the parton enters the low-virtuality transport regime, elastic collisions become the dominant mechanism for momentum exchange with the medium.

The elastic processes are described by the collision term $C_{\mathrm{el}}[f]$ in the Boltzmann equation~\eqref{boltzman},
\begin{align}
    C_{\mathrm{el}}[f_a] &= 
    \sum_{b,c,d}
    \frac{1}{2E_a}
    \int d\Gamma_b \, d\Gamma_c \, d\Gamma_d \,
    |\mathcal{M}_{ab\to cd}|^2 
    \nonumber \\
    &\quad \times 
    \left[
    f_c f_d (1 \pm f_a)(1 \pm f_b)
    - f_a f_b (1 \pm f_c)(1 \pm f_d)
    \right] \nonumber \\
    & \quad \times (2\pi)^4 \delta^{(4)}(p_a + p_b - p_c - p_d),
    \label{eq:boltzmann-elastic-kernel}
\end{align}
where $d\Gamma_i = d^3p_i / \left[(2\pi)^3 2E_i\right]$ is the Lorentz-invariant measure and $|\mathcal{M}_{ab\to cd}|^2$ is the squared QCD matrix element. The $(1\pm f)$ terms encode Bose enhancement and Pauli blocking of the thermal final states.
The scattering rate $\Gamma_{ab\to cd}$ determined from this elastic collision term defines the probability per unit time for a hard parton of type $a$ to undergo an elastic scattering with a thermal parton $b$, producing final-state partons $c$ and $d$.

A key component of elastic interactions in a hot medium is the treatment of~\emph{medium response} arising from the momentum transferred to thermal constituents. 
When a fast parton transfers momentum $k^{\mu}$ to a thermal constituent, the struck medium particle is promoted to the jet population as a \emph{recoil}, while a negative-energy \emph{hole} carrying momentum $q^\mu$ is recorded to represent the depletion of the corresponding thermal mode. 
This bookkeeping ensures exact local energy--momentum conservation and provides a microscopic description of the hydrodynamic wake generated by the jet. 
Such recoil and hole excitations have been shown to be essential for accurately modeling jet profiles, jet shapes, and reconstructed jet substructure in heavy-ion collisions~\cite{He:2015pra}.

The strength of elastic momentum exchange is highly sensitive to the underlying phase-space distribution of thermal medium partons. 
Consequently, any modification of this distribution alters both the accumulated broadening from elastic scatterings and the inelastic radiation rate that depends on it.
A modification of this distribution, in the hadronic phase of the collision, will be outlined in the subsequent section.


\section{Modified Distributions near Transition}
\label{sec:Reduced_Temperature}


In the preceding section, we outlined the basic working of the \textsc{Jetscape} framework with the \texttt{MATTER} and \texttt{LBT} generators combined to act on partons over a range of virtualities. 
The entirety of the interaction with the medium depends on the thermal distribution of medium partons that scatter off the jet partons. 
These partons, after recoiling off the jet partons become part of the evolving shower. 
The scattering rate off these medium partons, also generates the transverse momentum transport coefficient $\hat{q}$ [Eq.\eqref{eq:qhat-general}]. 
The consistent scattering and recoil formalism has been quite successful in describing a wide range of data from jet and leading hadron suppression~\cite{JETSCAPE:2022jer,JETSCAPE:2022hcb,JETSCAPE:2023hqn,JETSCAPE:2024nkj}. 
In this section, we will discuss a deviation from the thermal distribution of medium partons that allows the recoil framework to be extended to the hadronic phase. 

\subsection{The Reduced Distribution Model}
\label{subsec:reduced-dist-model}

A weakly interacting plasma of quarks, anti-quarks and gluons at a temperature $T$ will yield a momentum ($p$)-dependent distribution for a quark or an anti-quark as,  
\begin{align}
    \tilde{n}^q(\vec{p}) = \frac{d^q_s d^q_N}{e^{E_p/T} + 1}.  
\end{align}
In the equation above, $d^q_s = 2$ is the spin degeneracy, and $d^q_N = N_c$ (the number of colors) is the color degeneracy. Where $E_p = \sqrt{ |\vec{p}|^2 + m^2}$. Similarly, the gluon distribution may be expressed as, 
\begin{align}
    n^g(\vec{p}) = \frac{d^g_s d^g_N}{ e^{E_p/T} -1 },
\end{align}
where, $d^g_s = 2$, and $d^g_N = N_c^2 - 1$. In the equation above, The energy $E_p = \sqrt{p^2 + m^2} \simeq p$ for light flavors.

The distributions above represent the populations of quarks and gluons in full equilibrium, at weak coupling (ignoring thermal masses). Calculating the dimensionless jet transport coefficient $\hat{q}/T^3$ for a hard quark with an energy $E \gg T$, using the above distributions, we obtain,
\begin{align}
    \frac{\hat{q}}{T^3} = 16 C_F  C_A  \zeta(3) \pi \alpha_S(\Lambda_{ET}) \alpha_S(m_D) \ln\left( \frac{\Lambda_{ET}^2}{ m_D^2} \right), \label{eq:qhat-htl}
\end{align}
where, $\Lambda_{ET} = \mathcal{N} E T $, with $2 \lesssim \mathcal{N} \lesssim 6$. This produces a $\hat{q}/T^3$ that gradually increases with decreasing temperature until the expression is no longer valid in the vicinity of the phase transition. In the above expression, $m_D$ is the minimum momentum transfer, or the inverse screening distance in the medium. In a thermalized plasma, this is the Debye mass.

To be clear, the rise of $\hat{q}/T^3$ with dropping temperature, followed by a sudden cutoff at the transition temperature, is \emph{unphysical}. 
There are now calculations from Lattice-QCD that indicate that $\hat{q}/T^3$ may have a shallow peak followed by a drop as the temperature approaches $T_C$ and continues below it~\cite{Kumar:2020wvb}. 
Thus, there is every indication that $\hat{q}/T^3$ does not go to zero sharply at $T_C$ but instead gradually reduces to zero within the hadronic phase. 

In a QCD plasma, the coupling between in-medium quarks and gluons gets stronger as the temperature approaches the transition temperature from above, eventually becoming large enough that all quarks and gluons are confined within colorless hadrons. This has little bearing on the couplings used in the above expression, as the coupling between the hard parton and a parton in the medium can remain weak, even if the coupling between medium partons becomes stronger at lower temperature. 
In Hard Thermal Loop effective theory~\cite{Braaten:1989mz,Frenkel:1989br}, the dispersion relation (for $p \gtrsim T$) is modified as $E = \sqrt{ p^2 + m_T^2 }$, where $m_T$ is the thermal mass given as $ m_T = n gT$, where the value of $n$ depends on whether the parton is a quark or a gluon. In the HTL limit, the thermal mass gap tends to drop with temperature, i.e. partons with a momentum $p$ have a lower energy at a lower temperature. As a result, the inclusion of thermal masses (or terms proportional to $T$) will not lead to a reduction of the degrees of freedom at a given temperature.
Thus, with the use of either on-shell dispersion relations or with the inclusion of thermal masses, $\hat{q}/T^3$ will continue to rise at lower temperatures.

To mimic the effect of a reduction in the color carrying degrees of freedom with decreasing temperature, we introduce the model of the ``Reduced Temperature Distributions": The quark and gluon distributions are modified as, 
\begin{align}
    \tilde{n}^q_R(\vec{p}) &= \frac{d^q_s d^q_N}{e^{\frac{E_p}{T - a(T) } } + 1}, \nonumber \\
     n^g_R(\vec{p}) &= \frac{d^g_s d^g_N}{ e^{ \frac{E_p}{T - a(T) } } -1 },
     \label{eq:ModifiedDistribution}
\end{align}
where, $a(T) \ll T$ is a small reduction in the effective temperature of the bare quark and gluon distributions. We further impose the condition that $a(T)$ can at most be a constant or a decreasing function of $T$, i.e.,
\begin{align}
    a(T) = a_0 + \frac{a_1}{T} + \frac{a_2}{T^2} + \dots. 
\end{align}
As a result, $a(T)/T \rightarrow 0$ as $T\rightarrow \infty$, and the regular thermal distributions are recovered at high temperature.

Taking the simplest case of $a(T) = a_0$, and $a_0 \ll T$, we obtain a simpler form of the distributions by Taylor expansion, 
\begin{align}
    \tilde{n}^q_R(\vec{p}) &= \frac{d^q_s d^q_N}{e^{\frac{E_p}{T} + \frac{a_0 E_p}{T^2} } + 1}, \nonumber \\
     n^g_R(\vec{p}) &= \frac{d^g_s d^g_N}{ e^{ \frac{E_p}{T} + \frac{a_0 E_p}{T^2} } -1 }.
\end{align}
The above expressions are valid in the region where $T \gg a_0$. This will yield a consistency check for our calculations applied at temperatures $T < T_C$ (the phase transition temperature). 

In the simple form above, the modification can also be understood as a modification of the thermal dispersion relation where $ E(T) = \sqrt{p^2 + m^2} ( 1 + a_0/T) $. 
Such a modification is very different from modifications in the HTL approximation, where the modification can be understood as a series of additive contributions with positive powers of $T$, while here, we are expanding in negative powers of $T$. 
We remind the reader that at high temperatures, several times $T_C$, contributions from these negative powers of $T$ will diminish and have no effect. 
Thus, our corrections do not conflict with the HTL expansion. 
They are only significant near or below the transition temperature $T_C$. 
The reader may question why we did not use an additive contribution such as $ E(T) = \sqrt{p^2 + m^2} + a/T$. 
The main reason for this is \emph{simplicity}, the former multiplicative form allows for many simplifications in the expression of the scattering rate and transport coefficients, as we will demonstrate below. These will allow the reader to immediately ascertain the effect of the new factor on the energy loss of hard jets.

The inclusion of these reduced densities will allow energy loss calculations in the \texttt{MATTER}+\texttt{LBT} framework of \textsc{Jetscape} to be extended to temperatures much below the transition temperature. As a result, we call this model ``reduced \texttt{MATTER} + reduced \texttt{LBT} (r-\texttt{MATTER}+r-\texttt{LBT})". The affect of this modification of the distributions on the scattering rate and transport coefficients will be discussed in the subsequent subsections.

\subsection{Scattering Rate}

\begin{figure}[t]
    \centering

    \begin{subfigure}[t]{0.95\columnwidth}
        \centering
        \includegraphics[width=\linewidth]{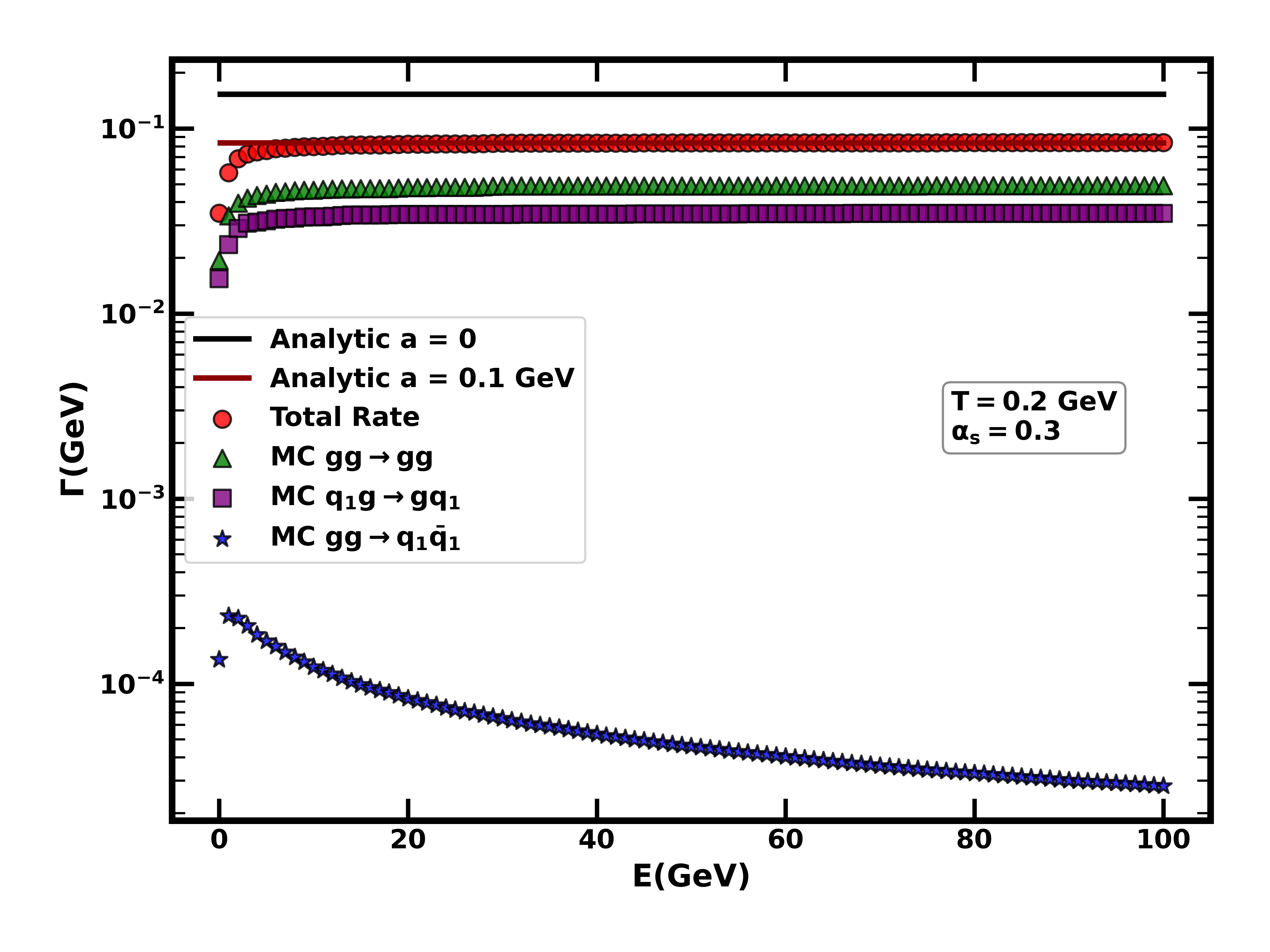}
        \label{fig:GluonRateT1}
    \end{subfigure}

    \vspace{-0.8cm}

    \begin{subfigure}[t]{0.95\columnwidth}
        \centering
        \includegraphics[width=\linewidth]{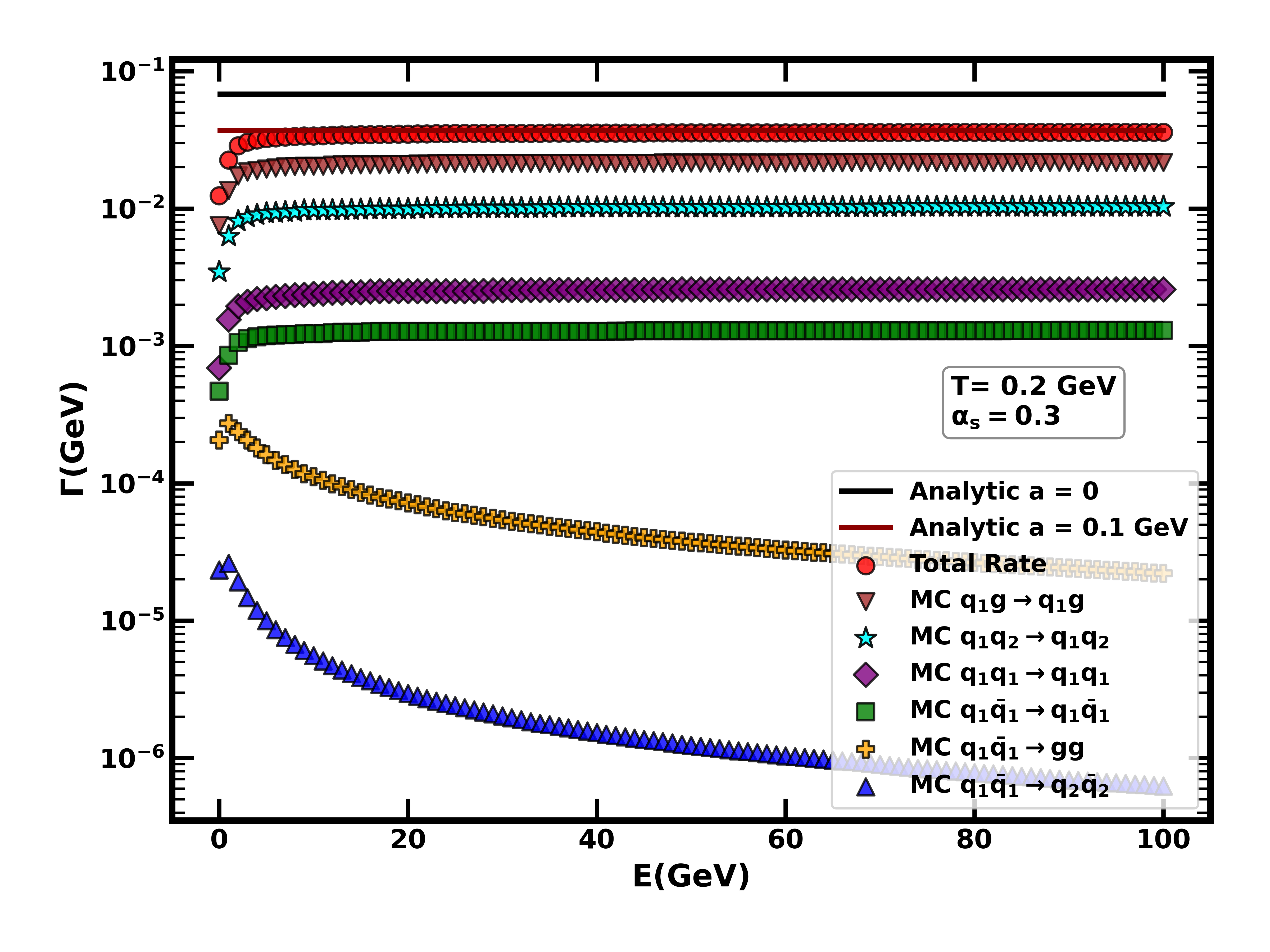}
        \label{fig:QuarkRateT1}
    \end{subfigure}

    \caption{
    Elastic scattering rates $\Gamma_{ab\rightarrow cd}$ for a gluon (top)
    and a quark or antiquark (bottom) as functions of projectile energy at
    fixed temperature $T=200\,\mathrm{MeV}$ (with $\alpha_s=0.3$).
    Solid curves show equilibrium rates; dark-red curves and Monte-Carlo
    points correspond to the modified recoil distribution of
    Eq.~\eqref{eq:ModifiedDistribution}.
    }
    \label{fig:rate_vs_energy}
\end{figure}
\begin{figure}[t]
    \centering

    \begin{subfigure}[t]{0.95\columnwidth}
        \centering
        \includegraphics[width=\linewidth]{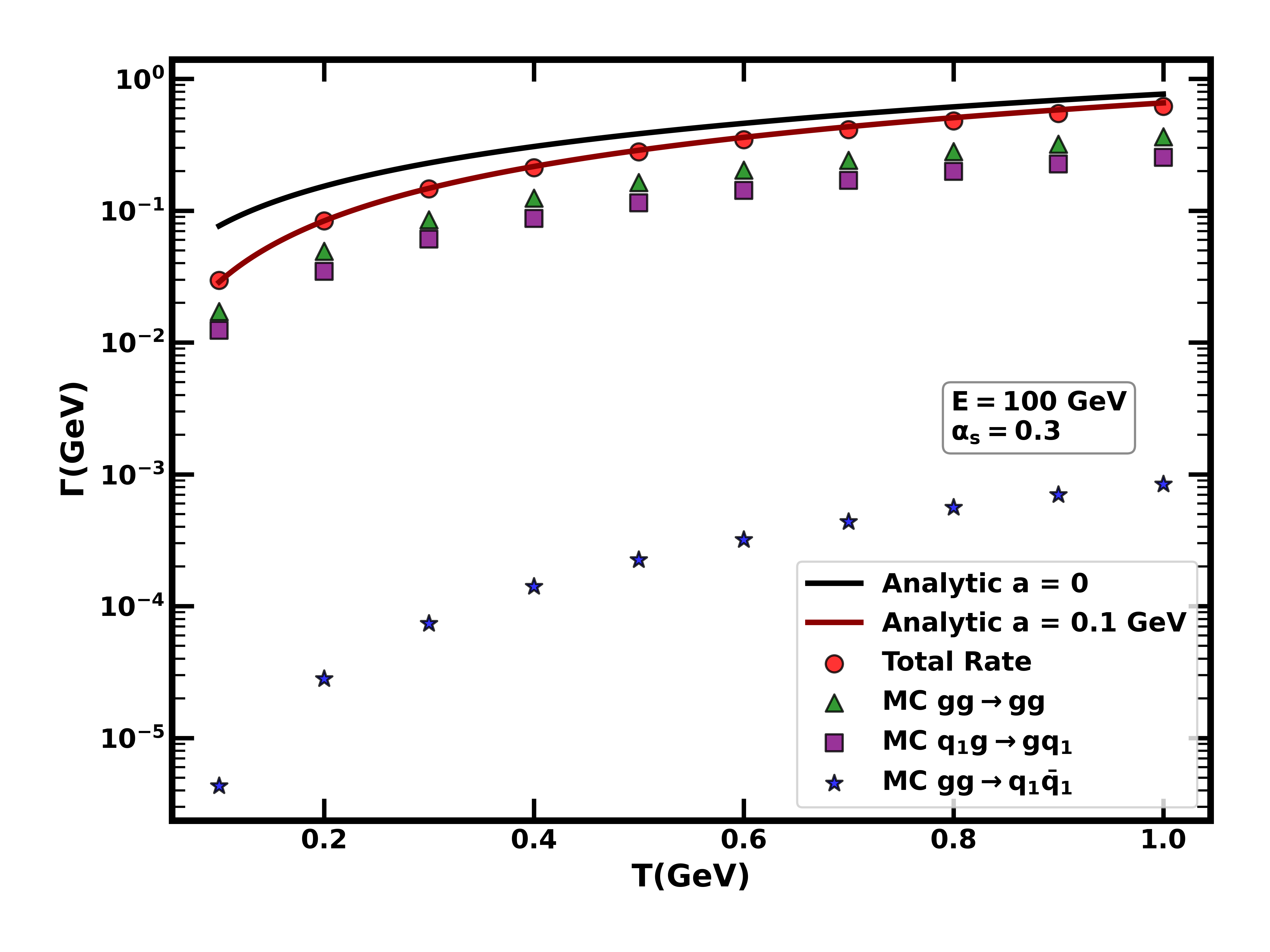}
        \label{fig:RateGluonT}
    \end{subfigure}

    \vspace{-0.8cm}

    \begin{subfigure}[t]{0.95\columnwidth}
        \centering
        \includegraphics[width=\linewidth]{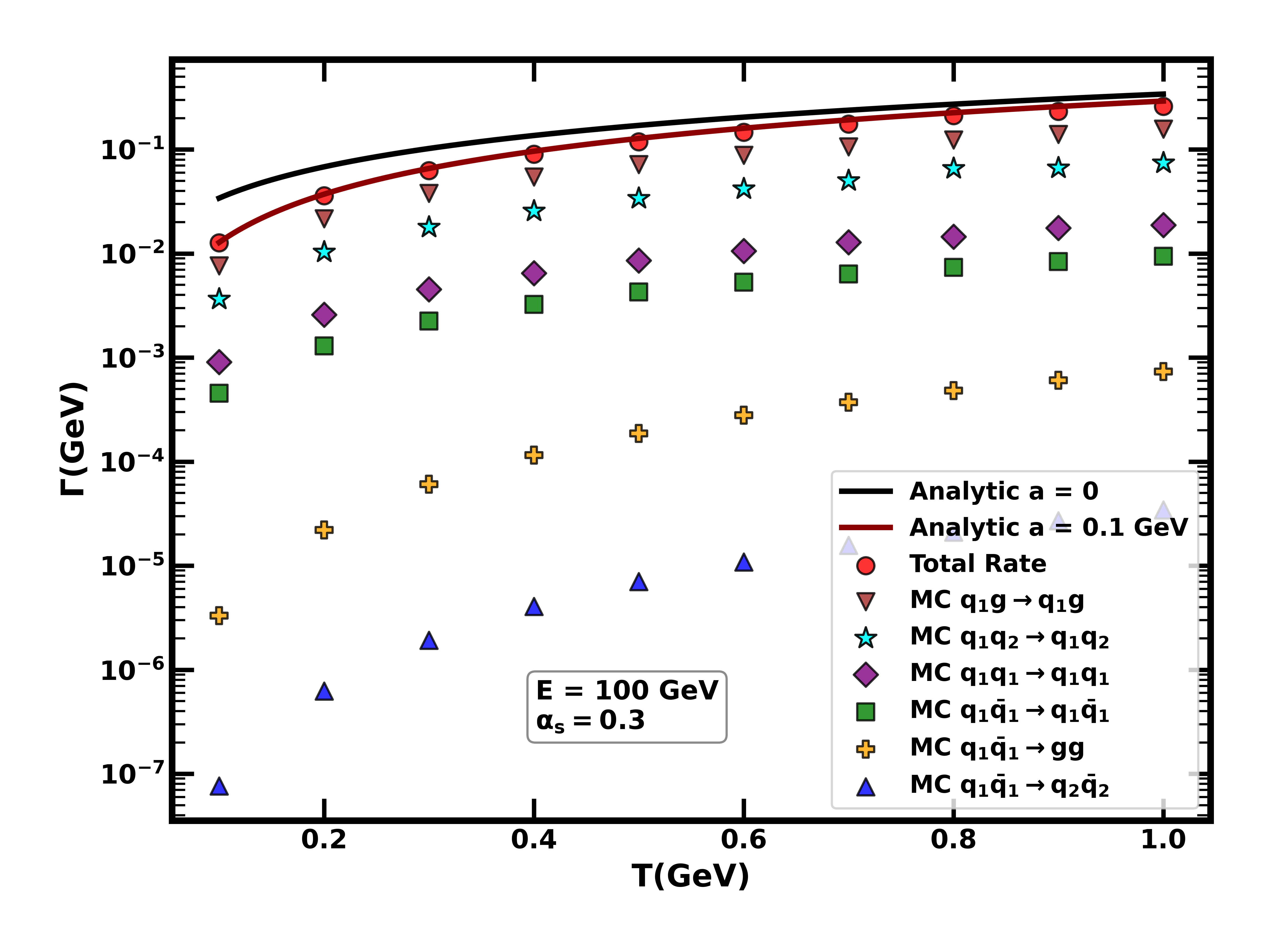}
        \label{fig:RateQuarkT}
    \end{subfigure}

    \caption{
    Elastic scattering rates $\Gamma_{ab\rightarrow cd}$ for a gluon (top)
    and a quark or antiquark (bottom) as functions of temperature at fixed
    projectile energy $E=100\,\mathrm{GeV}$ (with $\alpha_s=0.3$).
    Solid curves show total scattering rates of quark/gluon; dark-red curves and Monte-Carlo
    points correspond to the modified recoil distribution of
    Eq.~\eqref{eq:ModifiedDistribution}.
    }
    \label{fig:rate_vs_temperature}
\end{figure}

The elastic scattering rate for a hard parton $a$ interacting with a thermal
medium constituent $b$ can be obtained from simplifying Eq.~\eqref{eq:boltzmann-elastic-kernel} as
\begin{align}
\Gamma_{ab\to cd}(E_1,T)
&= \frac{g_b}{2E_1}
\int d\Gamma_b \, d\Gamma_c \, d\Gamma_d |\mathcal{M}_{ab\to cd}|^2 
\nonumber\\
&\times
f_b(p_2,T) S_2(s,t,u) \nonumber \\
&\times (2\pi)^4\delta^{(4)}(p_1+p_2-p_3-p_4).
\label{eq:Gamma-general}
\end{align}
Medium dependence enters this expression through the thermal distribution $f_b(p_2,T)$ and the
Debye screening mass $m_D(T)$, which regulates the momentum exchange.  
The regulator $S_2(s,t,u)$ enforces the Debye-screened kinematic constraints 
$m_D^2 \le |t| \le s - m_D^2$ together with the requirement $s \ge 2 m_D^2$, 
ensuring that only momentum transfers above the screening scale contribute~\cite{Zapp:2008gi,Auvinen:2009qm}.

For the modified distributions introduced in
Eq.~\eqref{eq:ModifiedDistribution}, the Debye mass becomes
\begin{equation}
m_{D,\mathrm{mod}}^2(T)
=
\frac{g^2 T^2}{3}
\left( N_c + \frac{N_f}{2} \right)
\frac{1}{(1 + a/T)^2},
\label{eq:mDmod}
\end{equation}
which is reduced relative to the equilibrium HTL value.  
Since both $f_b$ and $m_D$ enter the scattering kernel, the total rate is
suppressed when the modified distribution is used.  
A compact analytic estimate of the temperature dependence gives
\begin{equation}
\Gamma_{\mathrm{mod}}(E_1,T)
\simeq
\frac{\Gamma_{\mathrm{eq}}(E_1,T)}
     {1 + a/T},
\label{eq:Gammamod}
\end{equation}
indicating that the suppression increases toward lower temperatures as the medium cools and the thermal population becomes increasingly depleted.

The trends in Fig.~\ref{fig:rate_vs_energy} follow directly from the modified thermal distribution introduced in Sec.~\ref{subsec:reduced-dist-model}. Fig.~\ref{fig:rate_vs_energy} shows the scattering rates of gluons (top panel) and quarks (bottom panel) as a function of the energy of the hard parton, evaluated at a fixed temperature of $T = 0.2~\mathrm{GeV}$ and a constant coupling $\alpha_s = 0.3$. The total scattering rate, shown by the solid red circles, is obtained as the sum of contributions from the individual elastic scattering channels, which are indicated by different colors and marker styles. The solid red curve represents the analytic small-angle scattering limit evaluated using the same modified distribution with a modification parameter $a = 0.1~\mathrm{GeV}$, while the black solid curve shows the corresponding analytic estimate in the absence of the modification.

At fixed temperature, the suppression of the rates is essentially a constant shift: the thermal density and the screening scale are fixed, so the modified distribution simply reduces the overall normalization of the total rate and rate of every branch.  The saturation of the rate at high projectile energy is preserved even in the presence of the modified thermal distribution.  Although the modification reduces the Debye mass, the energy scale for saturation is determined by the condition $s \gg |t_{\min}| \sim m_D^2$.  Since the modified $m_D$ differs from the standard value only by an $\mathcal{O}(1)$ factor, the resulting shift in the saturation scale, $E_{\mathrm{sat}} \sim m_D^2/(2T)$, is numerically small compared with the large energies at which saturation occurs.  Consequently, the modified and unmodified rates approach their asymptotic high-energy plateau at essentially the same projectile energy.

Fig.~\ref{fig:rate_vs_temperature} shows the total and channel-resolved scattering rates as functions of temperature for a fixed hard parton of energy $100~\mathrm{GeV}$, using the same conventions and definitions as in Fig.~\ref{fig:rate_vs_energy}.  The modified rates increase more rapidly with temperature than in the unmodified case, implying that a hard parton encounters scattering more frequently during the early, high-temperature stage of its evolution relative to the later, cooler stage.  At sufficiently high temperatures, the modified distribution becomes indistinguishable from the Bose--Einstein or Fermi--Dirac form, and the rate smoothly approaches the usual HTL expression, ensuring continuity with the expected weak-coupling behavior of the medium.

\subsection{Transport Coefficients}
\begin{figure}[t]
    \centering
%
    \begin{subfigure}[t]{0.95\columnwidth}
        \centering
        \includegraphics[width=\linewidth]{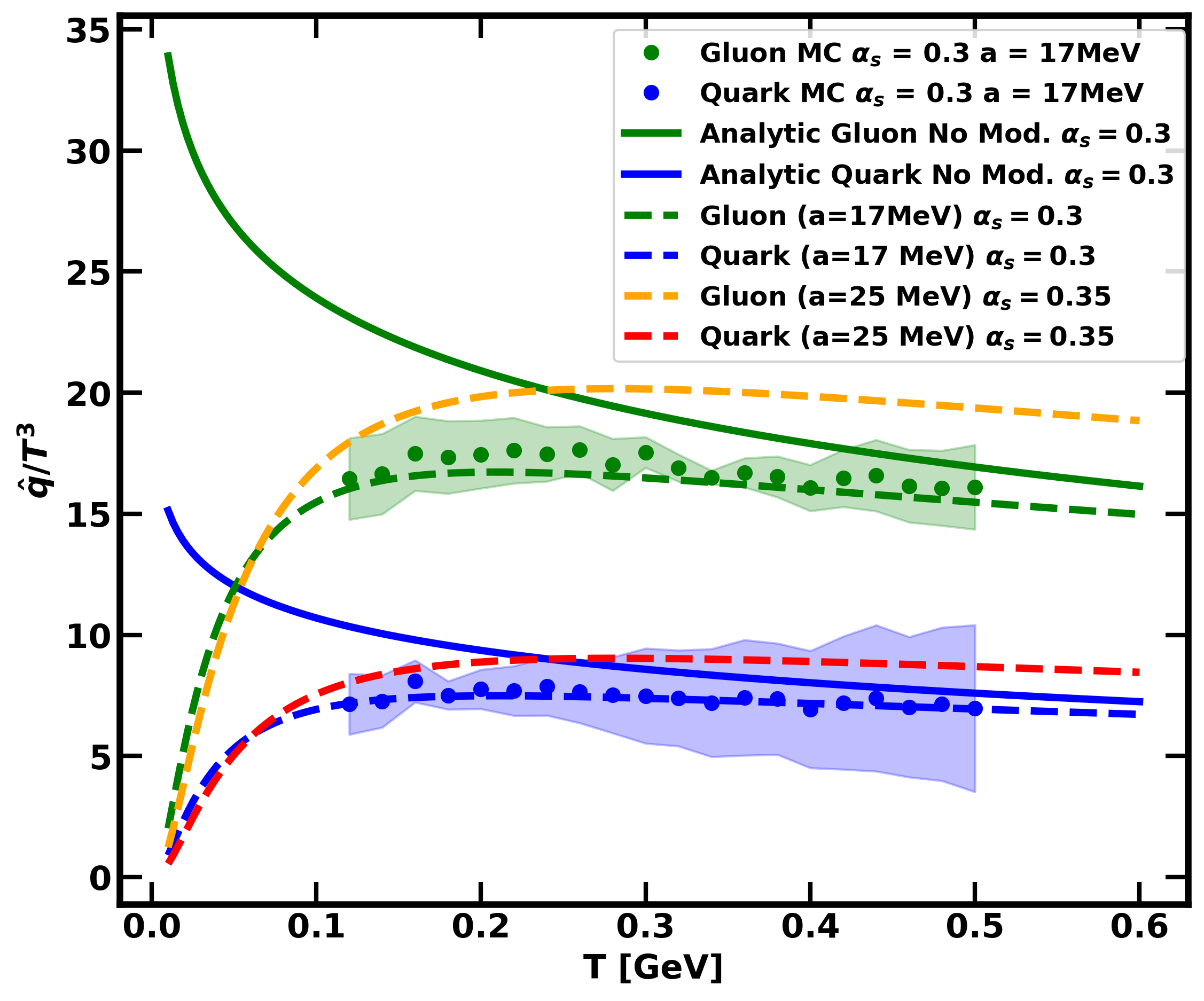}
        \label{fig:qhatT3vsT9}
    \end{subfigure}
    \caption{
    Ratio of the jet transport coefficient $\hat{q}/T^3$ as a function of temperature $T$. The solid lines show the high-temperature hard-thermal-loop (HTL) limits for gluons (green) and quarks (blue) with a fixed coupling $\alpha_s = 0.3$. The dashed green and blue lines represent the corresponding gluon and quark estimates including a medium-modification factor with $a = 17~\mathrm{MeV}$, while keeping the coupling unchanged. The yellow and red dashed lines illustrate the effect of changing the coupling to $\alpha_s = 0.35$ and the modification parameter to $a = 25~\mathrm{MeV}$. The Monte Carlo performance of the model is shown by the circular blue(quark) and green(gluon) points.
    }
    \label{fig:qhatT3vsT}
\end{figure}

\begin{figure}[t]
    \centering

    \begin{subfigure}[t]{0.95\columnwidth}
        \centering
        \includegraphics[width=\linewidth]{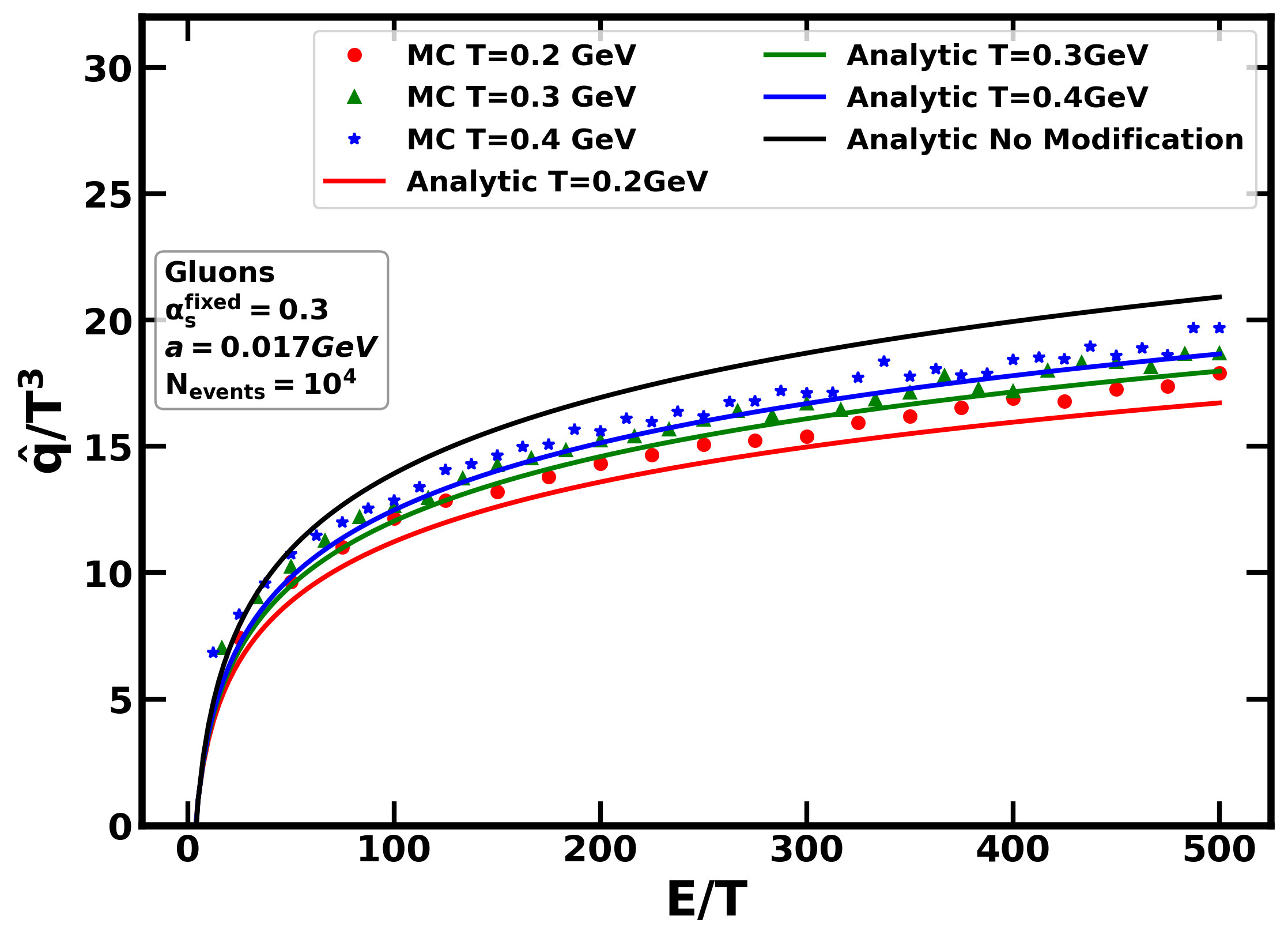}
    \end{subfigure}

    \vspace{-0.2cm}

    \begin{subfigure}[t]{0.95\columnwidth}
        \centering
        \includegraphics[width=\linewidth]{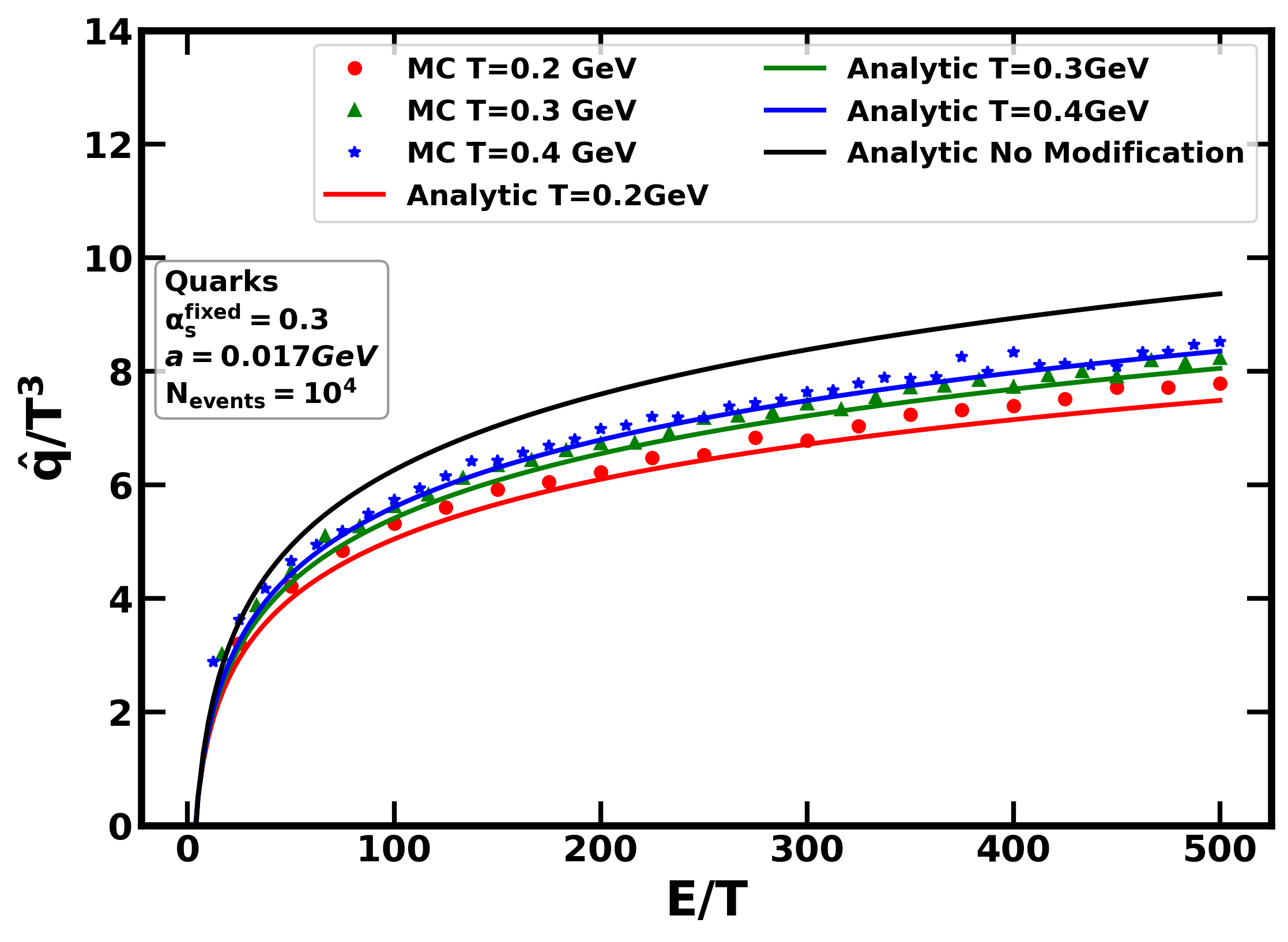}
    \end{subfigure}
    \caption{The jet transport parameter for an initial gluon~(top) or quark~(bottom) with different initial energies going through a single scattering~(solid symbols) in a uniform and static QGP medium at different temperatures. Solid lines are analytic results for a single scattering within the small-angle approximation.}
     \label{fig:qhatT3vsEbTQG}
\end{figure}
\begin{figure}[t]
    \centering

    \begin{subfigure}[t]{0.95\columnwidth}
        \centering
        \includegraphics[width=\linewidth]{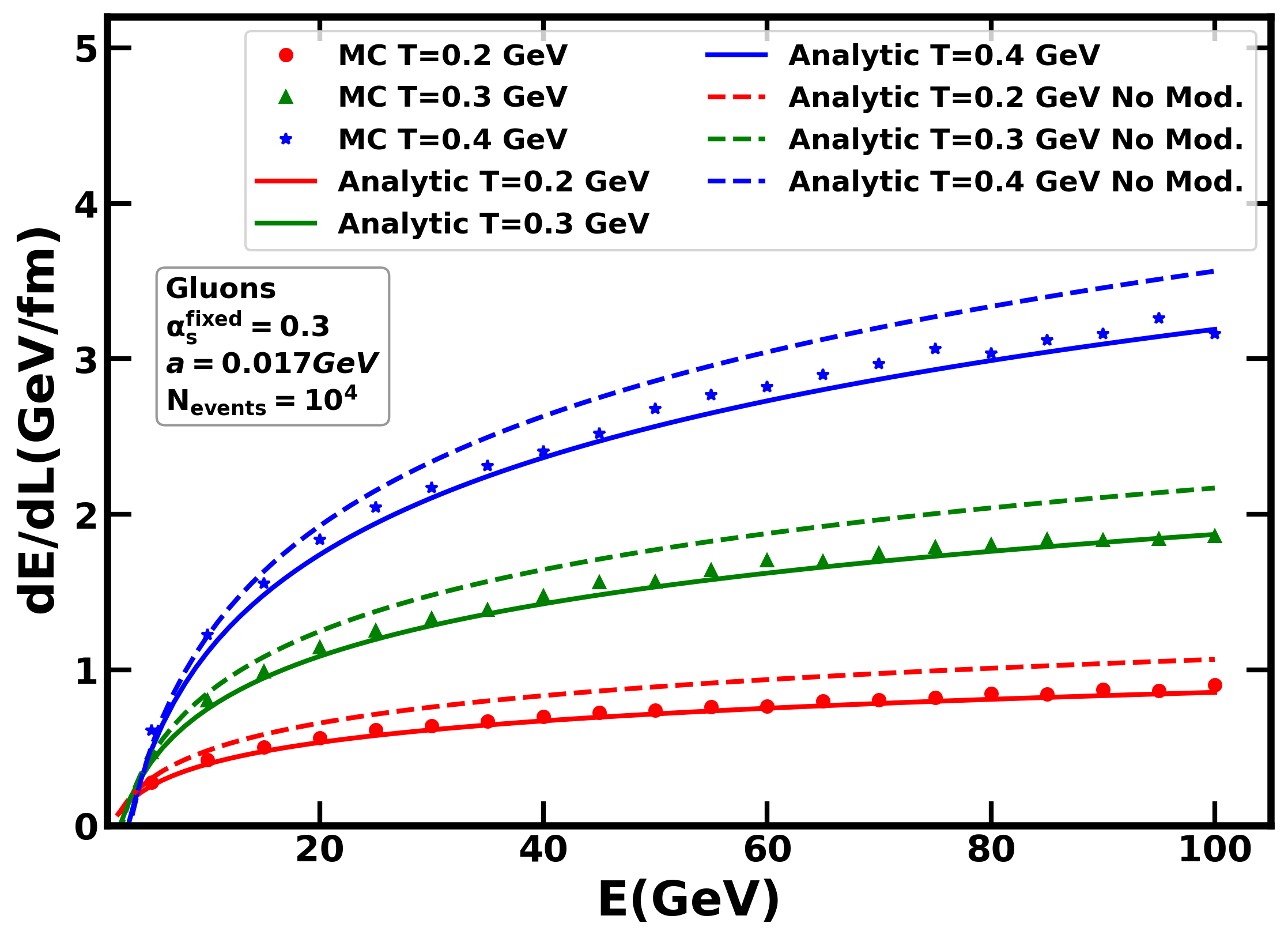}
    \end{subfigure}

    \vspace{-0.2cm}

    \begin{subfigure}[t]{0.95\columnwidth}
        \centering
        \includegraphics[width=\linewidth]{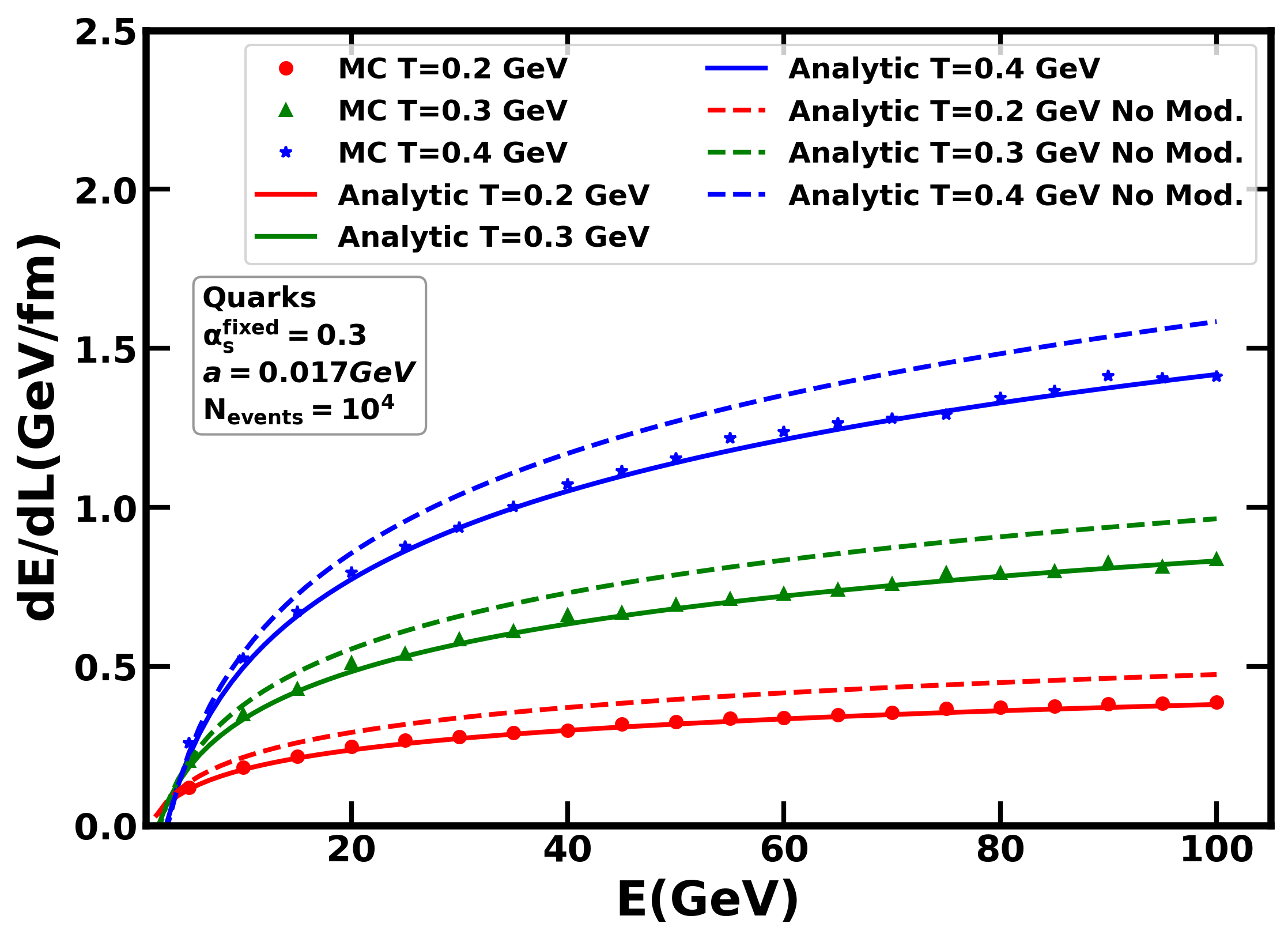}
    \end{subfigure}
    \caption{
    Elastic energy loss per unit length a gluon(top) or a quark(bottom) in a uniform and static medium with a temperature $T = 200, 300$ and $400$~MeV as a function of the initial energy from simulations of a single scattering~(solid symbols) as compared to analytic results~(solid lines) with a small-angle approximation. The dashed lines are analytic estimates for the case without reduced distributions.}
     \label{fig:qhatT3vsE}
\end{figure}

Within the small-angle approximation of the $2\!\rightarrow\!2$ cross sections, the transport coefficient $\hat{q}$, with the introduced modification, could be estimated from Eq.~\eqref{eq:qhat-general}, in the limit of small $L$ (single scattering) as, 
\begin{equation}
    \hat q \simeq \Gamma\,\langle q_\perp^2\rangle
    \simeq
    \frac{C_a}{b^3}\,\frac{42\,\zeta(3)}{\pi}\,\alpha_s^2\,T^3
    \ln\!\left(\frac{s^*}{4b\,m_{D,{\rm mod}}^2}\right),
    \label{eq:qhat_T3}
\end{equation}
with $b=1+a/T$. Thus, the new factor shows up within the $\log$ and as an overall factor. 

When $\hat q/T^3$ is plotted as a function of $T$, as in Fig.~\ref{fig:qhatT3vsT}, the modified expression produces a turnover as the temperature decreases (dashed green and dashed blue lines), rather than the purely monotonic rise obtained from the HTL-inspired form (solid green and solid blue lines). 
This behavior is broadly consistent with trends suggested by lattice-based extractions of $\hat q$, which show an enhancement at intermediate temperatures followed by a drop at lower $T$. 
It is worth noting that the HTL formalism is strictly justified only at very high temperatures where the coupling is weak; its extension to the full temperature range of the QGP, especially into the later, cooler stages of the evolution, is known to be quantitatively limited. 
At sufficiently high $T$, the modification becomes negligible, and the expression smoothly approaches the standard HTL limit, ensuring continuity with its expected domain of validity.

In Fig.~\ref{fig:qhatT3vsT}, we also present limited results from a Monte-Carlo simulation of a hard quark (blue dots) and a hard gluon (green dots) propagating through a QGP held at fixed temperature (brick simulation), and undergoing a single hard scattering (with an $\alpha_S = 0.3$ and $a=17$~MeV). 
These simulations are meant to coincide with the blue and green dashed lines, and yield a consistency check for our derived formula for $\hat{q}/T^3$ in Eq.~\eqref{eq:qhat_T3}, derived in the small angle approximation.

Fig.~\ref{fig:qhatT3vsT} also makes it clear that the reduced distributions for a given $\alpha_S$ produce a $\hat{q}/T^3$ that is always below the value generated from the un-reduced distributions. 
As a result, when reduced distributions are used in actual jet and hadron modification calculations, they will result in a reduced energy loss, unless compensated with a larger $\alpha_S$. 
An example of this is presented in Fig.~\ref{fig:qhatT3vsT}, where reduced curves for the $\hat{q}/T^3$ with an $a=25$~MeV and an $\alpha_S = 0.35$ for a quark and gluon are shown (dashed red and orange lines, respectively).

Figure~\ref{fig:qhatT3vsEbTQG} shows the jet transport coefficient $\hat{q}/T^3$ as a function of the scaled energy $E/T$ of the hard parton, with the top panel corresponding to gluons and the bottom panel to quarks. The black solid curve represents the analytic estimate obtained using the unmodified thermal distribution, while the colored solid curves (blue: $T=0.4$~GeV, green: $T=0.3$~GeV, red: $T=0.2$~GeV) show the corresponding analytic results including the modified distribution with a modification parameter $a = 17~\mathrm{MeV}$. The solid circular markers indicate results from the Monte Carlo simulation evaluated at the corresponding temperatures, providing a numerical realization of the same framework.
In the case with unreduced Fermi and Bose distributions, $\hat q/T^3$ collapses onto a nearly universal curve when plotted as a
function of $E/T$, indicating that no additional temperature dependence enters
beyond the explicit $T^3$ factor.  
With the reduced distribution, this
universality is lost: each temperature produces a distinct line, and the
spread between them reflects the extra $T$-dependence introduced by the
depleted thermal population. 
We demonstrate this temperature dependence both using a Monte-Carlo simulation and the semi-analytic estimate of Eq.~\eqref{eq:qhat_T3}.
These changes have direct implications for how a jet samples the medium. 
Because $\hat q$ is uniformly reduced, the total accumulated broadening along a given path is smaller than in the unmodified medium. 
Matching the same overall quenching therefore requires increasing $\alpha_s$.  
Once the coupling is re-tuned, the pattern of quenching shifts, making the shower parton being less quenched at early stages compared to the standard distribution, while at later stages of evolution, the shower partons receive enhanced momentum broadening compared to the standard distribution, shifting the balance of quenching across the jet’s  evolution. 

A similar suppression pattern appears in Fig.~\ref{fig:qhatT3vsE}, where the longitudinal energy loss per unit length from a single scattering is plotted as a function of $E$ for a hard parton propagating through an $8~\mathrm{fm}$ brick medium.  The suppression in longitudinal energy loss becomes more pronounced with increasing parton energy.  Its temperature dependence, however, differs from that of $\hat q$: for a fixed energy, the absolute decrease in $dE/d\lambda$ due to the modification is larger at higher temperatures, yet the relative deviation from the unmodified case diminishes as $T$ increases.  This trend mirrors the behavior of the $\hat q/T^{3}$ versus $E/T$ curves, where the impact of the modification is strongest at lower temperatures and 
gradually weakens toward the high-$T$ regime.

In both Fig.~\ref{fig:qhatT3vsT} and Fig.~\ref{fig:qhatT3vsEbTQG}, the Monte-Carlo results are generated by propagating a hard parton of fixed energy through a static, uniform brick medium of length 8~fm, at fixed temperature and vanishing chemical potential,  and extracting the mean transverse momentum squared per mean free path from  single-scattering events.


\section{Simulation Setup}
\label{sec:simulation}


In the preceding sections, we studied the effect of the reduced thermal distributions (or modified parton dispersion relations), on the scattering rate, the transport coefficient $\hat{q}$ and the energy loss per unit length $dE/dL$. At a given value of $\alpha_S$, the introduction of the multiplicative factor $(1+ a/T)$ leads to a temperature dependent reduction in each of the above three quantities. As a result, the quenching of both hadron and jet spectra will be reduced, leading to higher values of the nuclear modification factor $R_{AA}$.

With the inclusion of the reduction factor, calculations can now be extended deeper into the hadronic phase. 
Thus, the scattering and recoil framework does not stop in the vicinity of $T_C \simeq 165$~MeV, as in Refs.~\cite{JETSCAPE:2022jer,JETSCAPE:2022hcb,JETSCAPE:2023hqn,JETSCAPE:2024nkj}. 
In the current effort, the scattering and recoil framework is extended down to $T_0 \simeq 135$~MeV (This is the lowest temperature down to which the 2+1D hydro simulation can provide a spacetime temperature profile). 
The size of the reduction factor is controlled by the parameter $a$. 
Eventually, a full Bayesian calibration of the hard sector would have to be carried out with the inclusion of this $a$ factor and now also including the stopping temperature $T_0$. 
This is not the goal of the current effort. 
In this effort, we will make the simplest extension from the simulations of Ref.~\cite{JETSCAPE:2022jer}. 
Almost all the coefficients will be kept the same as in that effort. 
We will only change the stopping temperature of jet quenching to $T_0 =135$~MeV. 
We will use an $a=25$~MeV, which leads to $\hat{q}/T^3$ curves similar to the lattice QCD calculations of Ref.~\cite{Kumar:2020wvb} (which produce a shallow maximum around $T=200$~MeV). 
We will also include a standard nuclear shadowing prescription and only adjust the coupling in the medium $\alpha_S^{\rm fixed}$ to compensate for these additional effects. 
The main question explored in this paper is whether the addition of energy loss in the hadronic phase and shadowing in the initial distributions leads to a noticeable improvement in the description of the nuclear modification factor and the elliptic anisotropy. 

\subsection{Nuclear Shadowing}

We perform simulations using the multi-scale in-medium jet evolution model with \emph{reduced} \texttt{MATTER+LBT} to account for a modified distribution, allowing our calculations to proceed into the hadronic phase. 
In this study, we simulate jet events in Pb\,-\,Pb collisions at $\sqrt{s_{NN}} = 5.02~\mathrm{TeV}$. 
In order to account for the nuclear effects at hard scattering we use \texttt{EPS09} nuclear PDFs~\cite{Eskola:2009uj}. The space-time medium profile is obtained from a (2 + 1)-dimensional free-streaming evolution~\cite{Liu:2015nwa} followed by viscous hydrodynamics using \texttt{MUSIC}~\cite{Shen:2014vra}, with the \texttt{TRENTO} initial conditions~\cite{Moreland:2014oya}. While the soft-sector parameters follow the \texttt{JETSCAPEv3.5 AA22} tune, the $\alpha_S^{\rm fixed}$ parameter in the hard sector is re-tuned to reflect both the modified medium distribution and the inclusion of nuclear shadowing. All other hard sector parameters other than the stopping temperature $T_0$ and $\alpha_S^{\rm fixed}$ are fixed to the values from the \texttt{JETSCAPEv3.5 AA22} tune.

A useful way to characterize how nuclear-PDF effects modify the initial hard-parton spectrum is through its local power-law slope. At high transverse momentum, the inclusive parton spectrum may be approximated as
$d\sigma/dp_T \propto p_T^{-n}$, and a small energy loss would then modify the yield as (\emph{the following equation is only used for illustration and not used in any actual calculation}),
\begin{equation}
R_{AA}(p_T)
\simeq
\left(1 - \frac{\Delta E}{p_T + \Delta E}\right)^{n},
\end{equation}
a relation widely used in jet-quenching phenomenology
\cite{Renk:2014gwa}.  The exponent $n$ thus controls the sensitivity of the suppression to a given $\Delta E$: a steeper spectrum
(larger $n$) amplifies the impact of energy loss, while a flatter spectrum (smaller $n$) weakens it.

For jets at mid-rapidity the incoming partons carry
$x_{1}\!\approx x_{2}\!\approx 2p_T/\sqrt{s_{NN}}$, so the $p_T$ range relevant for this study ($20\!-\!200$~GeV at $\sqrt{s_{NN}}=5.02$~TeV) probes $x\!\sim\!10^{-3}$–$10^{-1}$, precisely the region where nuclear shadowing is significant in \texttt{EPS09}.  Because shadowing suppresses small-$x$ partons more strongly than moderate-$x$ partons, the initial $A$-$A$ hard-parton spectrum is tilted (compared to the $p$-$p$ spectrum): lower-$p_T$ modes are reduced more, while the higher-$p_T$ tail is affected less.  This deformation effectively reduces the slope $n$ of the spectrum, thereby decreasing its sensitivity to a fixed energy loss.  To achieve the same level of suppression in $R_{AA}$, the jet must therefore experience a larger effective $\Delta E$—implemented in practice as a higher value of the fixed in-medium coupling $\alpha_s^{\mathrm{fixed}}$.

In addition to the nuclear-PDF effects discussed above, the modification of the in-medium parton distribution also contributes to the need for a larger fixed coupling, as shown in the previous sections.  In our simulations we therefore use $\alpha_s^{\mathrm{fixed}} = 0.45$.  Finally, to ensure a consistent treatment of late-time interactions in the hadronic phase, the hydrodynamic evolution is extended down to a temperature of $135~\mathrm{MeV}$ for all centralities.

\subsection{Analysis Setup}
Unlike the nuclear modification factor, the jet azimuthal anisotropy is a statistics-intensive observable.  To obtain sufficient precision, we therefore  employ the $\hat{p}_T$–binned event generation method, which allows us to enhance statistics in the $\hat{p}_T$ intervals that contribute the largest  fluctuations.  Although this procedure yields an uneven distribution of events,  in particular with an over representation of $\hat{p}_T \!\lesssim\! 200~\mathrm{GeV}$  given the $p_T$ range of interest, the corresponding weights are properly accounted for in the final normalization.  A further subtlety arises when incorporating nuclear effects into this normalization.

\subsubsection{Nuclear modification factor}
\begin{figure}[t]
    \centering

    \begin{subfigure}[t]{0.95\columnwidth}
        \centering
        \includegraphics[width=\linewidth]{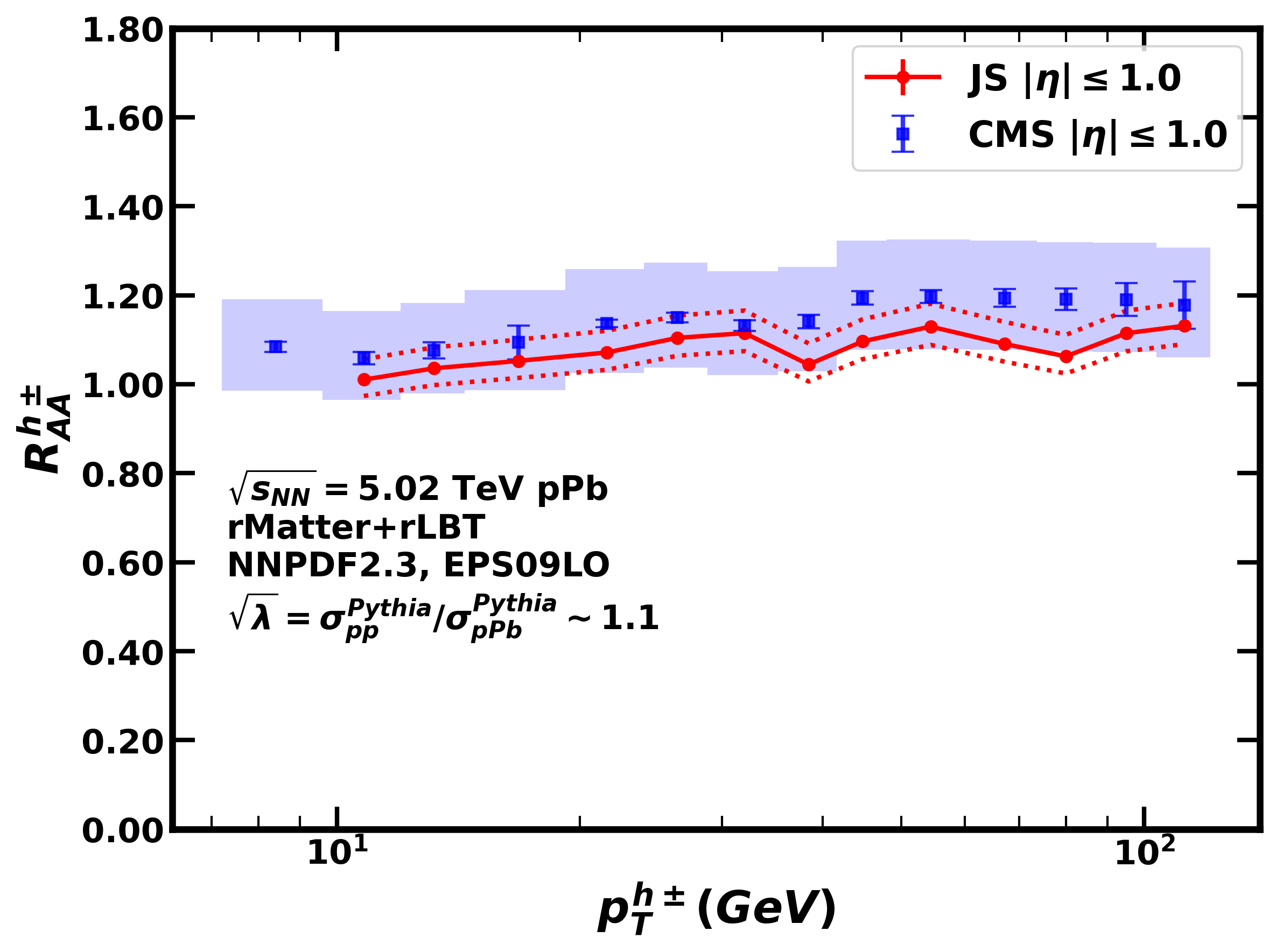}
    \end{subfigure}

    \vspace{0.1cm}

    \begin{subfigure}[t]{0.96\columnwidth}
        \centering
        \includegraphics[width=\linewidth]{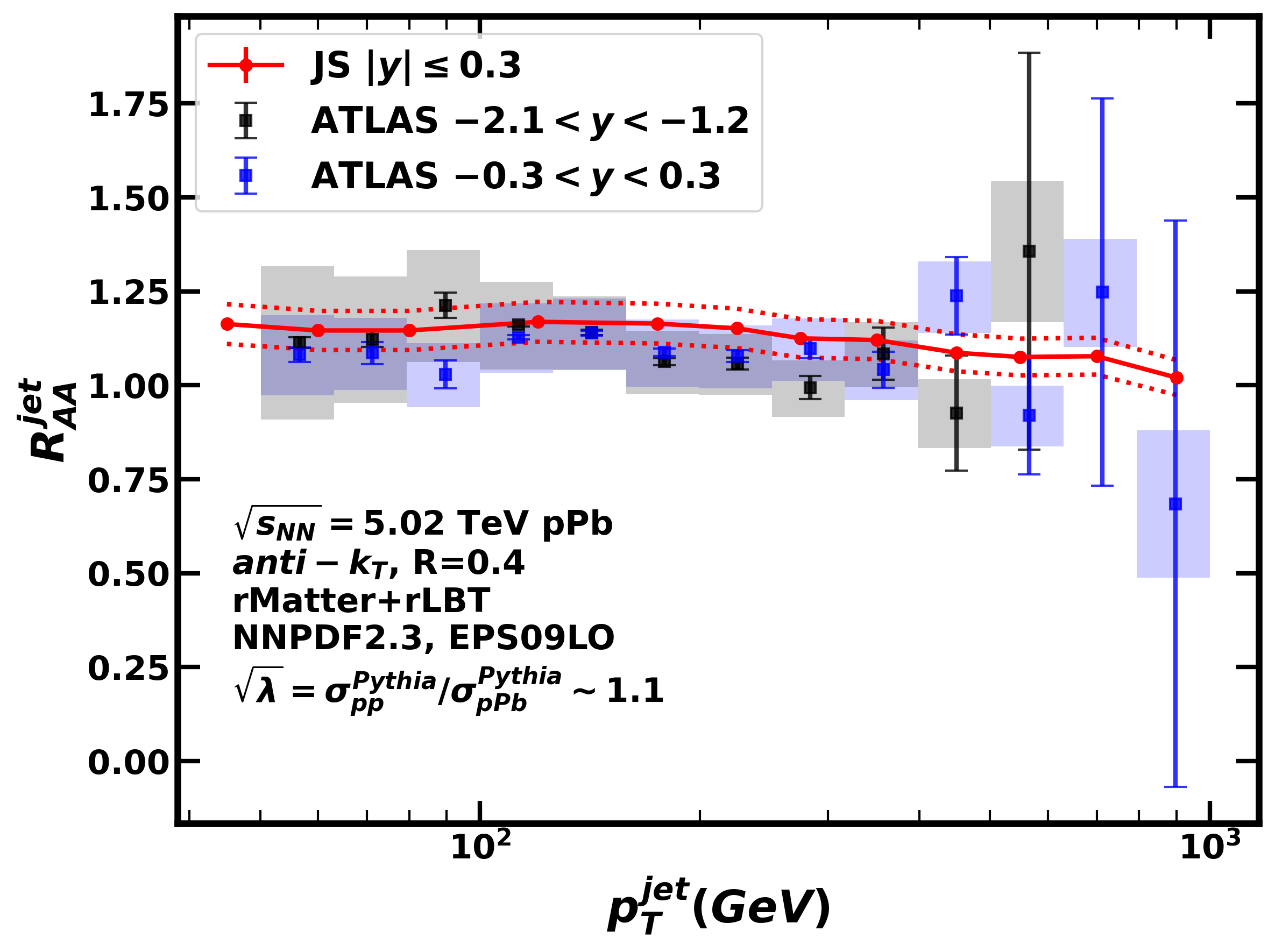}
    \end{subfigure}
    \caption{The nuclear modification factor $R_{p\mathrm{Pb}}$ for charged hadrons (top) and jets (bottom) is compared to CMS and ATLAS data (blue squares) from Refs.~\cite{CMS:2016xef,ATLAS:2014cpa} for minimum-bias $p$-$\mathrm{Pb}$ collisions at $\sqrt{s_{NN}}=5.02~\mathrm{TeV}$. The solid red curve corresponds to the ratio of differential cross sections scaled by a factor $\sqrt{\lambda}$, which serves as an effective parameter encoding nuclear shadowing effects in the total inelastic nucleon-nucleon cross section. In this study, $\sqrt{\lambda}$ closely matches the ratio of the total perturbative cross sections in $p$-$p$ and $p$-$\mathrm{Pb}$ collisions as obtained from \texttt{Pythia}. The dotted red curves indicate the uncertainty associated with $\sqrt{\lambda}$, arising from the tension between the CMS hadron and ATLAS jet data set.}
     \label{fig:pPbJetsHadrons}
\end{figure}
The nuclear modification factor ($R_{AA}$) is given as the ratio of the transverse momentum spectrum of jets/hadrons, produced in a heavy-ion ($A$-$A$) collision, to that in a proton-proton ($p$-$p$) collision scaled up by the number of expected binary collisions: 
\begin{equation}
   R_{AA}(p_T,b_{min},b_{max}) = \frac{ \int\limits_{b_{min}}^{b_{max}} \!\!d^2 b \, \frac{d^2 N_{AA}}{dp_T dy d^2b} }{ \langle N_{\text{coll}} (b_{min},b_{max}) \rangle \frac{ d^2 N_{pp}}{dp_T dy } } .
\end{equation}

In the above equation, $\langle N_{coll} (b_{min},b_{max})\rangle$ is the average number of binary nucleon-nucleon collisions in the heavy-ion system, in the chosen range of centrality or impact parameter ($b_{min} \rightarrow b_{max}$).

Within the $\hat{p}_T$ method, the evaluation of the differential cross sections are split up into multiple hard scattering (or $\hat{p}_T$) bins, with each hard scattering bin contributing hadrons (jets) to multiple $p_T$ ($p_T^{jet}$) bins. 
The yield from each hard scattering $\hat{p}_T$ bin has to be weighted by the cross section of that hard scattering bin. 
These perturbative hard cross-sections for each $\hat{p}_T$ bin are iteratively evaluated in the process of event generation, in both $p$-$p$ and Pb\,-\,Pb. 
Since each \textsc{Jetscape} event contains exactly one binary $p$-$p$ collision generated by the \texttt{PythiaGun} module, obtaining the correct Pb\,-\,Pb yield in the presence of
nuclear modifications requires the inelastic nucleon–nucleon cross section, which involves both perturbative and nonperturbative  contributions.

To quantify this factor, we calibrate against the $R_{p\mathrm{Pb}}$ of jets and hadrons by fitting the ratio of $p$-$\mathrm{Pb}$ to $p$-$p$ differential cross sections with a parameter $\sqrt{\lambda}$ as shown in Fig.~\ref{fig:pPbJetsHadrons}. The extracted $\sqrt{\lambda}$ closely matches the ratio of the perturbative hard cross sections for $p$-$p$ with and without nuclear-PDF modifications as computed in \texttt{Pythia} (for a minimum $\hat{p}_T \geq 4$~GeV).

Comparing to the $R_{p Pb}$ of jets and hadrons, one notes a slight tension between the hadron and jet data. 
(The hadron data are from CMS and the jet data are from ATLAS, and there is a well known tension between these data sets).
To accommodate this tension, we allow for an uncertainty of $\pm 5$\% (dashed lines) in the extracted $\sqrt{\lambda} \simeq 1.1$, represented by the solid red lines in Fig.~\ref{fig:pPbJetsHadrons}. 
For Pb\,-\,Pb, we apply the square of same scaling factor $\lambda$ uniformly across all centralities, since the \texttt{EPS09} nuclear PDFs do not include an impact-parameter dependence~\cite{Li:2001xa,Deng:2010xg} and therefore do not encode centrality variation. 
In fact, our results will indicate the need for centrality dependence in the shadowing function. 

\subsubsection{\texorpdfstring{High-$p_T$ Azimuthal Anisotropy}{High-pT Azimuthal Anisotropy}}

To quantify the azimuthal anisotropy of high-$p_T$ jets and hadrons, we extract $v_n^{\rm jet/h\pm}$ for $n=2$ using two standard flow-analysis techniques: the event-plane (EP) method and the scalar-product (SP) method.

In the EP method, the jet/$h\pm$ anisotropy is obtained by correlating the jet/$h\pm$ azimuthal angle $\phi^{\rm jet/h\pm}$ with the event plane angle $\Psi_n$ determined from soft hadrons in the same centrality class,
\begin{equation}
    v_n^{\rm jet/h\pm,EP} = \left\langle \cos \left[ n(\phi^{\rm jet/h\pm} - \Psi_n) \right] \right\rangle ,
\end{equation}
where the brackets denote an average over events.  
This method is straightforward but does not account for event-by-event fluctuations of the soft-sector flow within a given centrality interval.

The SP method incorporates these fluctuations explicitly by weighting the jet/$h\pm$–soft correlation with the soft-sector flow magnitude in each event,
\begin{equation}
    v_n^{\rm jet/h\pm,SP} =
    \frac{ \left\langle 
        v_n^{\rm soft} \cos\left[ n(\phi^{\rm jet/h\pm} - \Psi_n) \right]
        \right\rangle }
         { \sqrt{ \left\langle (v_n^{\rm soft})^2 \right\rangle } } ,
\end{equation}
where $v_n^{\rm soft}$ is the flow coefficient of the bulk hadrons. This method yields a fluctuation-robust measure of jet/$h\pm$ anisotropy. $v_n^{soft}$ and $\Psi_{n}$ information is retrieved from \texttt{MUSIC}, using the soft hadrons with $p_T\leq 3$GeV at forward rapidity. 
Both methods are used in our analysis, and the comparison between them provides a consistency check on the extracted high-$p_T$ azimuthal anisotropy.


\section{Results}
\label{sec:results}


\begin{figure*}[!t]
    \centering

    \begin{subfigure}[t]{0.48\textwidth}
        \centering
        \includegraphics[width=\linewidth]{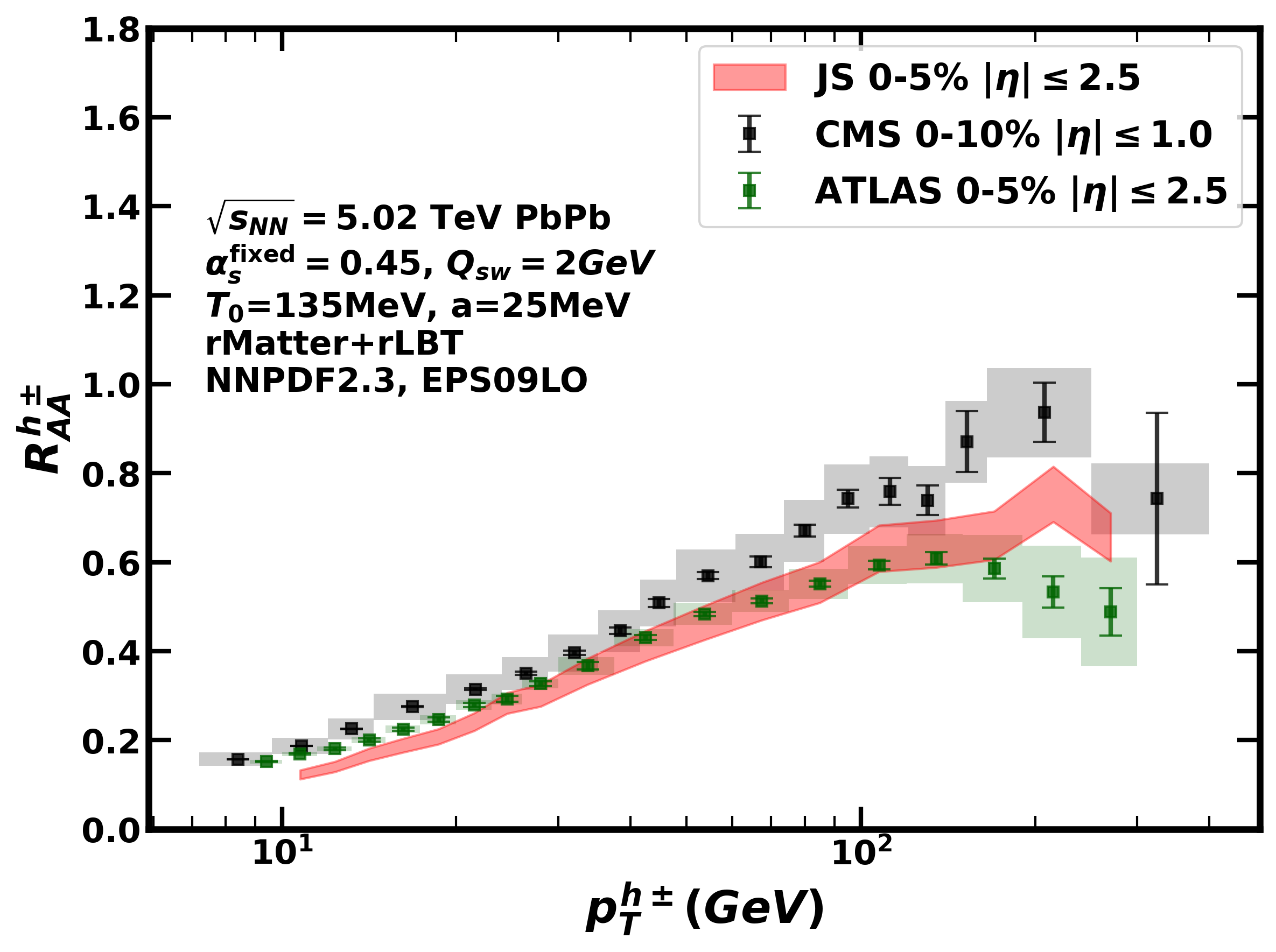}
    \end{subfigure}\hfill
    \begin{subfigure}[t]{0.48\textwidth}
        \centering
        \includegraphics[width=\linewidth]{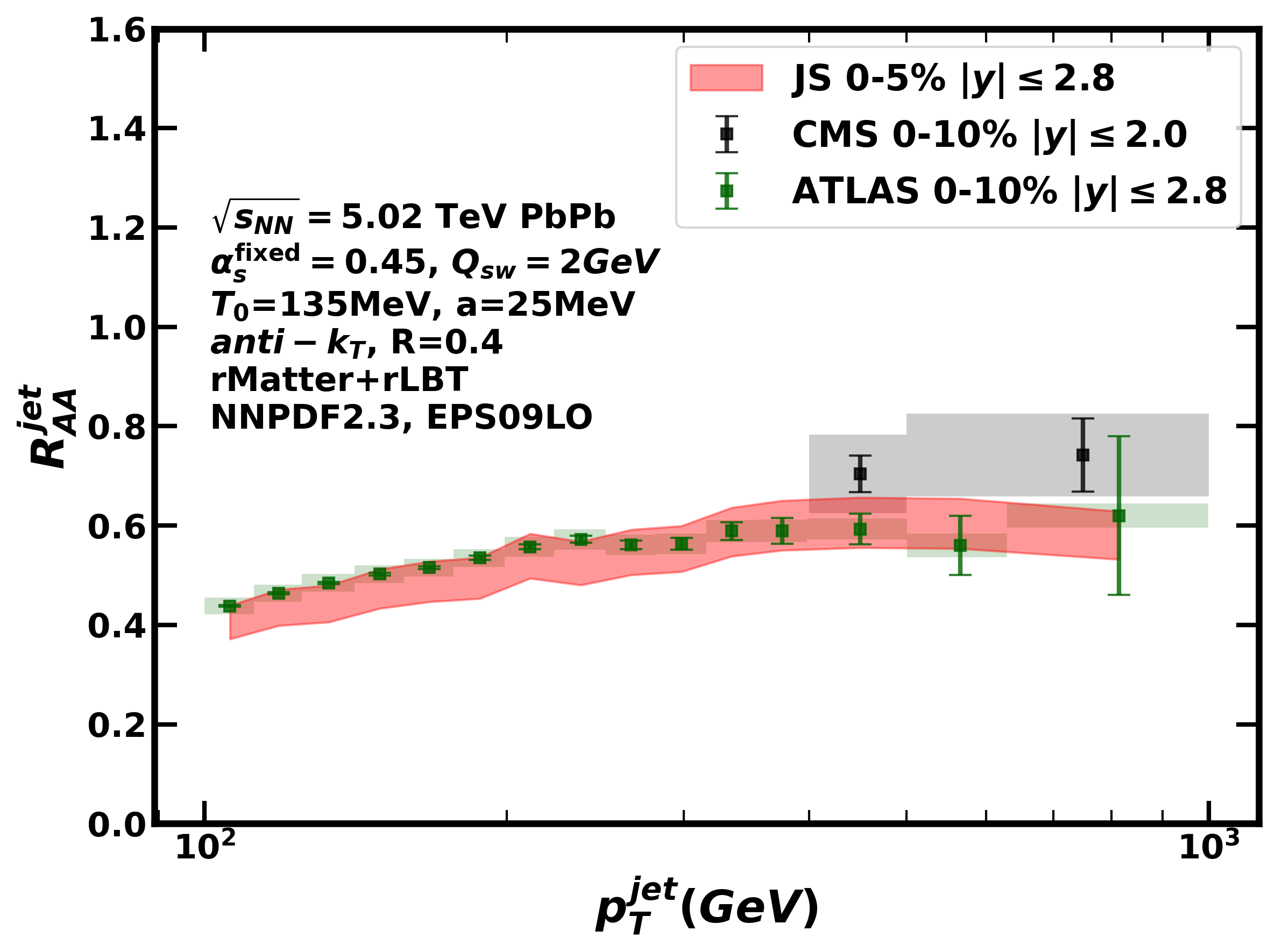}
    \end{subfigure}

    \vspace{-0.15cm}

    \begin{subfigure}[t]{0.48\textwidth}
        \centering
        \includegraphics[width=\linewidth]{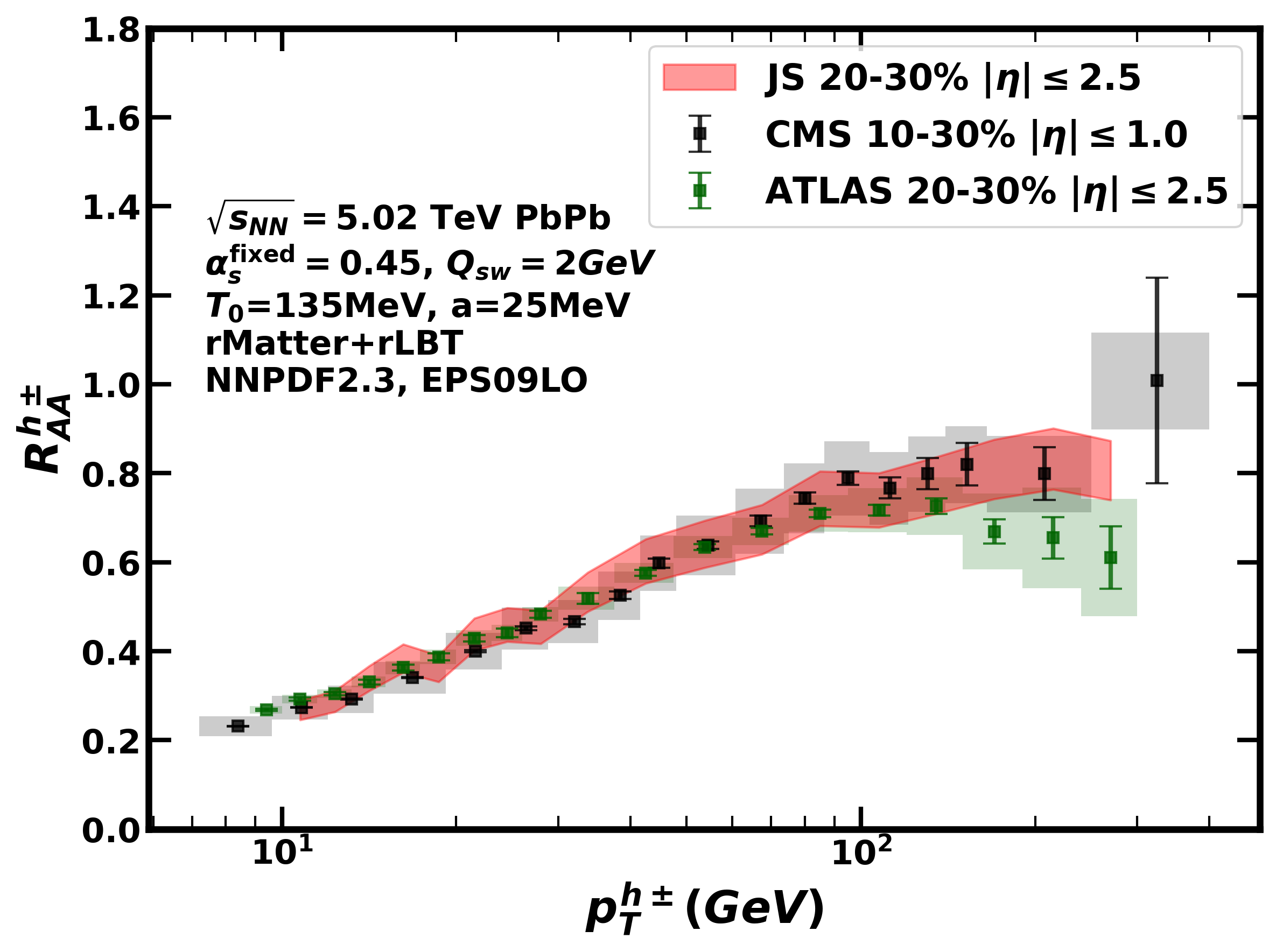}
    \end{subfigure}\hfill
    \begin{subfigure}[t]{0.48\textwidth}
        \centering
        \includegraphics[width=\linewidth]{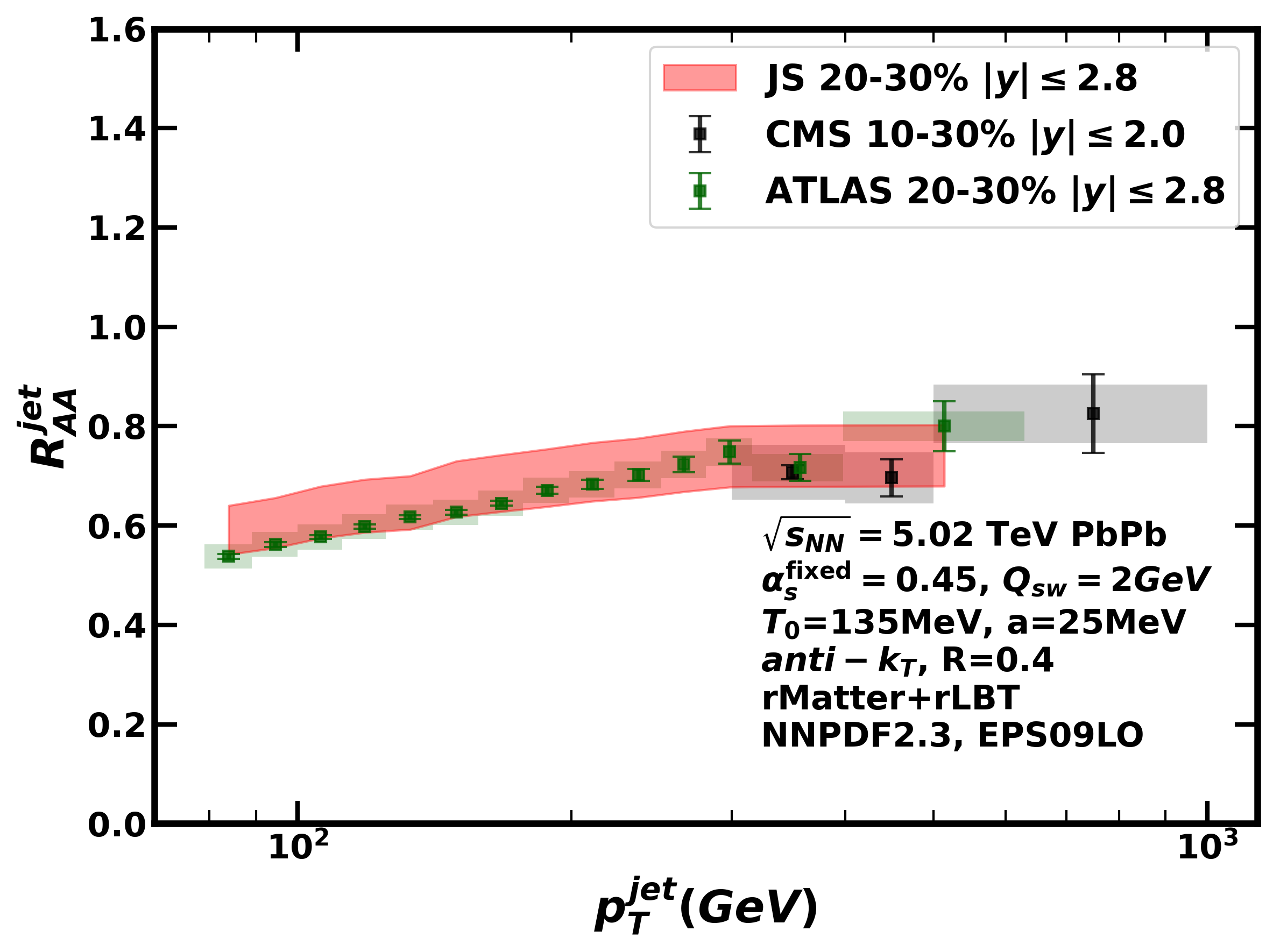}
    \end{subfigure}

    \vspace{-0.15cm}

    \begin{subfigure}[t]{0.48\textwidth}
        \centering
        \includegraphics[width=\linewidth]{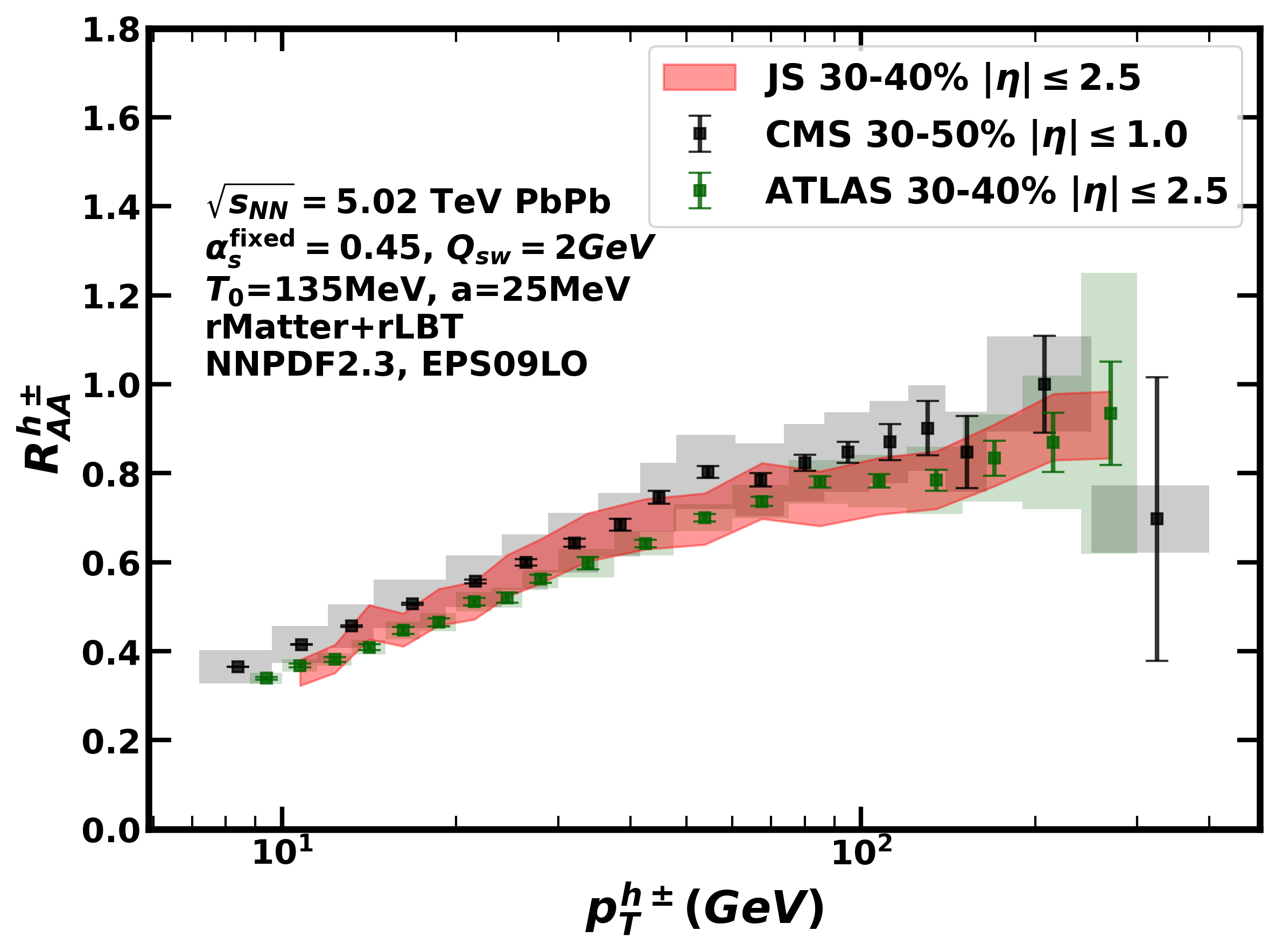}
    \end{subfigure}\hfill
    \begin{subfigure}[t]{0.48\textwidth}
        \centering
        \includegraphics[width=\linewidth]{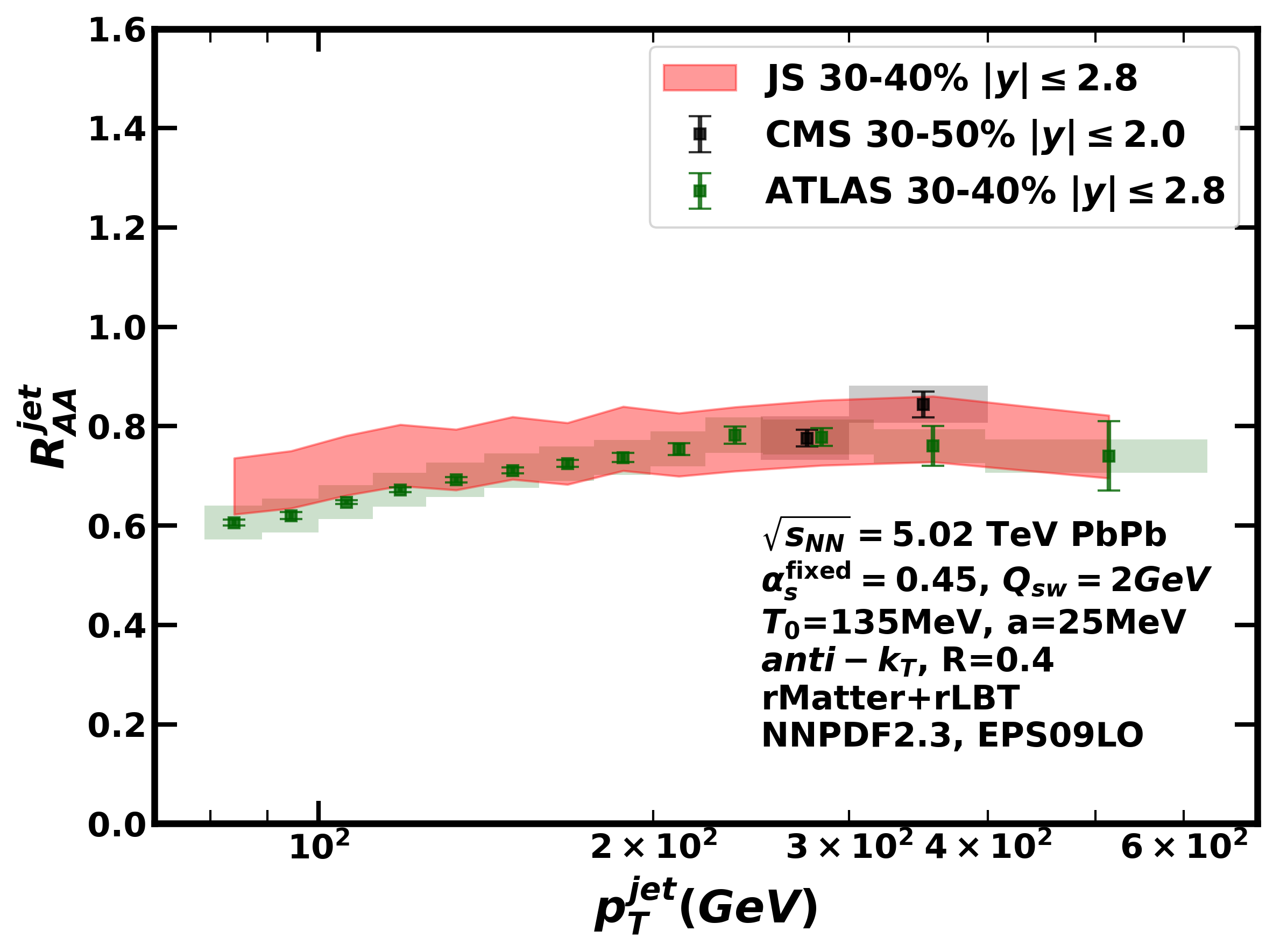}
    \end{subfigure}

    \caption{Left Column: Calculated central and semi-peripheral charged-hadron $R_{AA}$ (at $\sqrt{s_{\rm NN}}=5.02$~TeV, $|\eta | \leq 2.5$) are compared to ATLAS data (green squares)~\cite{ATLAS:2022kqu} with $|\eta|\leq 2.5$ and CMS data (black squares)~\cite{CMS:2016xef} with $|\eta|\leq 1.0$. Right Column: Calculated central and semi-peripheral jet $R_{AA}$ (at $\sqrt{s_{\rm NN}}=5.02$~TeV, $|y| \leq 2.8$) are compared to ATLAS data (green squares)~\cite{ATLAS:2018gwx} with $|y^{\mathrm{jet}}|\leq 2.8$ and CMS data (black squares)~\cite{CMS:2021vui} with $|y^{\mathrm{jet}}|\leq 2.0$, for $R=0.4$ jets. The red band includes statistical uncertainties from the reduced-\texttt{MATTER+LBT} (\textsc{Jetscape}) simulations with \texttt{EPS09LO} shadowing, uncertainties in reference cross sections, and the uncertainty associated with the parameter $\lambda$.}
    \label{fig:PbPb5020Central}
\end{figure*}

\begin{figure*}[!t]
    \centering

    \begin{subfigure}[t]{0.48\textwidth}
        \centering
        \includegraphics[width=\linewidth]{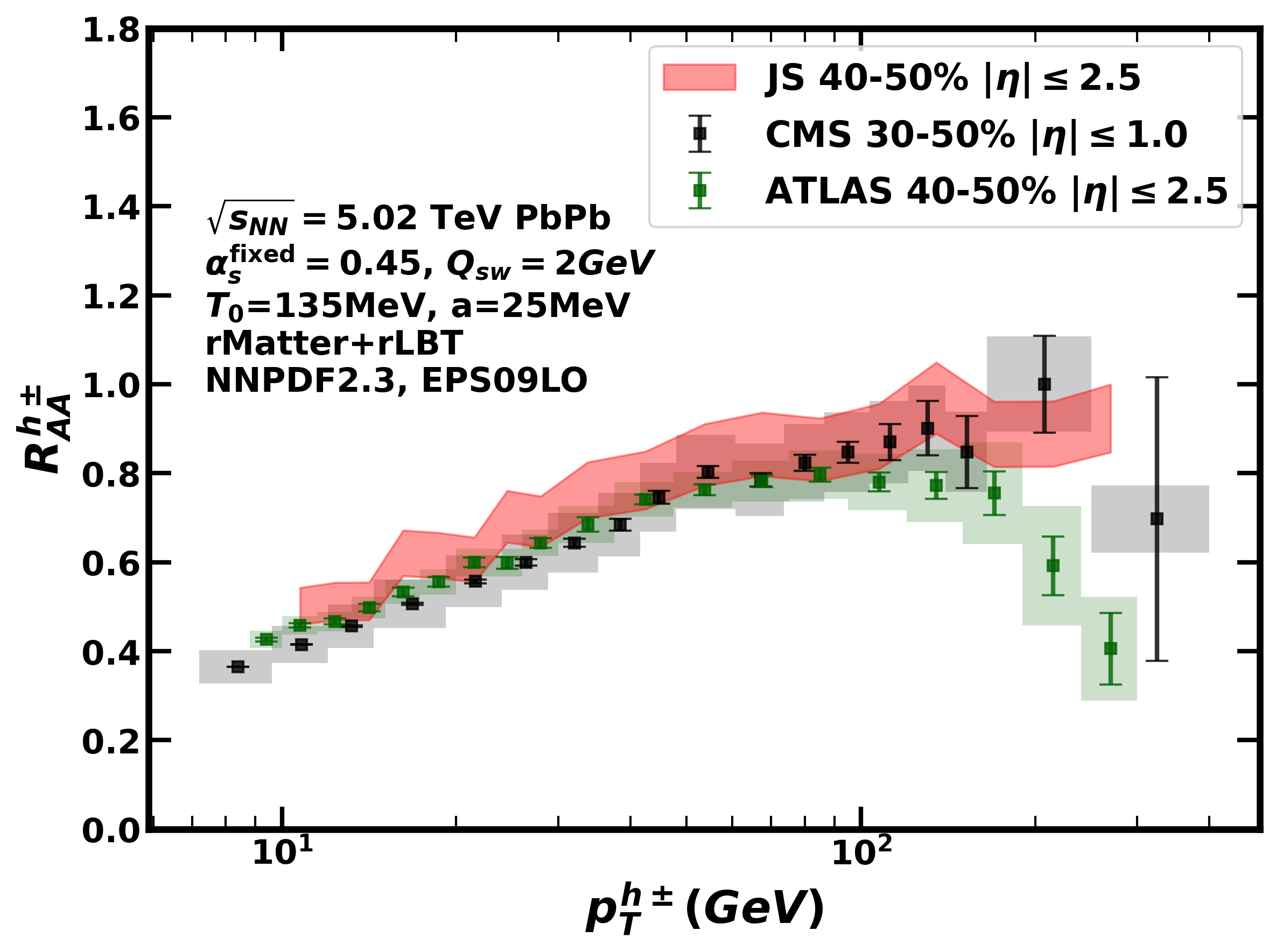}
    \end{subfigure}\hfill
    \begin{subfigure}[t]{0.48\textwidth}
        \centering
        \includegraphics[width=\linewidth]{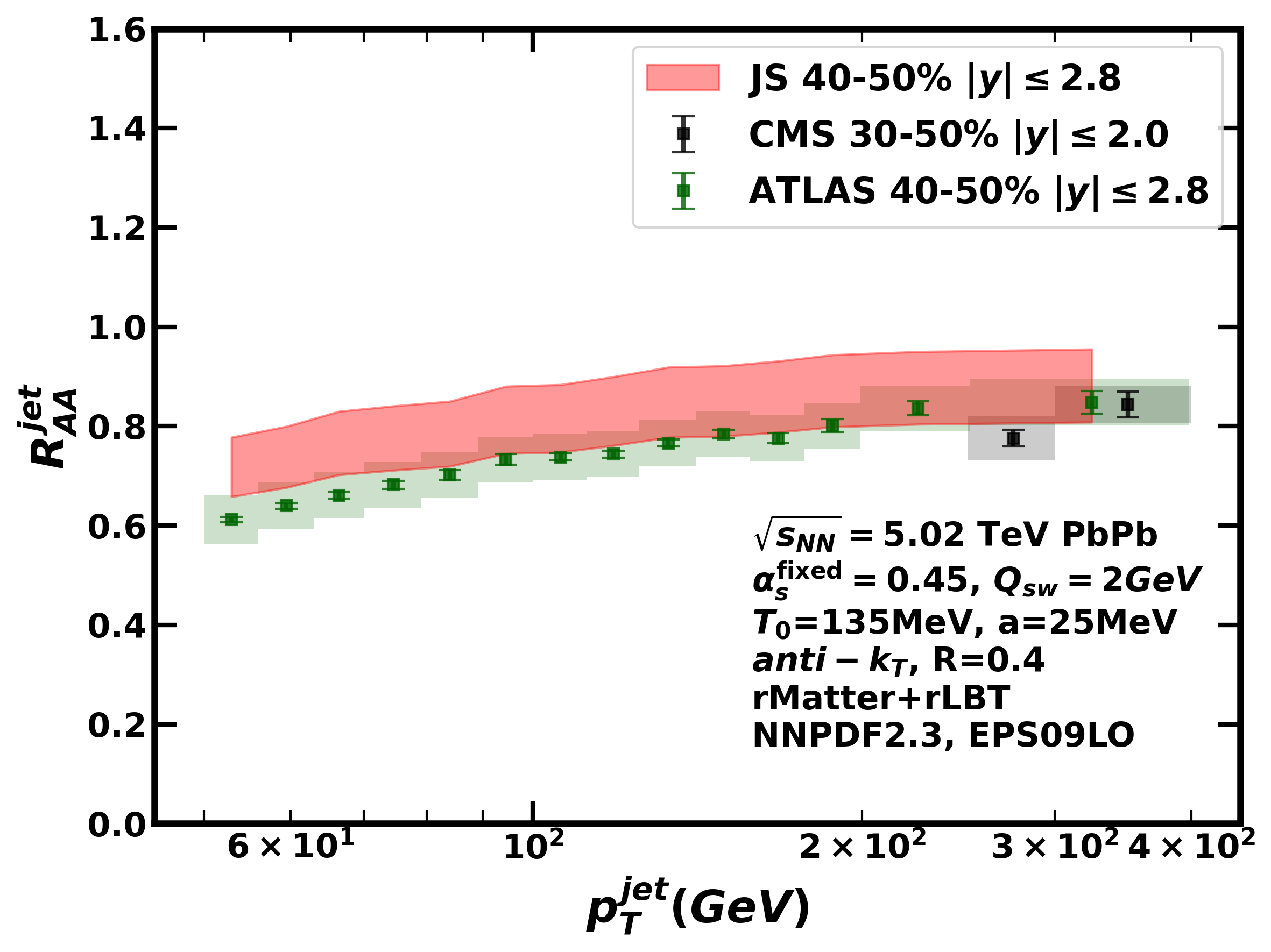}
    \end{subfigure}

    \vspace{-0.15cm}

    \begin{subfigure}[t]{0.48\textwidth}
        \centering
        \includegraphics[width=\linewidth]{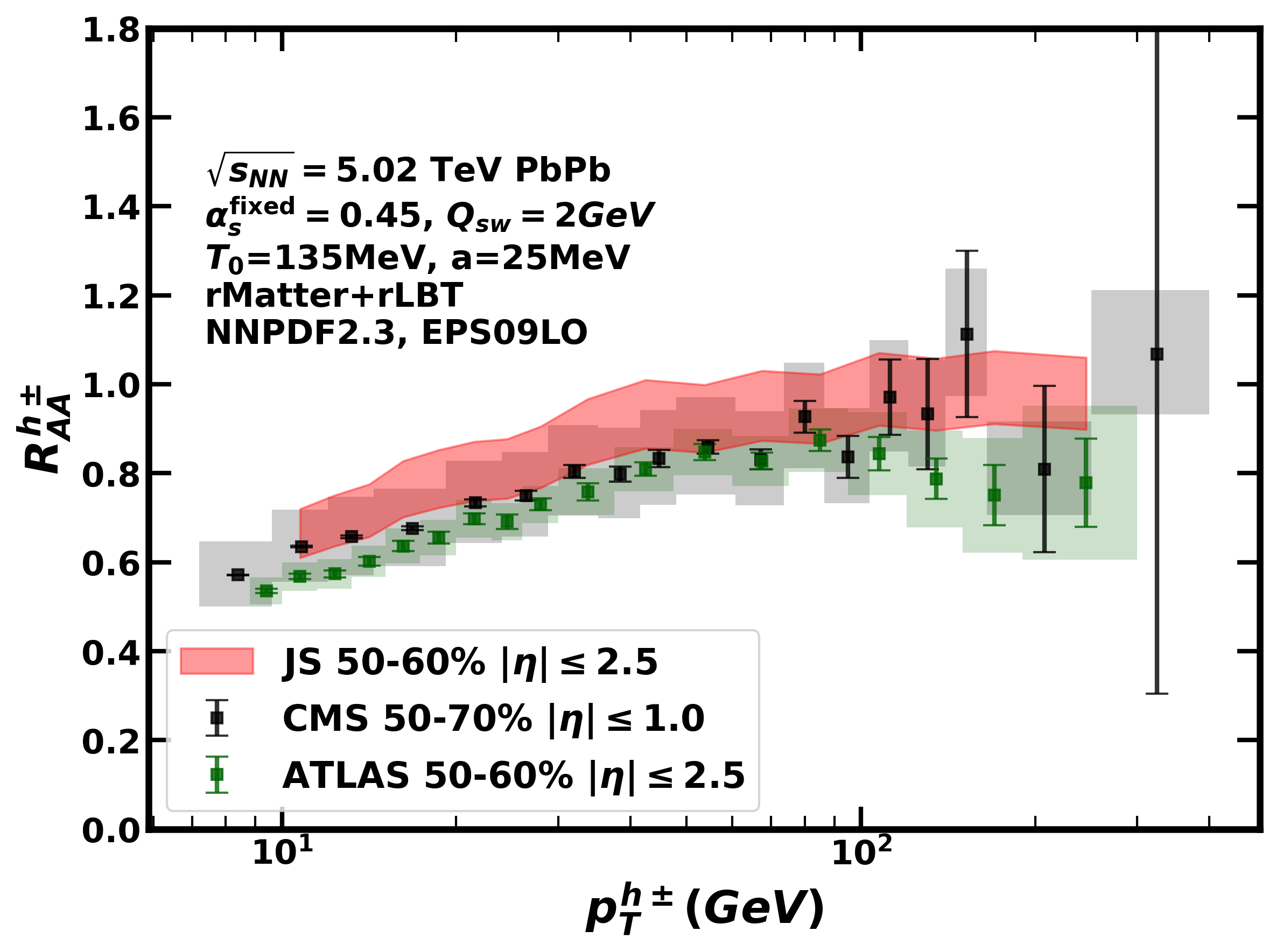}
    \end{subfigure}\hfill
    \begin{subfigure}[t]{0.48\textwidth}
        \centering
        \includegraphics[width=\linewidth]{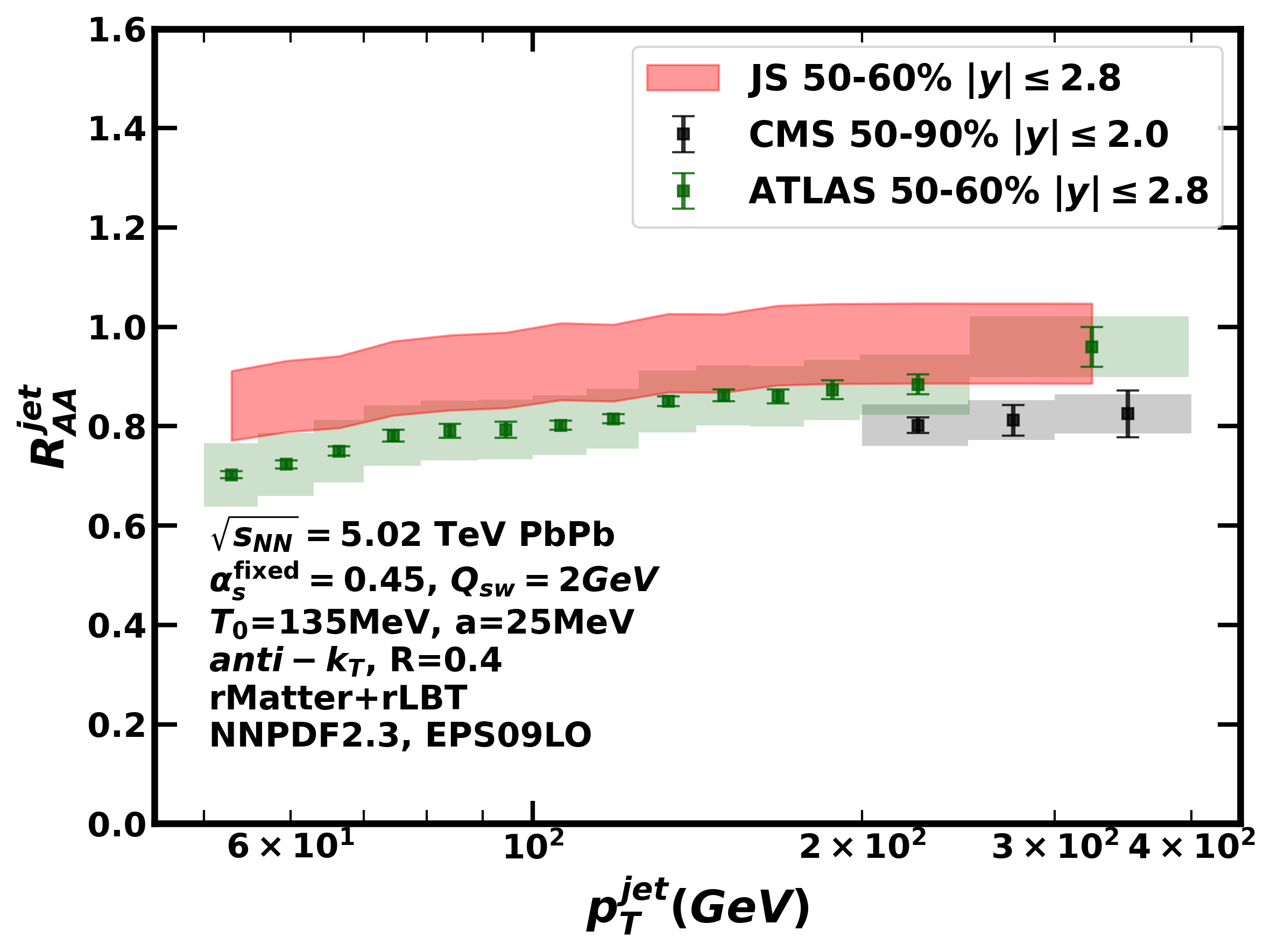}
    \end{subfigure}

    \caption{Left Column: Peripheral charged-hadron $R_{AA}$ results are compared to ATLAS data (green squares)~\cite{ATLAS:2022kqu} with $|\eta|\leq 2.5$ and CMS data (black squares)~\cite{CMS:2016xef} with $|\eta|\leq 1.0$. Right Column: Peripheral jet $R_{AA}$ results are compared to ATLAS data (green squares)~\cite{ATLAS:2018gwx} with $|y^{\mathrm{jet}}|\leq 2.8$ and CMS data (black squares)~\cite{CMS:2021vui} with $|y^{\mathrm{jet}}|\leq 2.0$, for $R=0.4$ jets. Calculational details and uncertainties of the reduced-\texttt{MATTER}+\texttt{LBT} model used are similar to those in Fig.~\ref{fig:PbPb5020Central}.}
    \label{fig:PbPb5020Peripheral}
\end{figure*}

\begin{figure*}[!t]

    \centering

    \begin{subfigure}[t]{0.48\textwidth}
        \centering
        \includegraphics[width=\linewidth]{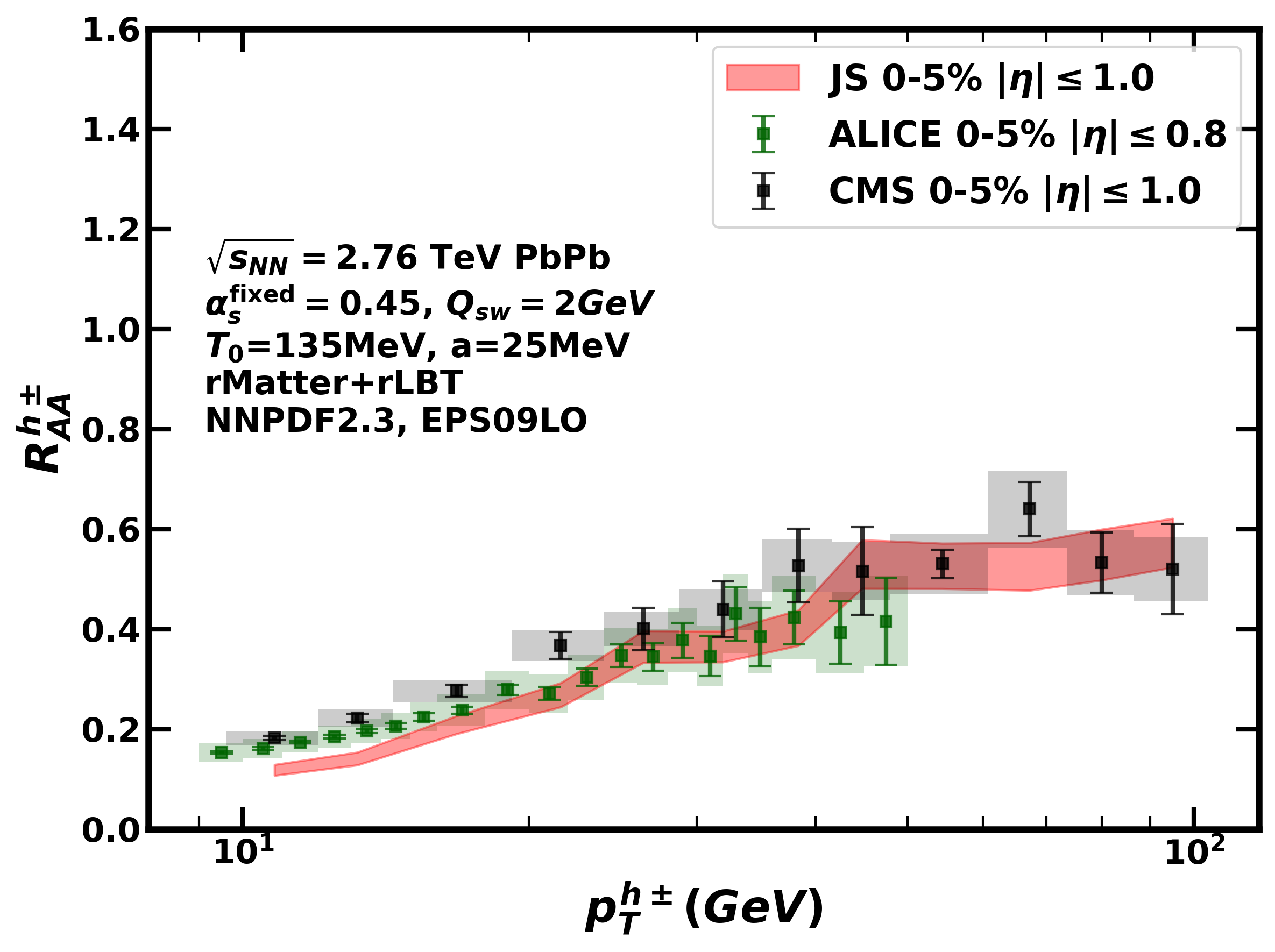}
    \end{subfigure}\hfill
    \begin{subfigure}[t]{0.48\textwidth}
        \centering
        \includegraphics[width=\linewidth]{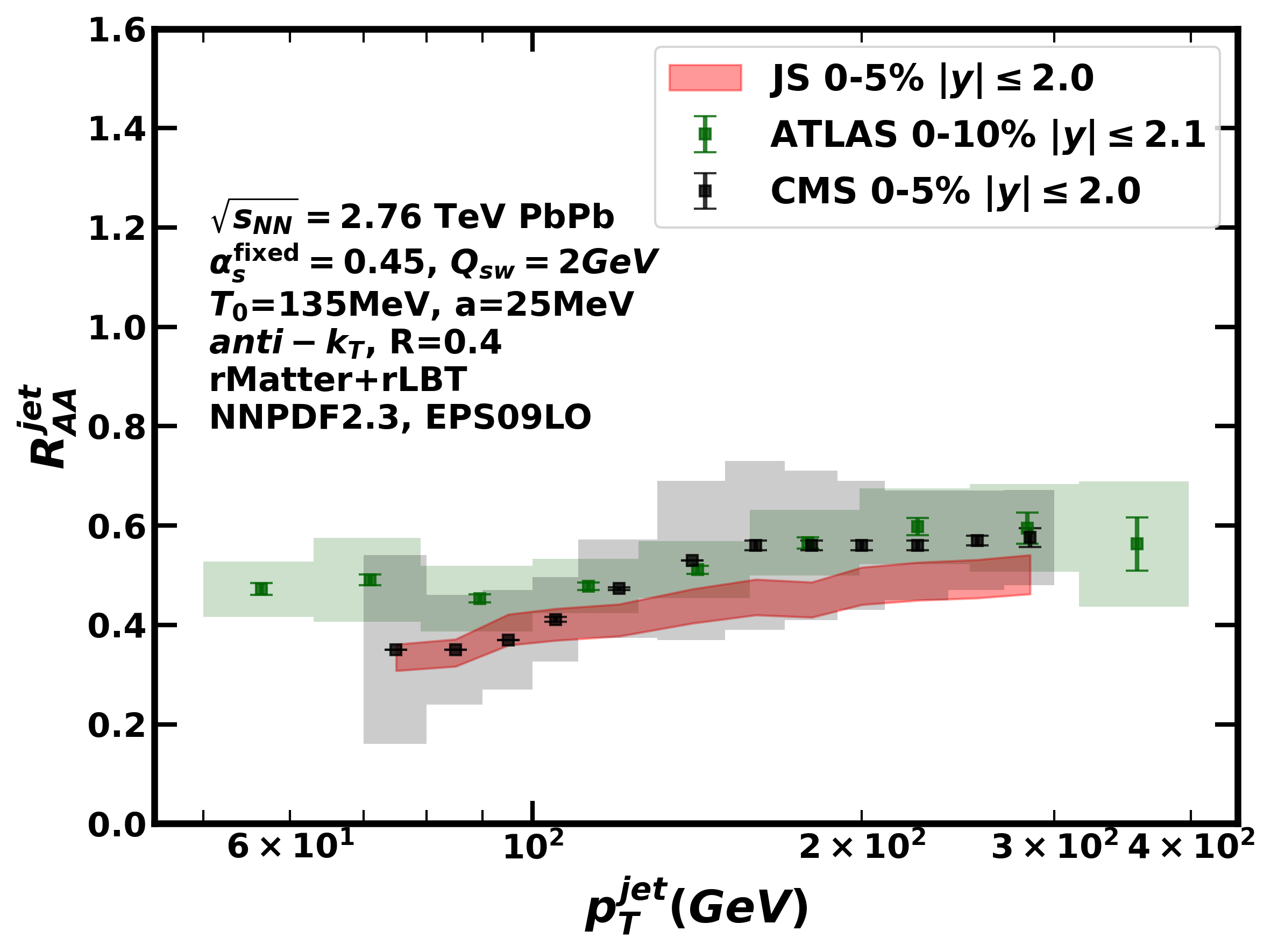}
    \end{subfigure}

    \caption{Left: Calculated charged-hadron $R_{AA}$ (at $\sqrt{s_{\rm NN}}=2.76$~TeV, 0--5\% most central, $|\eta| \leq 1.0$)  shown by the red band is compared to CMS 0--5\% data (black squares, $|\eta|\leq 1.0$)~\cite{CMS:2012aa} and ALICE 0--5\% data (green squares, $|\eta|\leq 0.8$)~\cite{ALICE:2012aqc}. Right: Calculated jet $R_{AA}$ (for 0--5\% most central, $|y| \leq 2.0$) are compared to CMS 0--5\% measurements (black squares, $|y|\leq 2.0$)~\cite{CMS:2016uxf} and ATLAS 0--10\% measurements (green squares, $|y|\leq 2.1$)~\cite{ATLAS:2014ipv}. Other than collision energy, all calculational details and uncertainties of the reduced-\texttt{MATTER}+\texttt{LBT} model used are similar to those in Fig.~\ref{fig:PbPb5020Central}. }
    \label{fig:PbPb2760Central}
\end{figure*}

In the preceding sections, we have outlined our model. 
In spite of the extended discussion, the current effort makes only two minor extensions to the work of Ref.~\cite{JETSCAPE:2022jer}: It introduces nuclear shadowing corrections to incoming PDFs, based on EPS09~\cite{Eskola:2009uj}; and it introduces reduced distributions which allow partonic energy loss based jet quenching to be extended into the hadronic phase, down to a minimum temperature of $T_0 \geq 135$~MeV (down from 160~MeV used in Ref.~\cite{JETSCAPE:2022jer}). 
These two minor corrections will yield a noticeable visual improvement in the comparison with the data for leading hadron and jet $R_{AA}$, which were considered in Ref.~\cite{JETSCAPE:2022jer}. 
In addition, these two simple extensions allow the model to address more peripheral collisions and for the first successful and simultaneous description of jet and hadron $v_2$, which has never been successfully described in combination with jet and hadron $R_{AA}$ before this. 
The only compensating factor that balances against the changes introduced,  will be the coupling in the medium $\alpha_S^{\rm fixed}$, which will be considerably larger than in Ref.~\cite{JETSCAPE:2022jer}.

We begin by presenting the nuclear modification factor $R_{AA}$ for charged hadrons and jets as a function of transverse momentum $p_T$ in Pb\,-\,Pb collisions at $\sqrt{s_{NN}} = 5.02~\mathrm{TeV}$ for different centrality classes. 
The results are shown in Fig.~\ref{fig:PbPb5020Peripheral}, where calculations from the reduced \texttt{MATTER+LBT} framework, including nuclear shadowing effects from \texttt{EPS09LO}, are represented by the red band. 
This band incorporates statistical uncertainties arising from the finite number of hadrons and jets in each $p_T$ bin, uncertainties in the reference cross sections, and the uncertainty associated with the parameter $\sqrt{\lambda}$, which accounts for the mild tension between the leading hadron and jet $R_{pA}$ measurements in Fig.~\ref{fig:pPbJetsHadrons}.

The simulation results are compared to charged-hadron measurements from ATLAS~\cite{ATLAS:2022kqu} and CMS~\cite{CMS:2016xef}, as well as jet measurements with radius $R = 0.4$ from CMS~\cite{CMS:2021vui} and ATLAS~\cite{ATLAS:2018gwx}. 
For central to mid--semi-peripheral collisions, the model describes the data with high accuracy for both charged hadrons and jets, with the simulation band largely overlapping the experimental points. 
Toward more peripheral collisions (40--50\% and 50--60\%), the charged-hadron $R_{AA}$ remains in good agreement with the data, while the jet suppression appears slightly underestimated. However, since nuclear shadowing effects are expected to weaken in peripheral collisions, the lower edge of the simulation band---corresponding to smaller values of $\lambda$, provides a closer description of the experimental measurements.

\begin{figure*}[!t]

    \centering

    \begin{subfigure}[t]{0.48\textwidth}
        \centering
        \includegraphics[width=\linewidth]{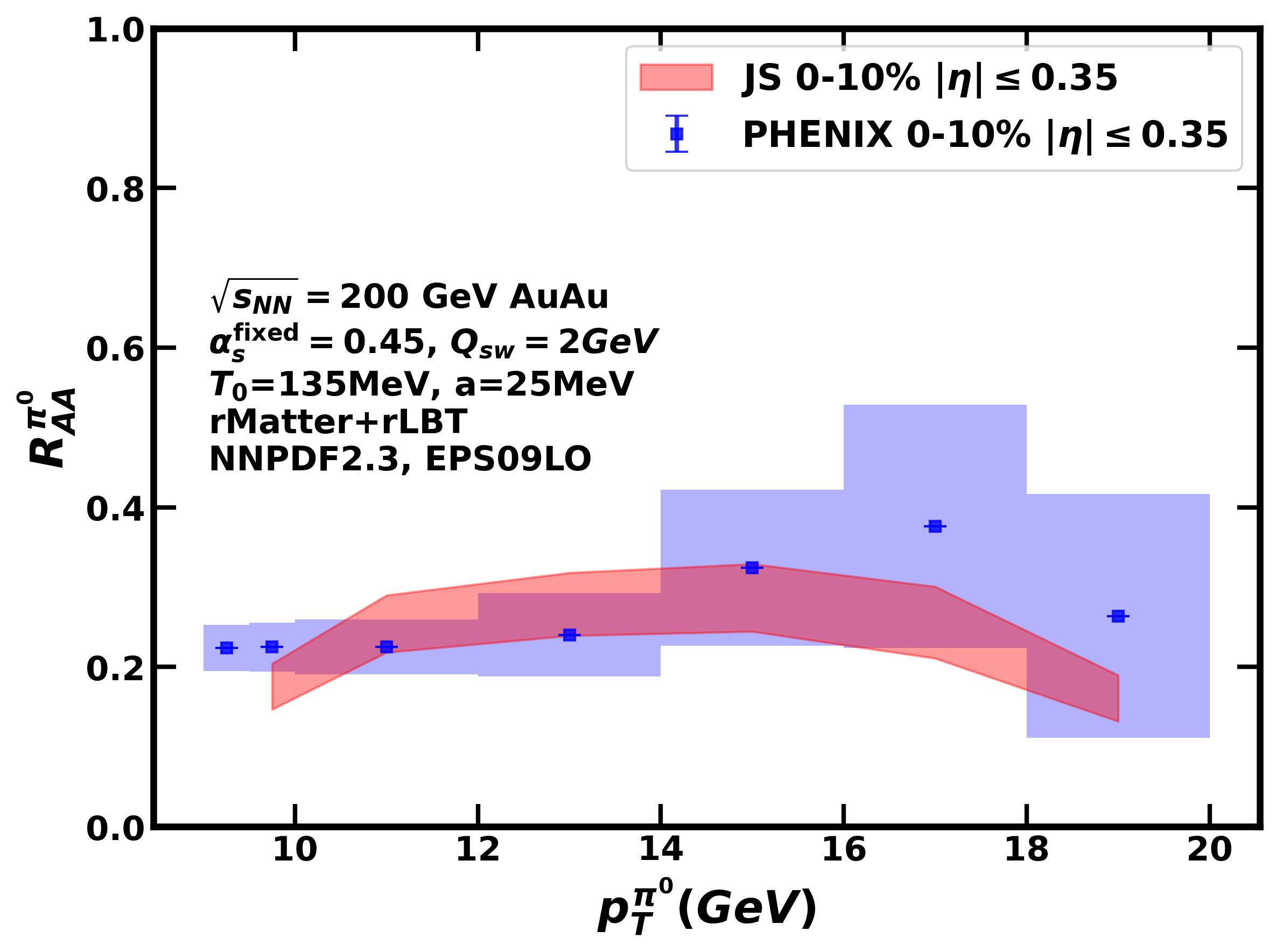}
    \end{subfigure}\hfill
    \begin{subfigure}[t]{0.48\textwidth}
        \centering
        \includegraphics[width=\linewidth]{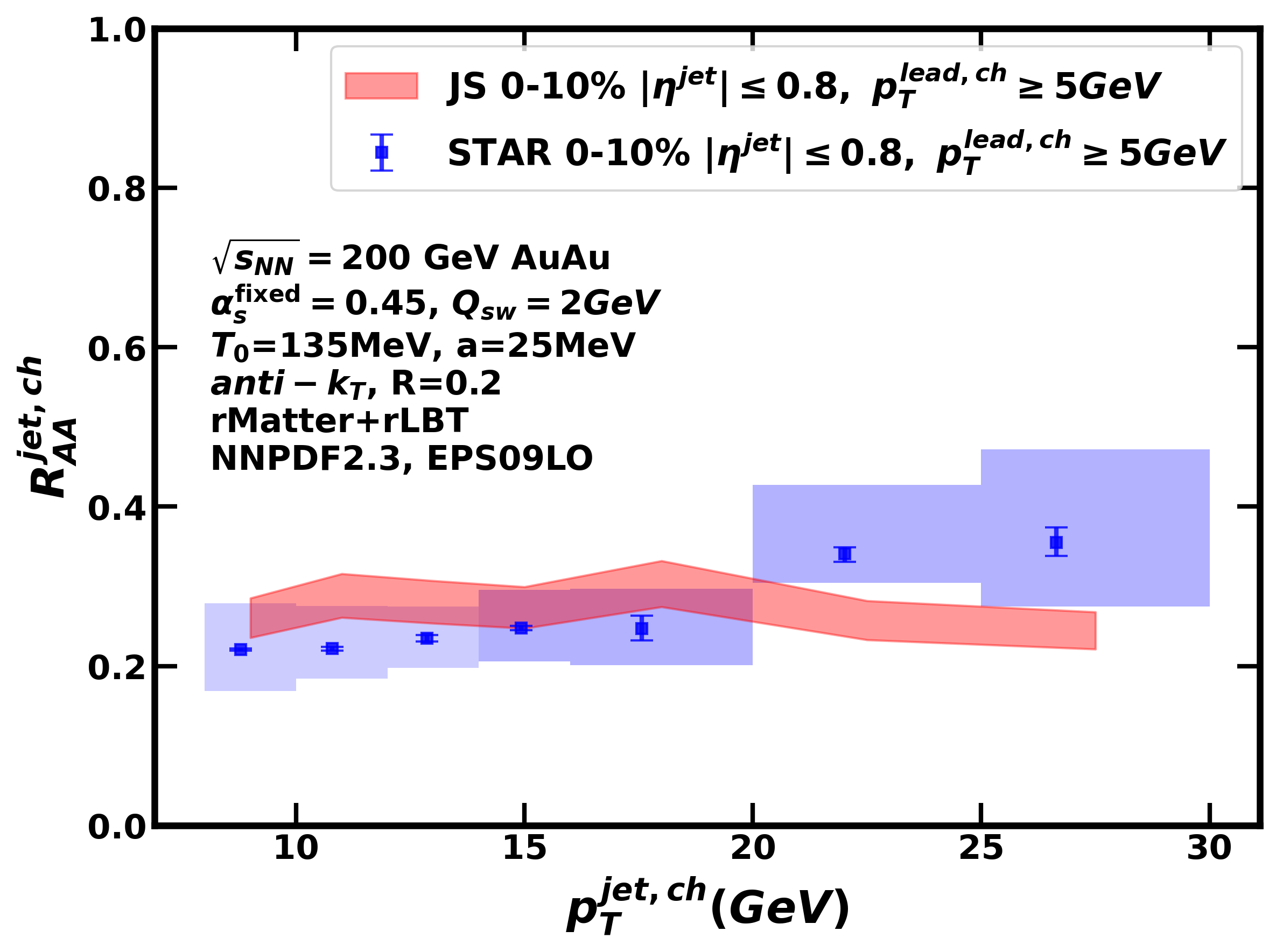}
    \end{subfigure}

    \caption{Left: Calculations of Neutral-pion $R_{AA}$ in red band at $\sqrt{s_{\rm NN}}= 200$~GeV (with $|\eta| \leq 0.35$) compared to measurements from PHENIX ($|\eta|\leq 0.35$)~\cite{PHENIX:2012jha} as blue squares. 
    Right: Calculations of charged-jet 
    $R_{AA}$ with $R=0.2$ ($|\eta^{\rm jet}| < 0.8$) and imposing a leading charged-hadron transverse momentum requirement of $p_T^{\mathrm{lead,ch}} \geq 5~\mathrm{GeV}$ compared to data with the same constraints from STAR~\cite{STAR:2020xiv} are shown as blue squares. Other than collision energy, all calculational details and uncertainties of the reduced-\texttt{MATTER}+\texttt{LBT} model used are similar to those in Fig.~\ref{fig:PbPb5020Central}.}
    \label{fig:200RHIC}
\end{figure*}

In addition to the 5.02~TeV results discussed above, Figs.~\ref{fig:PbPb2760Central} and~\ref{fig:200RHIC} present the charged-hadron and jet $R_{AA}$ obtained with the new changes in the framework, at lower collision energies. In the left panel of Fig.~\ref{fig:PbPb2760Central}, results for central (0--5\%) Pb+Pb collisions at $\sqrt{s_{NN}} = 2.76~\mathrm{TeV}$ for charged-hadrons are compared to charged-hadron measurements from ALICE ($|\eta|\leq 0.8$) and CMS ($|\eta|<1.0$). In the right panel,  inclusive jet measurements with $R=0.4$ from CMS ($|y|<2.0$) and ATLAS (0--10\%, $|y|<2.1$) are compared to calculations from the reduced-\texttt{MATTER+LBT} model. At RHIC energies, the corresponding results at $\sqrt{s_{NN}} = 200~\mathrm{GeV}$ for central (0--10\%) Au+Au collisions are compared to neutral-pion data from PHENIX ($|\eta|\leq 0.35$) and charged-jet measurements with $R=0.2$ from STAR ($|\eta|<0.8$), with a leading charged-hadron requirement of $p_T^{\mathrm{lead,ch}} > 5~\mathrm{GeV}$.

In all cases, the simulation results are shown as a red uncertainty band, constructed in the same manner as for the higher-energy results, with the parameter $\lambda$ determined from the ratio of perturbative cross sections obtained from \texttt{PYTHIA} at the corresponding collision energy. The uncertainty of $\lambda$ is kept the same as that at 5.02TeV. Taken together, the results shown in Figs.~\ref{fig:PbPb2760Central} and ~\ref{fig:200RHIC} demonstrate that the reduced \texttt{MATTER+LBT} along with nuclear shadowing provides a consistent description of charged-hadron and jet suppression across a wide range of collision energies. Thus, the level of simultaneous agreement with jets and leading hadrons, across all energies, obtained with the earlier \textsc{Jetscape} tuning~\cite{JETSCAPE:2022jer}, is not diminished. 
At the same time the agreement with more peripheral events, especially at lower $p_T$ has been improved. 

While $R_{AA}$ constrains the average energy loss experienced by hard probes, it is largely insensitive to the azimuthal structure of the medium. Additional information on the path-length dependence and the geometric response of the medium is provided by the elliptic flow coefficient $v_2$ at high transverse momentum. Calculations of the elliptic anisotropy of jets and leading hadrons using the same simulations as in Ref.~\cite{JETSCAPE:2022jer}, showed a lower anisotropy than the data~\cite{Park:2019sdn,JETSCAPE:2020hue}. The current model shows considerable improvement in simultaneously addressing the jet and hadron $v_2$ at high-$p_T$. 

\begin{figure*}[!t]
    \centering

    \begin{subfigure}[t]{0.48\textwidth}
        \centering
        \includegraphics[width=\linewidth]{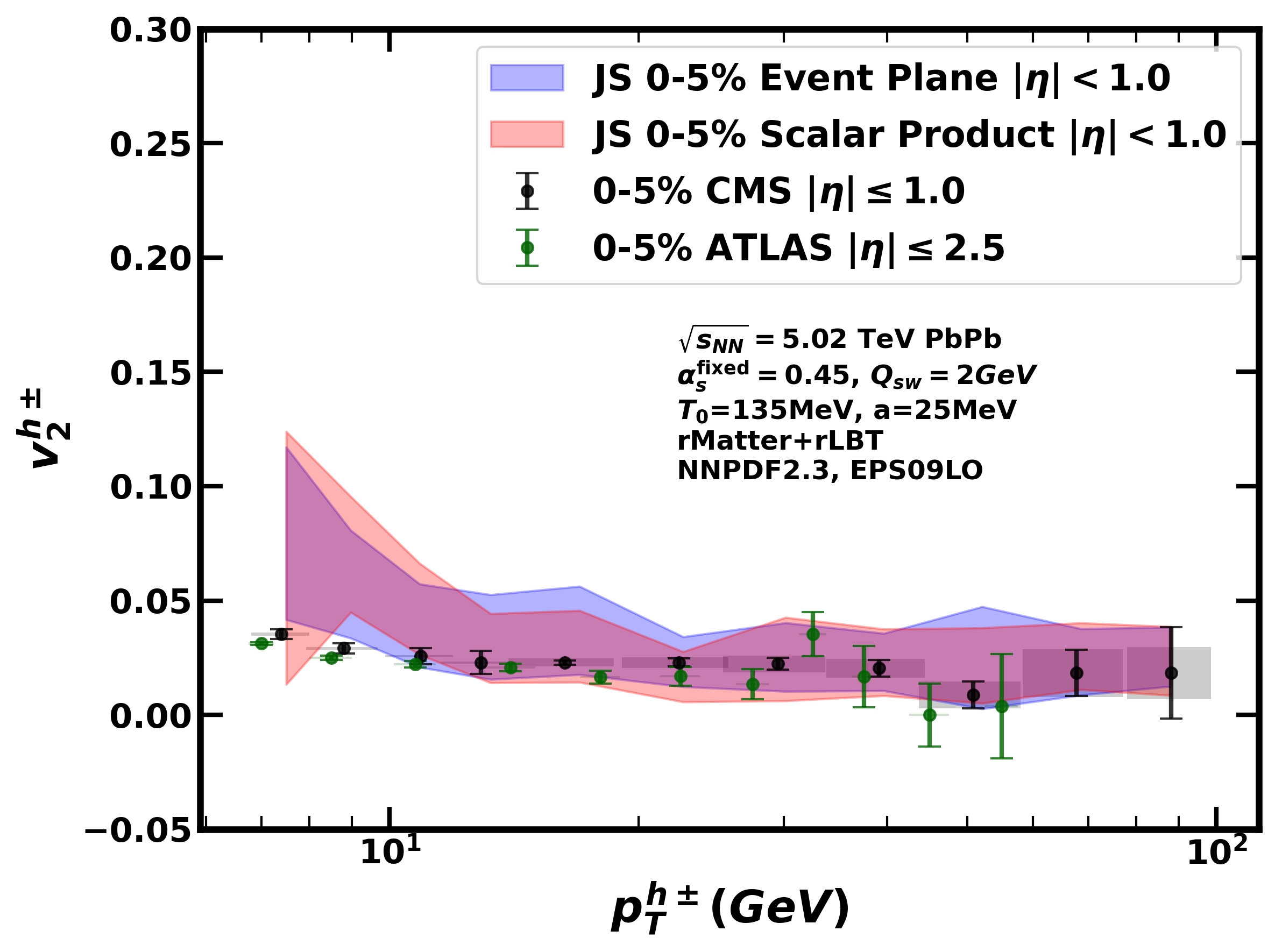}
    \end{subfigure}\hfill
    \begin{subfigure}[t]{0.48\textwidth}
        \centering
        \includegraphics[width=\linewidth]{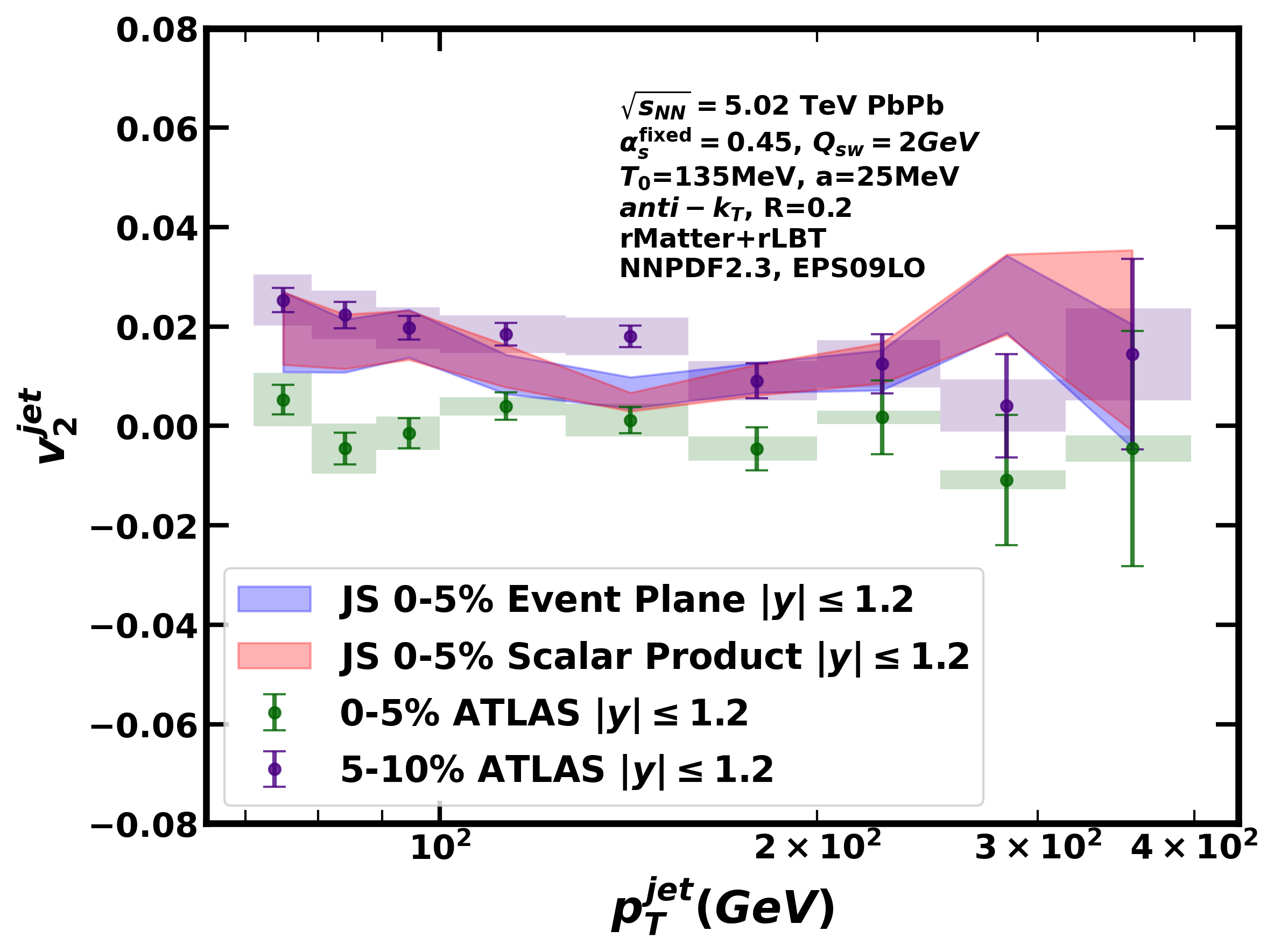}
    \end{subfigure}

    \vspace{-0.15cm}

    \begin{subfigure}[t]{0.48\textwidth}
        \centering
        \includegraphics[width=\linewidth]{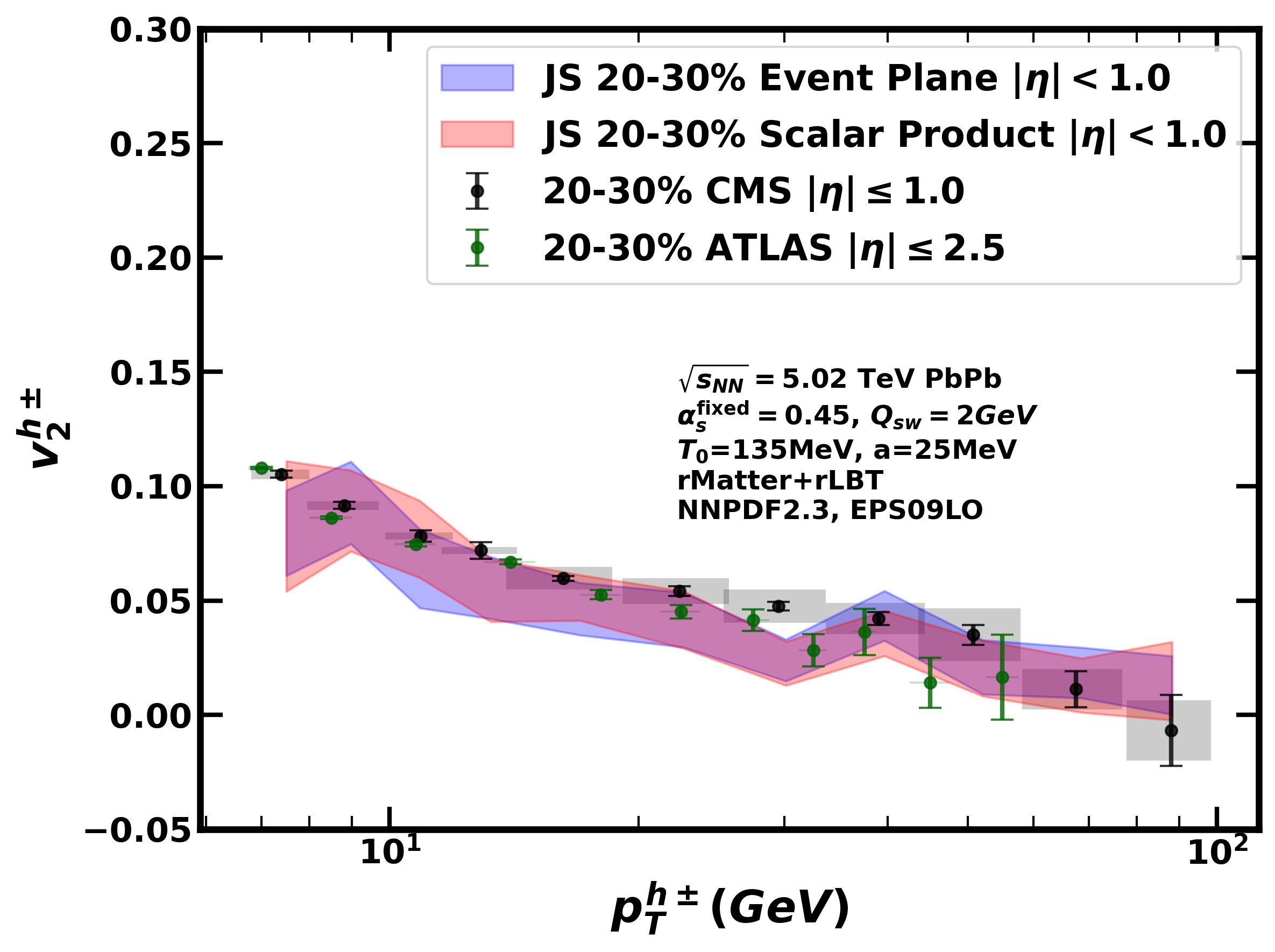}
    \end{subfigure}\hfill
    \begin{subfigure}[t]{0.48\textwidth}
        \centering
        \includegraphics[width=\linewidth]{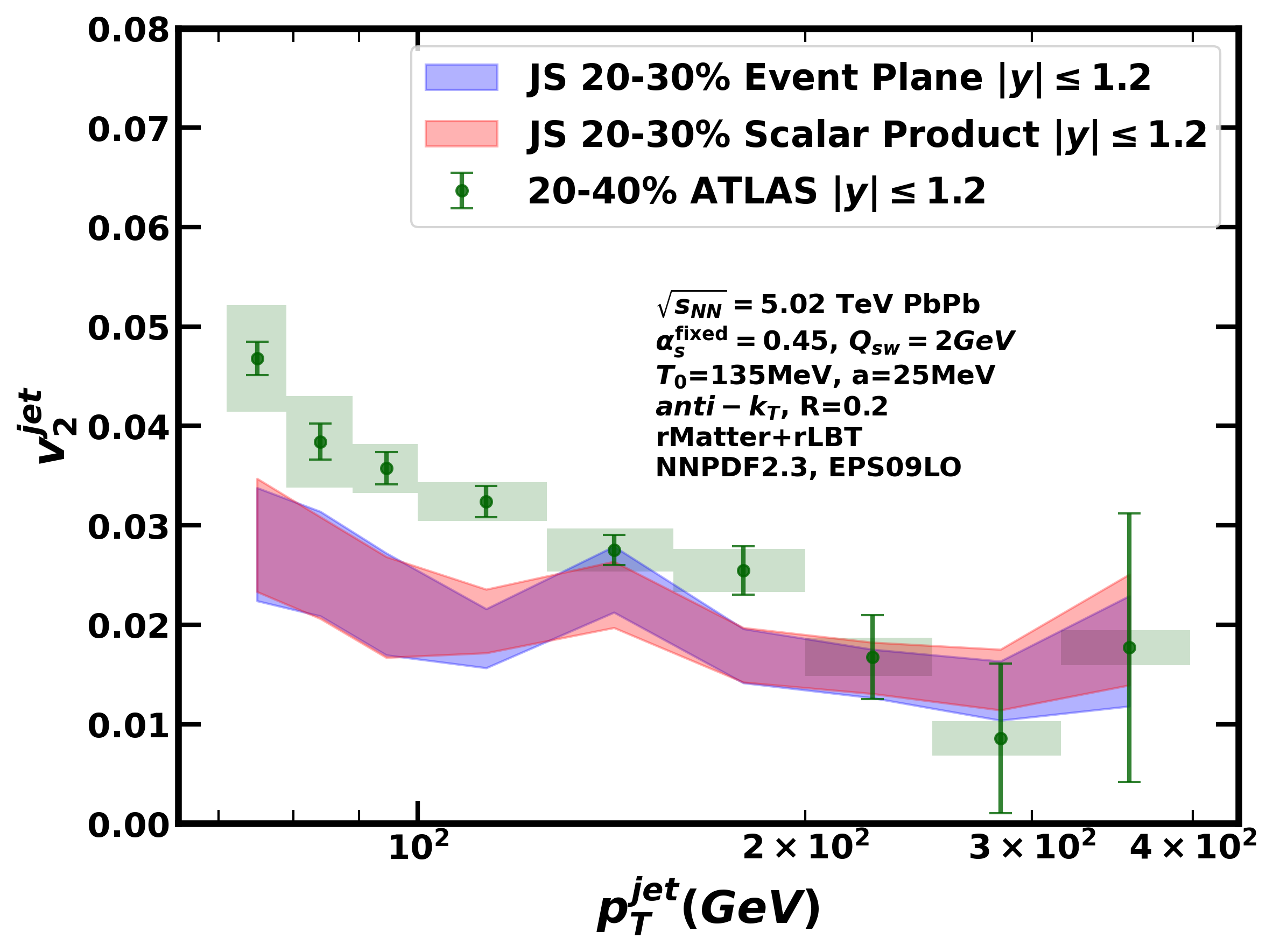}
    \end{subfigure}

    \vspace{-0.15cm}

    \begin{subfigure}[t]{0.48\textwidth}
        \centering
        \includegraphics[width=\linewidth]{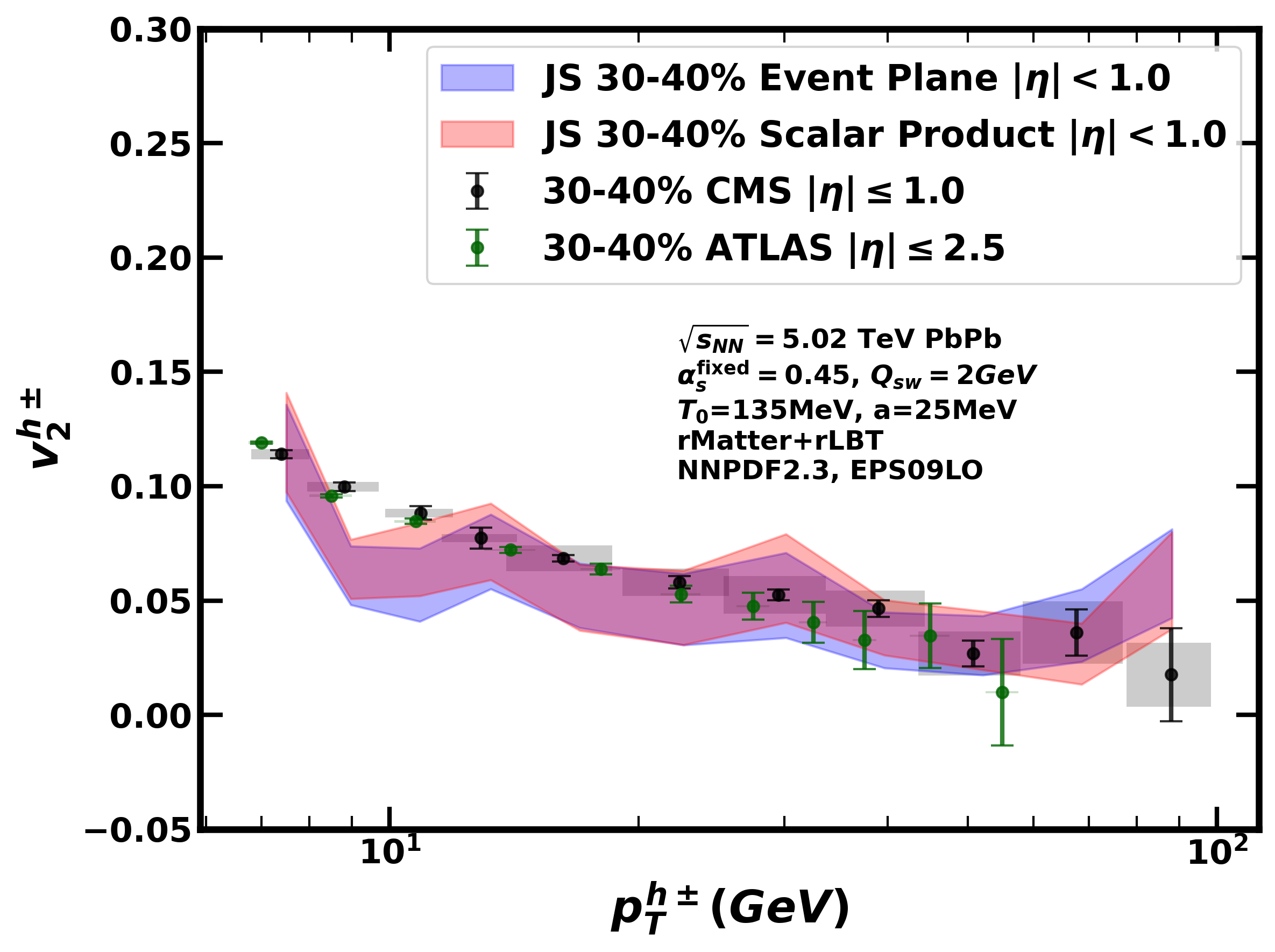}
    \end{subfigure}\hfill
    \begin{subfigure}[t]{0.48\textwidth}
        \centering
        \includegraphics[width=\linewidth]{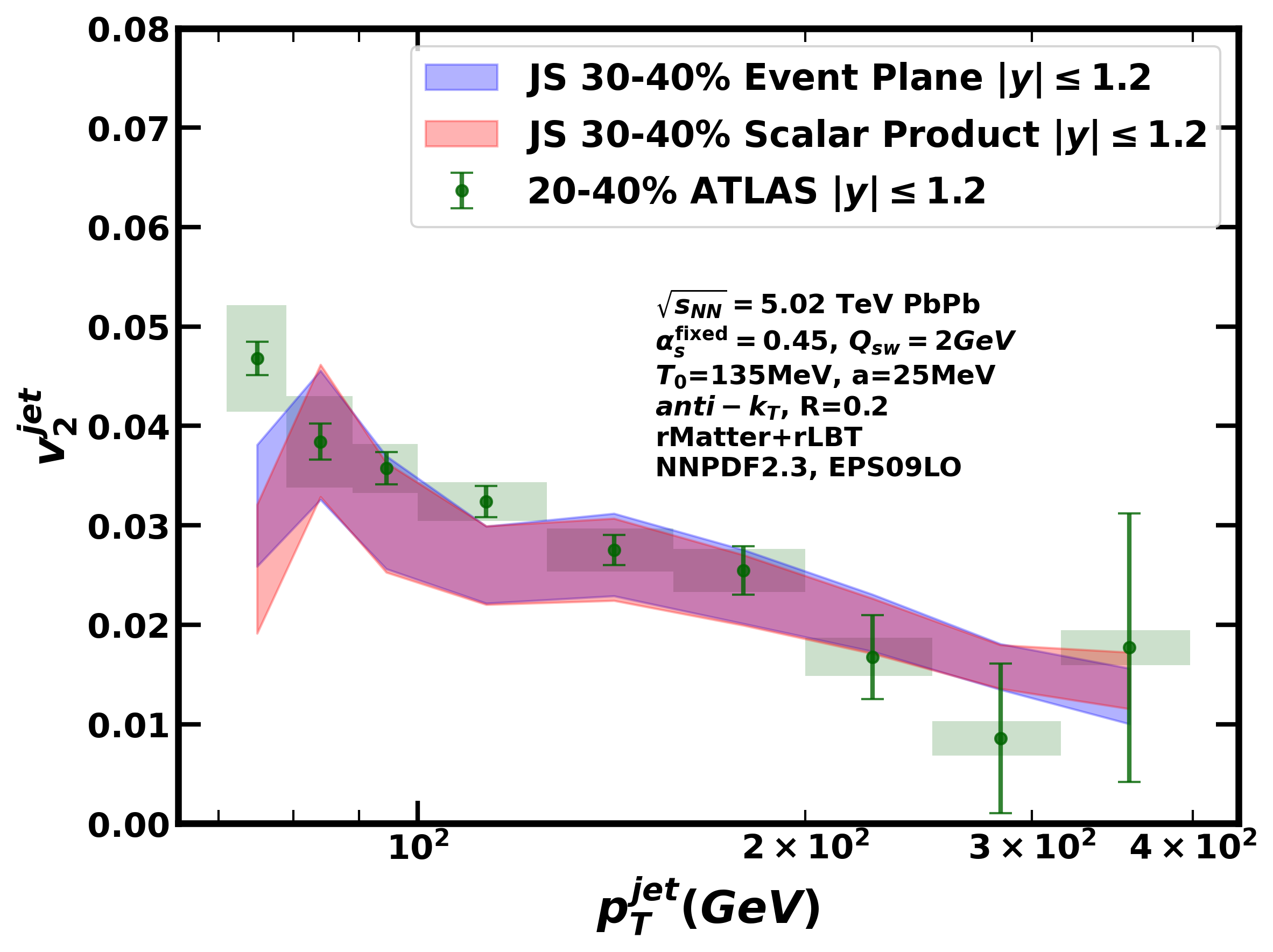}
    \end{subfigure}

    \caption{Elliptic anisotropy coefficient $v_2$ of high-$p_T$ charged hadrons (left) and  $R=0.2$ jets (right) as a function of $p_T$, computed using the event-plane (blue band) and scalar-product (red band) methods within the \textsc{Jetscape} reduced (\texttt{MATTER}+\texttt{LBT}) framework + EPSO9LO nuclear shadowing. The results are compared to ATLAS~\cite{ATLAS:2021ktw, ATLAS:2018ezv} (green circles) and CMS~\cite{CMS:2017xgk} (black circles) measurements in 0--5\% (top-left), 20--30\% (middle-left)  and 30--40\% (bottom-left) centrality classes (for charged hadrons) and 0--5\%, 5--10\% (top-right) and 20--40\% (middle and bottom right) centrality classes (for jets) in Pb\,-\,Pb collisions at $\sqrt{s_{NN}} = 5.02~\mathrm{TeV}$. See text for more details.}
    \label{fig:PbPbv2Central}
\end{figure*}

\begin{figure*}[!t]
    \centering

    \begin{subfigure}[t]{0.48\textwidth}
        \centering
        \includegraphics[width=\linewidth]{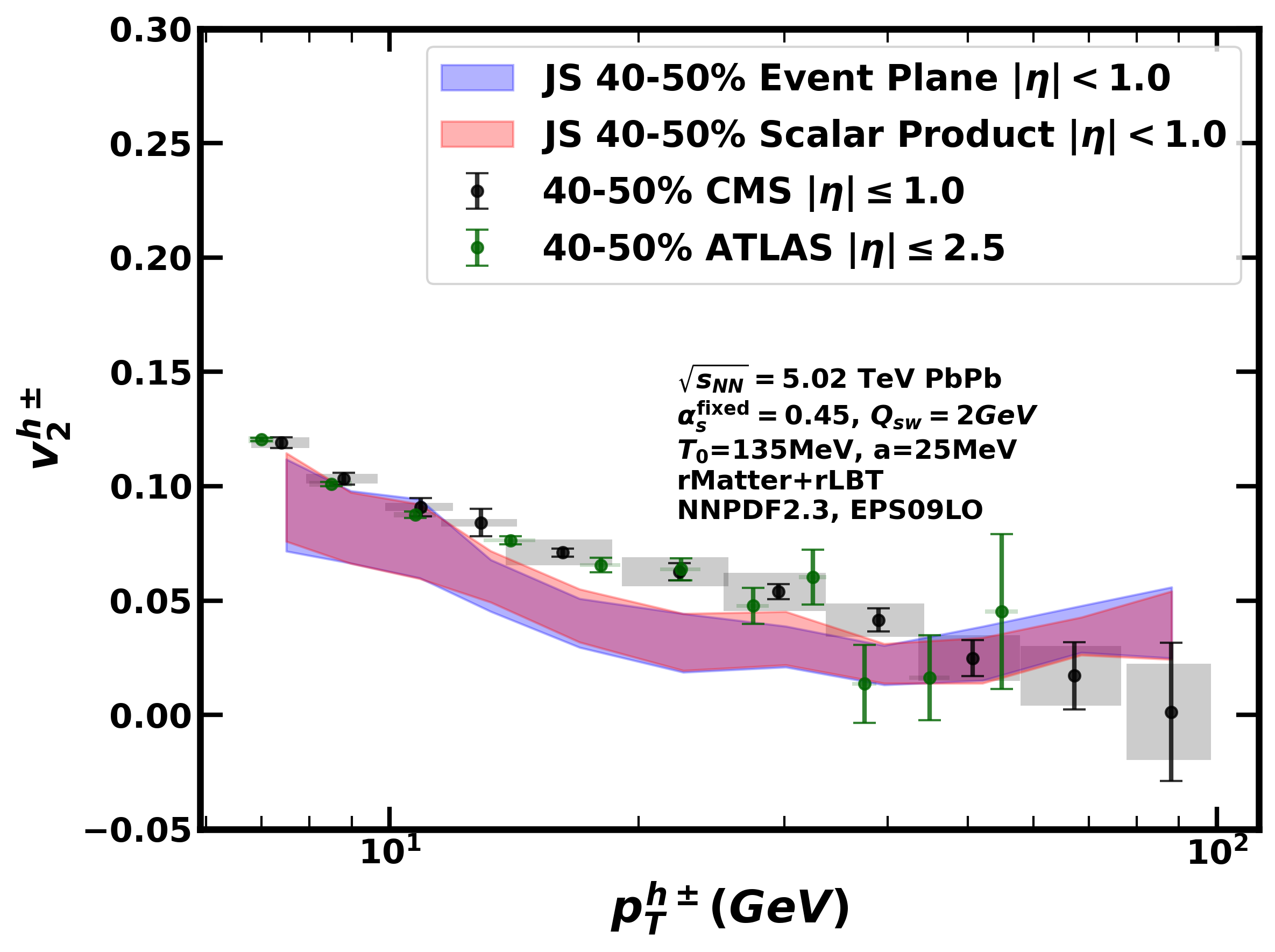}
    \end{subfigure}\hfill
    \begin{subfigure}[t]{0.48\textwidth}
        \centering
        \includegraphics[width=\linewidth]{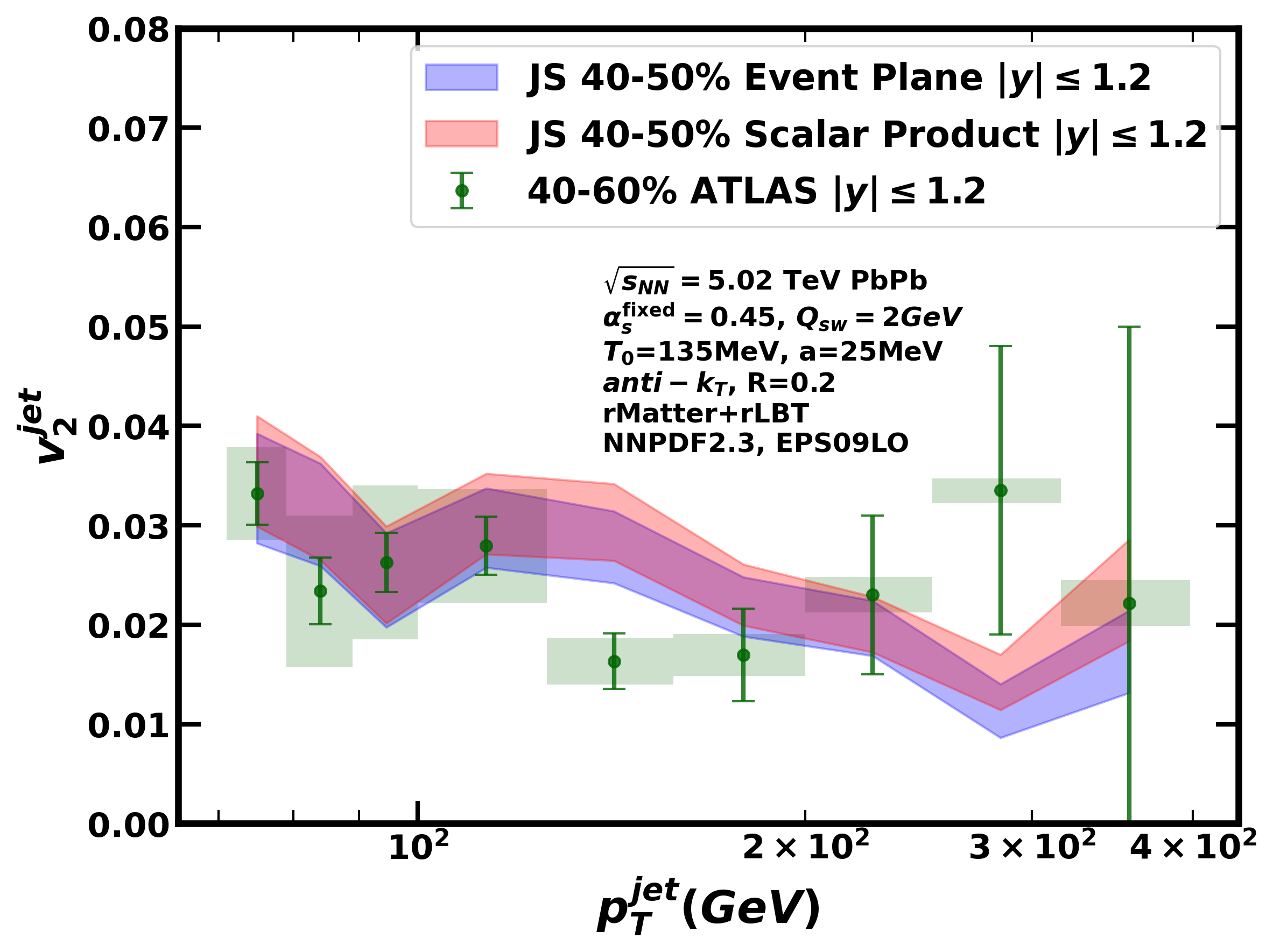}
    \end{subfigure}

    \vspace{-0.15cm}

    \begin{subfigure}[t]{0.48\textwidth}
        \centering
        \includegraphics[width=\linewidth]{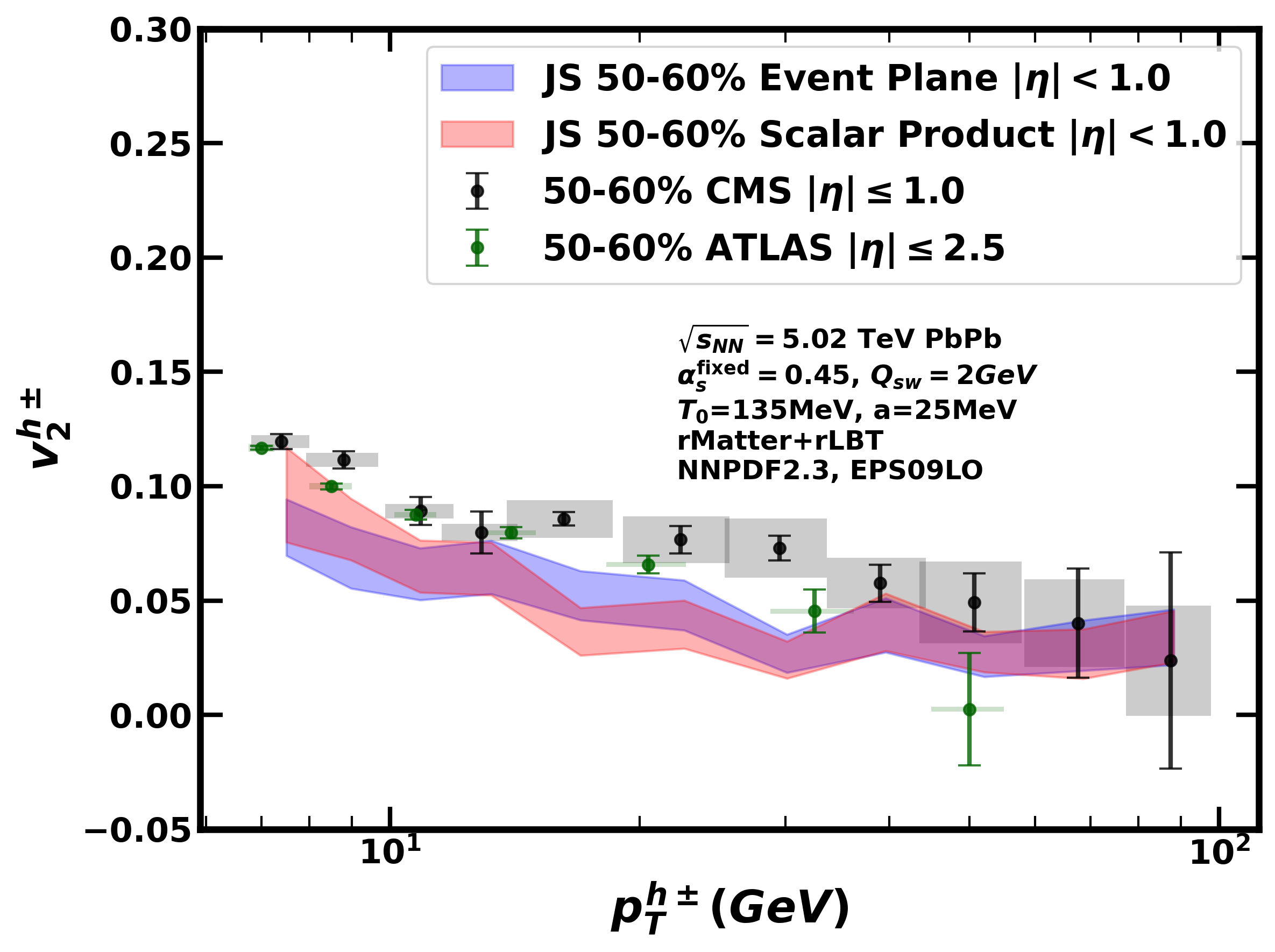}
    \end{subfigure}\hfill
    \begin{subfigure}[t]{0.48\textwidth}
        \centering
        \includegraphics[width=\linewidth]{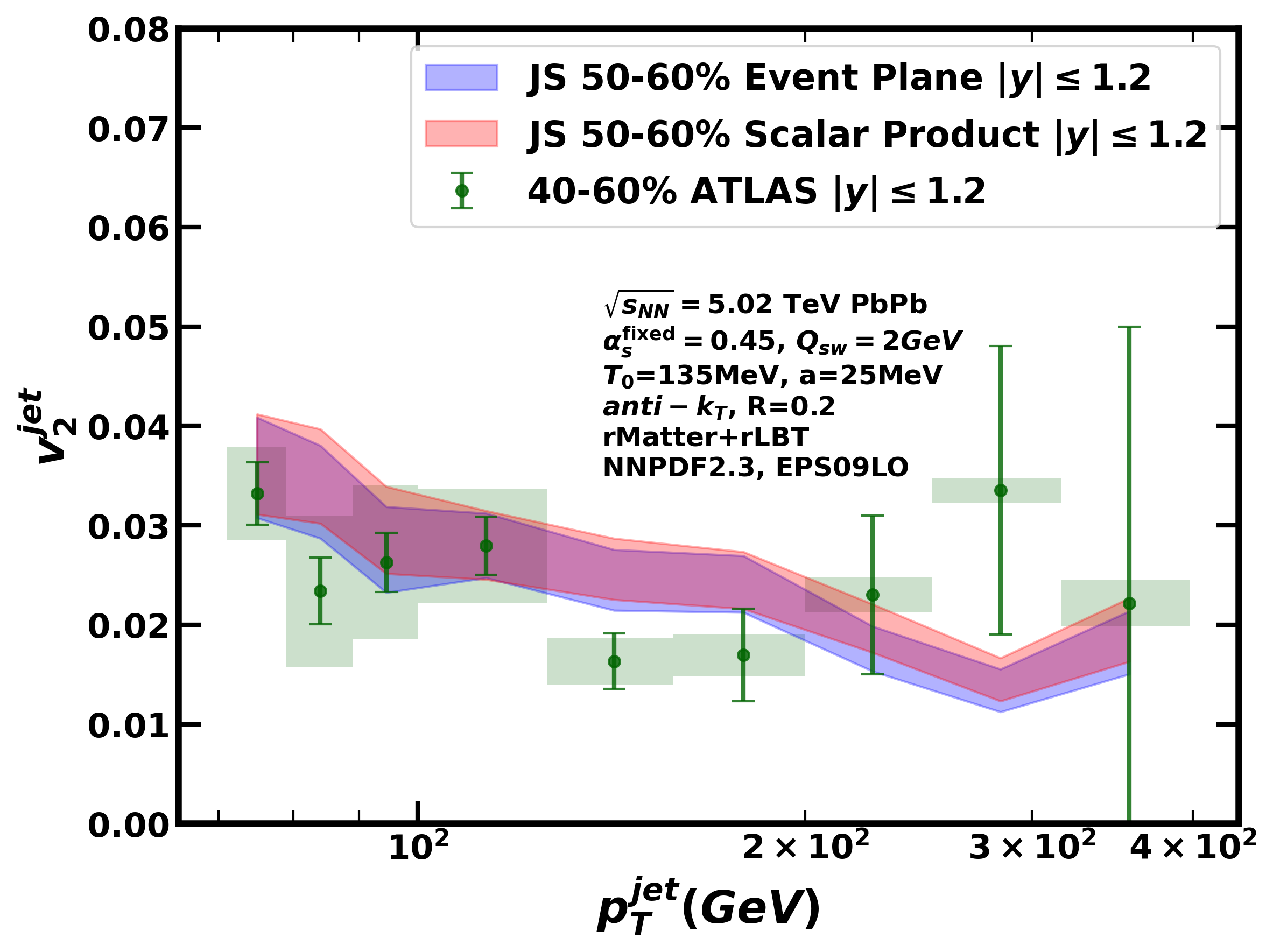}
    \end{subfigure}

    \caption{Elliptic anisotropy coefficient $v_2$ of high-$p_T$ charged hadrons (left) and jets (right, $R=0.2$) as a function of $p_T$, computed using the event-plane (blue band) and scalar-product (red band) methods within the \textsc{Jetscape} reduced (\texttt{MATTER+LBT}) framework + \texttt{EPSO9LO} nuclear shadowing. The results are compared to ATLAS~\cite{ATLAS:2021ktw, ATLAS:2018ezv} (green circles) and CMS~\cite{CMS:2017xgk} (black circles) measurements in 40--50\% (top-left), and 50--60\% (bottom-left) centrality classes (for charged hadrons) and 40--60\% (top and bottom right) centrality classes (for jets) in Pb\,-\,Pb collisions at $\sqrt{s_{NN}} = 5.02~\mathrm{TeV}$. See text for more details.}
    \label{fig:PbPbv2Peipheral}
\end{figure*}

Figure~\ref{fig:PbPbv2Central} shows the elliptic anisotropy coefficient $v_2$ of high-$p_T$ charged hadrons and jets with radius $R=0.2$, in 0--5\%, 20--30\% and 30--40\% events, and Fig.~\ref{fig:PbPbv2Peipheral} shows the $v_2$ of 40--50\% and 50--60\% events, as a function of transverse momentum $p_T$, calculated using both the event-plane (blue) and scalar-product (red) methods. 
The uncertainty band represents the square root of the combined variance in the determination of the elliptic anisotropy from the simulation statistics and the uncertainty associated with the parameter $\sqrt{\lambda}$, which accounts for the assumption of centrality-independent nuclear shadowing. 
The results are compared to measurements from ATLAS (green circles)~\cite{ATLAS:2021ktw, ATLAS:2018ezv} and CMS (black circles)~\cite{CMS:2017xgk} for the multiple centrality classes.

Experimentally, the high-$p_T$ charged-hadron and jet elliptic anisotropy exhibits a non-monotonic dependence on centrality: $v_2$ increases from central to semi-peripheral collisions and decreases again toward more peripheral events. This behavior, arising from the interplay between the geometric anisotropy of the QGP medium and the overall size and density of the system, is well reproduced by the simulation using both the event-plane and scalar-product methods.

In Fig.~\ref{fig:PbPbv2Central}, the left column of plots represents the $v_2$ of charged hadrons starting from most central at the top and more peripheral going down. 
The right column contains the plots of jet-$v_2$. 
In central collisions (0--5\% and 5--10\% most central events), the non-zero elliptic anisotropy generated by event-by-event fluctuations in the initial density profile of the medium is well reproduced for charged hadrons. For jets, the calculated $v_2$ in the 0--5\% centrality class is found to be closer to the experimental values reported for the 5--10\% class. 

In semi-peripheral collisions (20--30\% and 30--40\% most central events), where the path-length difference between in-plane and out-of-plane jet propagation is maximal, the reduced \texttt{MATTER}+\texttt{LBT} framework with nuclear shadowing provides a good description of the measured $v_2$ for both charged hadrons and jets. We point out that experimental data on jet-$v_2$ is only available in the combined centrality class of 20--40\%. Thus, the ATLAS data points in the 2$^{\rm nd}$ and 3$^{\rm rd}$ panel on the right have the same data, while the simulation results in the 2$^{\rm nd}$ panel are for 20--30\% events, while those in the bottom right panel are for 30--40\% most central events. The reader may take a straight average of these two curves to obtain the simulated $v_2$ for 20--40\% events. 

Towards more peripheral collisions (40--50\% and 50--60\% events), in Fig.~\ref{fig:PbPbv2Peipheral}, although the geometric eccentricity of the medium remains large, the smaller system size and reduced medium density lead to a weaker overall energy loss, resulting in a decrease of $v_2$. 
This trend is largely captured by the simulation for both hadrons and jets across the explored $p_T$ range. 
Once again, the charged-hadron $v_2$ is on the left, and the jet-$v_2$ is on the right. We point out again that experimental data on jet-$v_2$ is only available in the combined centrality class of 40--60\%. 
Thus, the ATLAS data points in the top and bottom panel on the right have the same data, while the simulation results in the top right panel are for 40--50\% events, while those in the bottom right panel are for 50--60\% most central events. 
The reader may average over these curves to obtain the simulated $v_2$ for 40--60\% events. 

It is worth noting that the present calculations employ \texttt{colorless} hadronization (a \texttt{Pythia} based module where partons are assigned colors based on their momentum and remnants close to beam rapidity are generated to make color singlets), which is expected to be a good approximation at high transverse momentum. At lower $p_T$ ($p_T \lesssim 10~\mathrm{GeV}$), additional mechanisms such as quark coalescence and recombination~\cite{JETSCAPE:2025wjn}, as well as hadronic rescattering and hadron energy loss in the afterburner phase, are known to play a role in shaping the observed elliptic flow~\cite{Greco:2003xt, Fries:2003vb, Takeuchi:2015ana}, especially for lower $p_T$ hadrons. These effects are not included in the current framework and may contribute to the residual differences observed in the low-$p_T$ region for jets and hadrons below $10$~GeV.


\section{SUMMARY AND OUTLOOK}
\label{sec:summary}


In this work, we investigate the combined effects of nuclear shadowing and a reduced in-medium parton distributions in the \texttt{MATTER+LBT} module within the  \textsc{Jetscape} framework. 
The study is carried out consistently across a wide range of collision energies and system sizes, including Pb\,-\,Pb collisions at $\sqrt{s_{NN}} = 5.02$ and $2.76~\mathrm{TeV}$ at the LHC, as well as Au\,-\,Au collisions at $\sqrt{s_{NN}} = 200~\mathrm{GeV}$ at RHIC. 
For each system and centrality, the space--time evolution of the medium is provided by 2+1D hydrodynamic profiles, which have been previously calibrated to bulk observables, at these energies.

The initial hard scattering processes are generated using \texttt{Pythia8}, which provides the multiparton interaction structure of the event and supplies the hard partons entering the jet-quenching evolution. These partons are first evolved in the high-virtuality stage using the \texttt{MATTER} module, where they are assigned a transverse-momentum–limited virtuality and undergo virtuality-ordered splittings governed by the Sudakov form factor. During this stage, elastic scatterings with the medium can induce additional radiation. Once a parton virtuality drops below $Q_{\rm SW} = 2~\mathrm{GeV}$, the evolution transitions smoothly to the \texttt{LBT} module, where the dynamics becomes scattering and stimulated emission dominated. 

A central element of this work is the consistent modification of the medium distribution entering both the \texttt{MATTER} and \texttt{LBT} stages. 
Instead of the standard Boltzmann (Bose--Einstein or Fermi--Dirac) form, we employ a reduced distribution constructed through an expansion in powers of $1/T$. 
The modification is designed to smoothly recover the standard Boltzmann limit at high temperatures, ensuring continuity with perturbative expectations. 
The modification parameter is tuned to reproduce the qualitative temperature dependence of the jet transport coefficient inferred from lattice QCD~\cite{Kumar:2020wvb}, which exhibits a non-monotonic behavior with a peak around $T \sim 200~\mathrm{MeV}$ followed by a decrease at lower temperatures. 
This behavior contrasts with the hard-thermal-loop expectation, where $\hat{q}/T^3$ continues to rise as the temperature decreases, a regime in which the perturbative description is not expected to be reliable.

An important consequence of the reduced medium distribution is that it naturally enables a consistent treatment of parton interactions in the hadronic phase. 
Partons that retain sufficient virtuality during their evolution are allowed to propagate into the confined phase and interact with partonic degrees of freedom inside hadrons. 
To model this effect numerically, the hydrodynamic evolution is extended below the critical temperature $T_c \simeq 155~\mathrm{MeV}$ to a lower switching temperature of $T_0 = 135~\mathrm{MeV}$. 
At this temperature, the hadron resonance gas remains sufficiently dense to satisfy the Knudsen criterion. 
This switching temperature is applied uniformly across all collision energies and centralities, assuming a universal interaction criterion. 
In rare cases, partons may propagate beyond the $135~\mathrm{MeV}$ hypersurface and continue to radiate, an effect that is already dynamically included in the default \textsc{Jetscape} framework (within the \texttt{MATTER} module). 
These late-stage interactions become particularly relevant in peripheral collisions, where the QGP phase is short-lived and the system spends a comparatively larger fraction of its evolution in the hadronic phase.

From a microscopic perspective, the reduced medium distribution may also be viewed as an effective modification of the parton dispersion relation, entering as a multiplicative factor $(1 + a/T)$ in the exponential of the Boltzmann-like distribution. The impact of this modification is reflected in the scattering rates, channel-wise contributions, longitudinal energy loss, and transport coefficients. Since the same modified distribution is used consistently in both the elastic scattering and medium-induced radiation kernels, the resulting changes affect the full interaction pattern of hard partons with the medium, particularly in the temperature range most relevant for jet quenching.

Following the partonic evolution, the shower partons—together with the recoils and holes introduced to ensure energy--momentum conservation—are hadronized using the colorless Lund string fragmentation model, as implemented in the default \textsc{Jetscape} tune. In parallel, nuclear effects are incorporated at the level of the initial hard scattering through nuclear parton distribution functions. Nuclear shadowing for Pb and Au systems is implemented using \texttt{EPS09LO}. In the transverse-momentum range explored in this work, the modification of the initial hard spectra due to shadowing necessitates an increase in the effective in-medium coupling $\alpha_s$. Moreover, because the reduced distribution suppresses the medium density at lower temperatures and reduces the average path-integrated $\hat{q}$, a further increase in the in-medium $\alpha_s$ is required to reproduce the observed suppression.

To obtain sufficient statistical precision for high-$p_T$ observables, particularly for the elliptic flow coefficient $v_2$, events are generated using the $\hat{p}_T$-binning method to adequately populate the hard-scattering phase space. To account for nuclear shadowing effects in the overall normalization, a parameter $\lambda$ is introduced to scale the total inelastic cross section from $p$-$p$ in vacuum to nucleon--nucleon collisions within large nuclei. The value of $\lambda$ is constrained using $p$-Pb jet and hadron $R_{AA}$ measurements. The associated uncertainty in $\lambda$ (caused due to a slight tension between CMS hadrons and ATLAS jet data in $p$-Pb) is estimated through a simultaneous description of charged-hadron and jet data in $p$-Pb collisions at $\sqrt{s_{\rm NN}} = 5.02~\mathrm{TeV}$.

Taken together, the extension of the reduced \texttt{MATTER+LBT} framework with modified medium distributions and nuclear effects provides a unified and consistent description of charged-hadron and jet suppression across both LHC and RHIC energies. The combined effects of increased $\alpha_s$ due to shadowing and the inclusion of hadronic-phase interactions introduce additional path-length dependence for high-$p_T$ probes that was absent in the default tune. This enhanced sensitivity to the medium geometry plays a central role in achieving a simultaneous description of charged-hadron and jet $v_2$ (and $R_{AA}$) at $\sqrt{s_{\rm NN}} = 5.02~\mathrm{TeV}$.

The current effort, should be seen as an extension of Ref.~\cite{JETSCAPE:2022jer}, to more peripheral events, and including jet and hadron $v_2$ simultaneously with jet and hadron $R_{AA}$. 
Almost all details of the simulation, and all except 2 of the parameters used in Ref.~\cite{JETSCAPE:2022jer} are left unchanged. 
The new parameter $a$ is dialed to obtain a $\hat{q}/T^3$ similar to that in Ref.~\cite{Kumar:2020wvb}, the minimum temperature of the medium is lowered from the vicinity of $T_c$ to $T_0 = 135$~MeV, and the in-medium coupling $\alpha_S^{\rm fixed}$ is dialed to fit one data set in central Pb\,-\,Pb at $\sqrt{s_{\rm NN}} = 5.02~\mathrm{TeV}$. All other plots, with varying centrality and collision energy, for jet and hadron $R_{AA}$ and $v_2$, are predictions.
Looking ahead, the present framework provides a natural starting point for more systematic constraints on medium properties using modern statistical techniques. In particular, a Bayesian calibration of the model parameters, including the modification strength ($a$) and effective coupling, using a combined set of charged-hadron, jet, and anisotropy observables across multiple collision energies would allow for a quantitative assessment of parameter correlations and uncertainties. Such an approach would also help further disentangle the relative roles of medium modification, nuclear shadowing, and hadronic-phase interactions.

As mentioned earlier, the current work does not include the effects of hadronization modifications at intermediate $p_T$ from recombination effects, and does not include hadronic energy loss in the hadronic medium. The addition of these mechanisms via the new hybrid hadronization~\cite{JETSCAPE:2025wjn}, and hadronic cascade~\cite{SMASH:2016zqf} modules, within the \textsc{Jetscape} framework, will allow for an extension of jet and hadron fits, and any future Bayesian calibration, to lower $p_T$. 

Yet another aspect, highlighted but not explored in this work, is the role of impact parameter dependent shadowing~\cite{Deng:2010xg,Li:2001xa}. We have used \texttt{EPS09LO} shadowing throughout this work, which does not have any impact parameter dependence. Our plots, typically the most peripheral $R_{AA}$, seem to indicate a slight role of impact parameter dependence in shadowing. A complete study of this aspect is beyond the scope of the current effort. Our goal was to demonstrate that the inclusion of nuclear shadowing, and a reduced thermal distribution allowing for partonic energy loss in the hadronic phase, leads to a marked improvement in the simultaneous description of jet and hadron $R_{AA}$ and $v_2$.

An additional and particularly sensitive probe of the modified medium description considered in this work may be provided by thermal photons and dileptons. It is expected that photon and dilepton yields, spectra, and azimuthal correlations will be influenced by modifications to the medium distribution in this temperature range. In this context, photon anisotropies and photon--hadron correlations offer a complementary avenue to test the interplay between medium dynamics and partonic interactions around the transition temperature. A simultaneous description of hadronic observables and electromagnetic measurements would therefore provide a stringent and independent validation of the reduced distribution conjecture.

All calculations presented in this paper can be reproduced using code that can be cloned from
\href{https://github.com/JETSCAPE/JETSCAPE/tree/jetmod}{\texttt{https://github.com/JETSCAPE/JETSCAPE/tree/jetmod}}. Instructions for installing and running the \textsc{Jetscape} package, required to run this code, are available at \href{https://github.com/JETSCAPE}{\texttt{https://github.com/JETSCAPE}}. 

\begin{acknowledgments}
 The authors thank Chun Shen for providing the fluid dynamical profiles used in this work. Both authors are members of the \textsc{Jetscape} collaboration and have benefited from extensive discussion with other members of the collaboration. 
 This work was supported in part by the U.S. Department of Energy (DOE) under grant number DE-SC0013460. The work of R. D. was also partially supported by the National Science Foundation (NSF) under grant number OAC-2004571, within the framework of the \textsc{Jetscape} collaboration. 
\end{acknowledgments}

\bibliography{refs}
\end{document}